\documentclass[]{aa} 
\usepackage{graphicx,txfonts,amssymb,natbib,epsfig}
\renewcommand{\d}{\mathrm{d}} 
\authorrunning{M. Meneghetti et al.}
\titlerunning{Weighing simulated galaxy clusters using lensing and X-ray} 

\begin{document}

\title{Weighing simulated galaxy clusters using lensing and X-ray}

 \author{Massimo Meneghetti\inst{1,2}, Elena Rasia\inst{3},  Julian Merten \inst{5}, Fabio Bellagamba \inst{7} ,Stefano Ettori\inst{1,2}, Pasquale Mazzotta \inst{4}, Klaus Dolag \inst{6}}

\institute {$^1$ INAF-Osservatorio
  Astronomico di Bologna, Via Ranzani 1, 40127 Bologna, Italy \\ 
  $^2$ INFN-National Institute for Nuclear Physics, Sezione di Bologna, Viale Berti Pichat 6/2. 40127, Bologna, Italy \\
  $^3$ {\it Chandra} Fellow, Fellow of Michigan Society of Fellows
  $^4$ Department of Physics, University of Michigan, 450 Church St., Ann Arbor, MI 48109-1120, USA \\
  $^4$ Dipartimento di Fisica, Universit\`a ``Tor Vergata", Via della Ricerca Scientifica 1, 00133 Roma, Italy \\
  $^5$ Institut f\"ur Theoretische Astrophysik, Zentrum f\"ur Astronomie der Universit\"at Heidelberg, Albert \"Uberle Str. 2, 69120, Heidelberg, Germany 
  \\
  $^6$ Max-Planck-Institute f\"ur Astrophysik, POBox 1317, 85741, Garching b. M\"unchen, Germany \\
  $^7$ Dipartimento di Astronomia, Universit\`a di Bologna,
  Via Ranzani 1, I-40127 Bologna, Italy\\
  }

\date{\emph{Astronomy \& Astrophysics, submitted}}

\abstract {Measuring the mass of galaxy clusters is a key issue in cosmology. Among the methods employed to achieve this goal, the techniques based on lensing and X-ray analyses are perhaps the most widely used. However, the comparison between these mass estimates is often difficult and, in several clusters, the results apparently are inconsistent.} {We aim at investigating potential biases in lensing and X-ray methods to measure the cluster mass profiles.}  {We do so by performing realistic simulations of lensing and X-ray observations that are subsequently analyzed using observational techniques. The resulting mass estimates are compared among them and with the input models. {  Three clusters obtained from state-of-the-art hydrodynamical simulations, each of which has been projected along three independent lines-of-sight, are used for this analysis.}} {We find that strong lensing models can be trusted over a limited region around the cluster core. Extrapolating the strong lensing mass models to outside the Einstein ring can lead to significant biases in the mass estimates{ , if the BCG is not modeled properly for example}. Weak lensing mass measurements can be largely affected by substructures, depending on the method implemented to convert the shear into a mass estimate. Using non-parametric methods which combine weak and strong lensing data, the projected masses within $R_{200}$ can be constrained with a precision of $\sim10\%$. De-projection of lensing masses increases the scatter around the true masses by more than a factor of two due to cluster triaxiality. X-ray mass measurements have much smaller scatter (about a factor of two smaller than the lensing masses) but they are generally biased low by $5-20\%$. This bias is entirely ascribable to bulk motions in the gas of our simulated clusters. Using the lensing and the X-ray masses as proxies for the true and the hydrostatic equilibrium masses of the simulated clusters and  by averaging over the cluster sample we are able to measure the lack of hydrostatic equilibrium in the systems we have investigated.} {Although the comparison between lensing and X-ray masses may be 
difficult in individual systems due to triaxiality and substructures, using a large number of clusters with both lensing and X-ray observations may lead to important information about their gas physics and allow to use lensing masses to calibrate the X-ray scaling relations.}


\maketitle

\section{Introduction}
Galaxy clusters are highly important test sites for cosmology. They are the most massive gravitationally bound structures in the universe and, in the framework of the hierarchical structure formation scenario, they are also the youngest systems formed  to date. For this reason, the interplay between baryons and dark-matter, and its effects on the cluster internal structure, is less important in these than in smaller and older objects. Thus, they are ideal systems for testing the predictions of the cold-dark-matter paradigm on the internal structure of dark-matter halos \citep{2000ApJ...544L..87Y,CL04.1,2004ApJ...606..819M,2008ApJ...674..711S}. Moreover, their mass function is highly sensitive to cosmology, since its evolution traces with exponential magnification the growth of the linear density perturbations \citep{PR74.1,SH02.1,JE01.1,2006ApJ...646..881W}.
    
For these reasons galaxy clusters are being used to constrain the cosmological parameters. 
Cluster masses are used to measure the time evolution of the mass function, which is compared to the theoretical predictions to constrain the contribution of the components of the universe to the overall density parameter, the equation of state of dark energy and the normalization of the power spectrum of the initial density fluctuations \citep[e.g][]{2009ApJ...692.1060V,2009arXiv0911.1788M,2009arXiv0909.3098M}. The measurements can be compared with those obtained from the observation of the universe on larger scales, and combined with CMB \citep[see e.g.][]{2009ApJS..180..330K}, Supernovae Ia \citep[e.g.][]{RI98.1,RI04.1, PE99.1}, and Baryonic-Acoustic-Oscillations observations \citep[e.g.][]{2005ApJ...633..560E,2007MNRAS.381.1053P}, in order to put tighter limits to the values of the cosmological parameters.
A different approach consists of measuring the concentration-mass relation of clusters and of comparing it with the theoretical predictions \citep[e.g.][]{2005A&A...435....1P,2005ApJ...628..655V,2007ApJ...664..123B,2007MNRAS.379..209S,2007MNRAS.379..190C}.  For example, Buote et al. (2007) find that the concentration-mass relation measured with a sample of 39 clusters is good agreement with the expectations in the framework of the concordance model.

\cite{2008MNRAS.383..879A} use the cluster gas fraction  to constrain the time evolution of the dark energy component of the universe. The gas fraction measured within a given angular radius is proportional to the distance of the cluster to the power 1.5. Thus, the evolution of the gas fraction can be used to measure the cosmic acceleration \citep{2004MNRAS.353..457A}. Combining the results from the $f_{gas}$ technique with CMB and SNIa data sets, they find that the time evolution of the dark energy equation of state is compatible with the cosmological constant paradigm \citep[see also][]{2003A&A...398..879E}.

These techniques rely on scaling relations which link the mass to X-ray observables, such as temperature, pressure and  luminosity of the X-ray emitting intra-cluster gas. The scaling relations are predicted to have limited scatter in mass. For example, simulations suggest that the mass-$T_X$ and the mass-$Y_X$ relations have just $15\%$ and $8\%$ scatter in mass at fixed $T_X$ and $Y_X$ \citep[e.g.][]{1996ApJ...469..494E,2006ApJ...650..128K}. Despite these encouraging predictions, it is clearly essential to measure and calibrate scaling relations empirically. To this purpose, we need to accurately measure the mass profiles of an as large as possible sample of galaxy clusters.

There are several methods to derive the masses of clusters. Two widely used approaches are based on X-ray and lensing observations. X-ray observations allow to derive the cluster mass profiles assuming that these systems are spherically symmetric and that the emitting gas is in hydrostatic equilibrium \citep[e.g.][]{1986ApJ...302..287H,1988xrec.book.....S,2002A&A...391..841E}. This method has the advantage that, since the X-ray emissivity is proportional to the square of the electron density, it is weakly sensitive to projection effects due to masses along the line of sight to the clusters. However, it is still not well established how safely the hydrostatic equilibrium approximation can be made.
 
Being the largest mass concentrations in the universe, galaxy clusters are the most efficient gravitational lenses on the sky. Their matter distorts background-galaxy images with an intensity which increases from the outskirts to the inner regions. Strong distortions, leading to the formation of ``gravitational arcs'' and/or to the formation of systems of multiple images of the same source, occur in the cores of some massive galaxy clusters. Weak distortions, which can be measured only statistically, are impressed on the shape of distant galaxies which lay on the sky at large angular distances from the cluster centers \citep[e.g.][]{BA01.1}.
Both these lensing regimes can be used to map the mass distribution in galaxy clusters. Determining the masses and the density profiles using lensing offers several advantages compared to X-ray observations. First, lensing directly probes the cluster total mass, including the dark matter component, without need to make strong assumptions on the equilibrium state of the lens. Second, mass profiles can be measured over a wide range of scales, from $\lesssim 100$ kpc out to the virial radius. The biggest disadvantage is that lensing measures the projected mass instead of the three-dimensional mass. It is much more sensitive than X-ray methods to projection effects, such triaxiality and additional concentrations of mass along the line of sight. Given the pros and cons of each method, we can conclude that lensing and X-ray are complementary to each other in many ways. In particular, the comparison of these two mass estimates can greatly help to improve the accuracy of the measurements and to understand the systematic errors.

The picture arising from the comparison of lensing and X-ray mass estimates is puzzling. The two estimates are often discrepant, which implies that we may be missing important ingredients for fully understanding the properties of galaxy clusters. A systematic discrepancy has
been revealed in the sense that masses derived from strong lensing are
typically larger by a factor 2-3 than masses derived from weak lensing
and from the X-ray emitting intra-cluster-medium {  (ICM)} \citep[e.g.][]{WU96.1,OT04.1,2005MNRAS.359..417S,2009ApJ...693.1570R}. The comparison between weak lensing and X-ray mass estimates is less problematic \citep[see e.g.][]{2003A&A...398L...5E,2001MNRAS.324..877A} but still the lensing masses are on average larger than the X-ray masses by $\sim 15-20\%$ \citep[e.g.][]{2007MNRAS.379..317H,2008A&A...482..451Z}.  
Explaining this discrepancy was found to be difficult. It was proposed
that X-ray masses could be systematically biased because of mergers or bulk motions of the gas that alter the state of hydrostatic equilibrium \citep{BA96.2,AL98.1,2008MNRAS.384.1567M,2008A&A...491...71P,2009MNRAS.394..479A,2009ApJ...705.1129L}. As stated above, lensing masses could also be affected by significant uncertainties  \citep[e.g.][]{BA95.2,2007A&A...461...25M}.  

In this paper, we study the systematic effects in mass measurements done with standard lensing and X-ray techniques. Our work extends some previous works on the systematics on X-ray based mass measurements \citep[see e.g.][]{2006MNRAS.369.2013R,2007ApJ...655...98N}, which used synthetic X-ray observations of clusters obtained from hydrodynamical simulations to recover their mass distribution under the assumption of hydrostatic equilibrium. Here, we simulate both optical and X-ray observations {\em of the same simulated clusters}, in order to be able to compare the mass estimates obtained from both kind of observations. {  We use three clusters obtained from hydrodynamical simulations and study them along three independent lines-of-sight.} 

The paper is organized as follows. In Sect.~\ref{sect:numsim} we describe the sample of numerically simulated galaxy clusters, as well as the techniques used to simulate lensing and X-ray observations. In Sect.~\ref{sect:analyses}, we introduce the methods through which the mass estimates are derived from the mock data. In Sect.~\ref{sect:resu} we show the results of the analyses. Finally, in Sect.~\ref{sect:summary} we summarize and discuss the major findings of our study.

\section{Simulations}
\label{sect:numsim}
{  In this Sect. we describe the numerical methods used in the simulations. We start by considering the N-body/hydrodynamical simulations from which the cluster models are obtained. Then, we explain the lensing and the X-ray simulation pipelines.}

\subsection{N-body/SPH simulations} 

The clusters used in this work are $g1$, $g51$, and $g72$ from the sample of numerical hydrodynamical simulations presented by \cite{2006MNRAS.373..397S}. These objects have been already used in several other studies \citep{DO05.1,PU05.1,2007A&A...461...25M,2008A&A...482..403M,2006MNRAS.369.2013R,2008ApJ...674..728R}. Thus, we refer the reader to these papers for more details. They are extracted from a parent simulation of only dark matter \citep{YO01.1} with a box size of $479 h^{-1}$ Mpc of a flat $\Lambda$CDM model with $\Omega_m=0.3$ for the present matter density parameter, $h=0.7$  for the Hubble constant in units of 100 km $s^{-1}$ Mpc$^{-1}$, $\sigma_8=0.9$ for the rms fluctuation within a top-hat sphere of $8 h^{-1}$Mpc radius, and $\Omega_b=0.04$ for the baryon density parameter.  {  Each of them is subsequently re-simulated at higher mass and spatial resolution using the Zoomed initial condition method \citep{TO97.2}.} 
{  The re-simulations are carried out with 
{\small GADGET-2} 
\footnote{http://www.MPA-Garching.MPG.DE/gadget/} \citep{SP05.1}. In the code we include (i) a description of several physical processes in the ICM (see \citealt{2006MNRAS.373..397S}), (ii) a numerical scheme to suppress artificial viscosity far from the shock regions \citep[see][]{DO05.1}, and (iii) a treatment of chemical enrichment from both SNIa , SNII, as well as from low and intermediate mass stars \citep{2004MNRAS.349L..19T,2007MNRAS.382.1050T}.
Our simulations 
assume the 
power-law shape for initial stellar mass function, as proposed by 
\cite{1955ApJ...121..161S}, and galactic ejects with a speed of 500 km s$^{-1}$. 
They start with a gravitational softening length fixed at $\epsilon=30 h^{-1}$kpc comoving (Plummer-equivalent) and switch to a physical softening length of $\epsilon=5 h^{-1}$kpc at $z=5$. The final masses of the DM and gas particles are set to $m_{\rm DM} = 1.13\times 10^{9} h^{-1} M_\odot$ and $m_{\rm GAS} = 1.7\times 10^{8} h^{-1} M_\odot$, respectively. }

During the simulation 92 time slices are saved from redshift 60 to 0. These are equidistant in time.  For the current work, we use the snapshots corresponding to redshift $z_{\rm l} = 0.297$ for $g1$ and $g72$ and to redshift $z_{\rm l}=0.2335$ for $g51$. These redshifts are optimal for both the X-ray and the lensing analyses, since 1) the X-ray surface brightness of these objects is high, and 2) the strength of the lensing signal for sources at $z_{\rm s} \sim 1$, which is approximately the median redshift of the sources in our simulations, is maximal at $z_{\rm l} \sim 0.3$. All clusters are massive, with masses $M_{200}$ ranging between $\sim 7\times10^{14}\;h^{-1}M_\odot$ and $\sim 1.2\times10^{15}\;h^{-1}M_\odot$ (see Table~\ref{tab:mainprop}). {  $M_{200}$ is the mass enclosed by the radius $r_{200}$, i.e. the radius within which the average density is 200 times the critical density,} 
\begin{equation}
	\rho_{\rm cr}	= \frac{3 H^2(z)}{8 \pi G} \;,
\end{equation}
where $H(z)$ is the Hubble parameter.

\begin{figure*}[t!]
 \includegraphics[width=1.0\hsize]{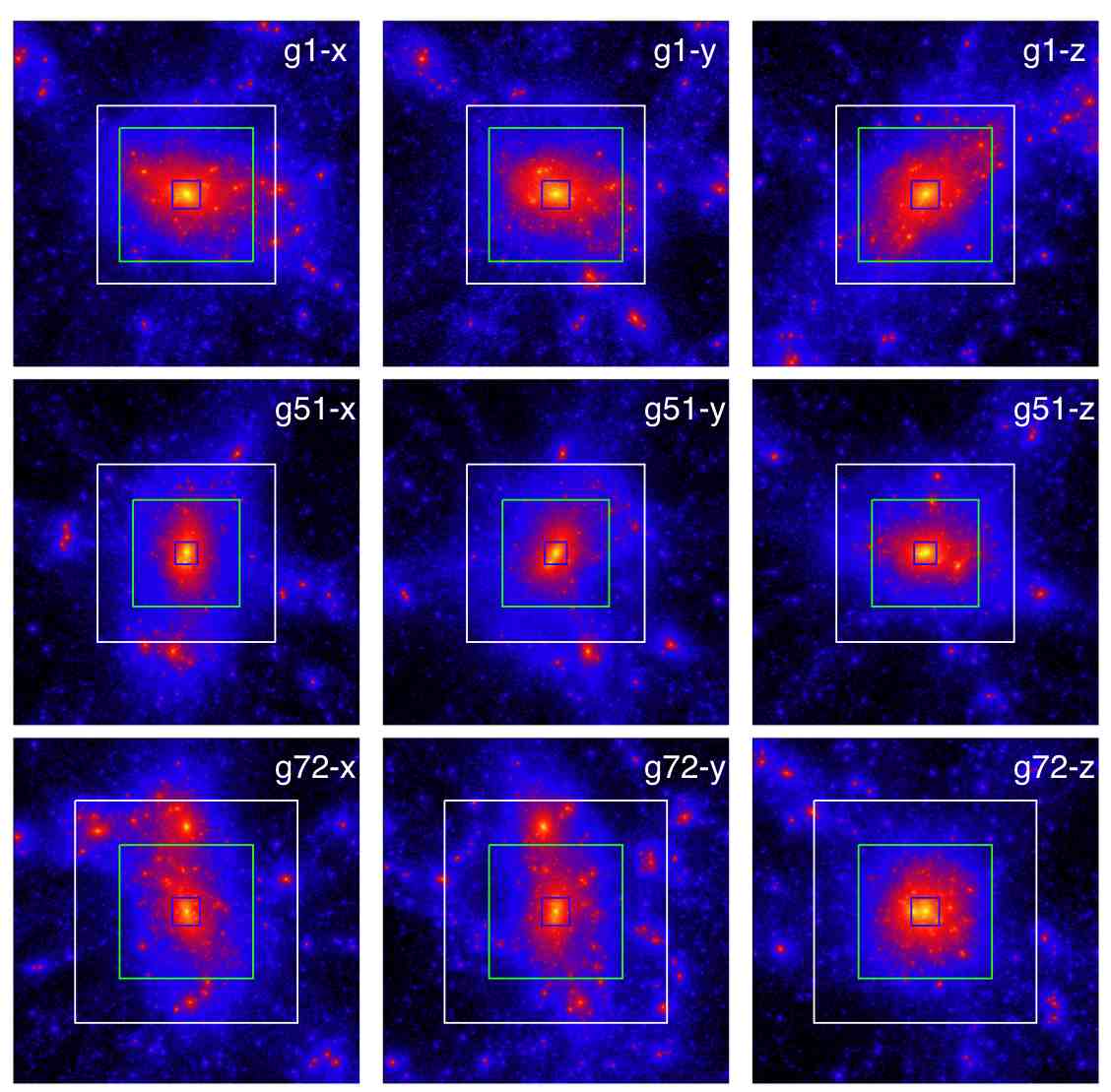}
  \caption{Surface density maps of the three projections of the clusters $g1$ (top panels), $g51$ (middle panels) and $g72$ (bottom panels) used in this work. The size of each map is $10\;h^{-1}$Mpc comoving. {  Such a scale corresponds to $\sim51.85$ arcmin for $g51$ and to $\sim41.47$ arcmin for $g1$ and $g72$}. In each panel, the inner, middle, and outer boxes indicate the fields-of-view used for the HST, {\it Chandra}, and SUBARU simulated observations, respectively (see text for more details).}
\label{fig:sdmaps}
\end{figure*}  

\begin{figure*}[t!]
  \includegraphics[width=1.0\hsize]{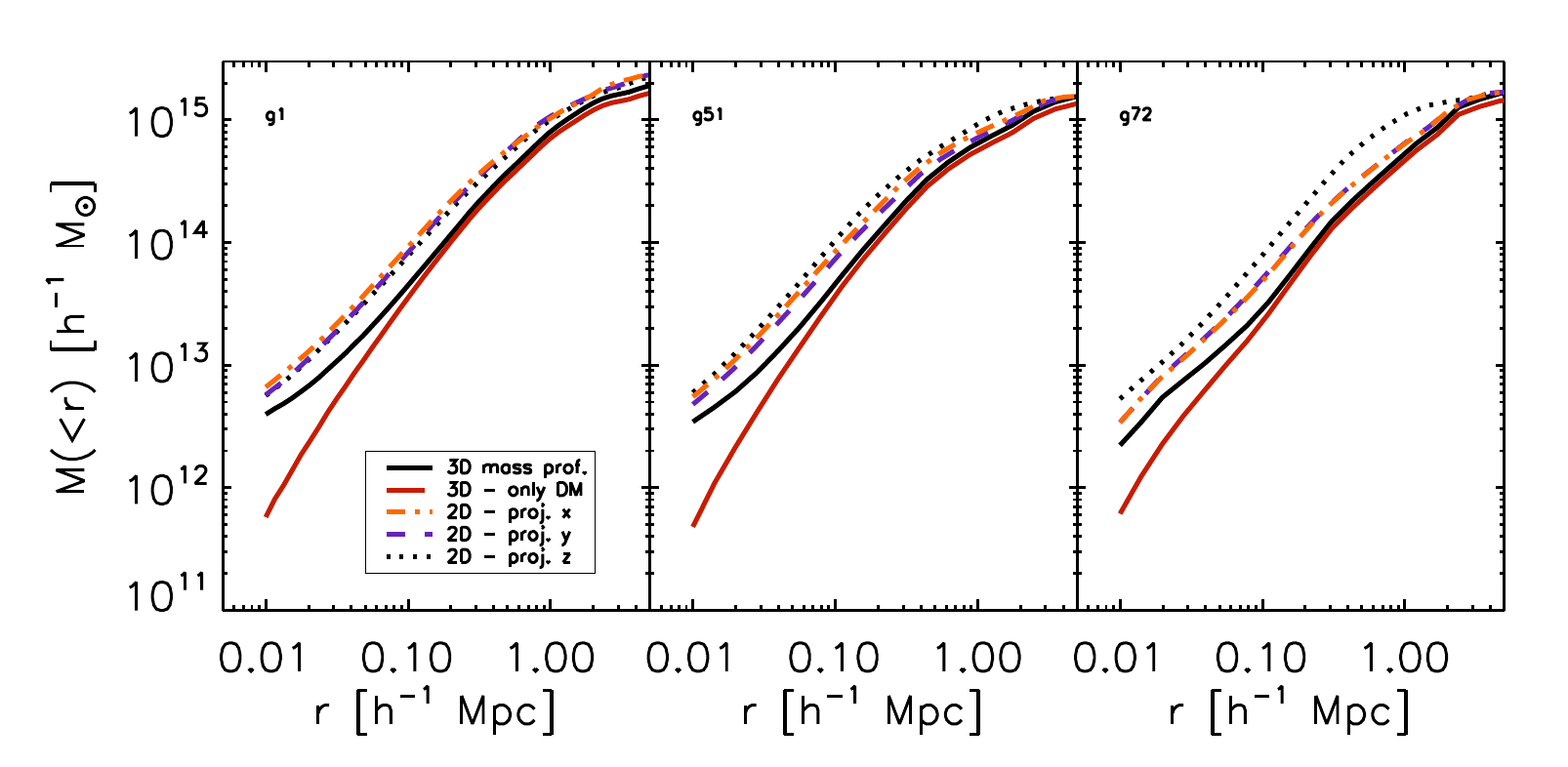}
  \caption{Mass profiles of the clusters $g1$, $g51$, and $g72$. The solid black and red lines indicate the total and DM only 3D-mass profiles, respectively. The total 2D-mass profiles corresponding to the $x$, $y$, and $z$ projections of each cluster are given by the dotted, dashed, and dash-dotted lines.}
\label{fig:mprofs}
\end{figure*}

For each cluster, we select a cube of $20 h^{-1}$Mpc side length (comoving), centered on the most bound dark matter particle. This is sufficiently large to cover a wide field of view, needed for weak lensing simulations. All the matter contained into this box is projected along three orthogonal lines of sight, in order to produce three lens planes for each cluster. The surface density maps corresponding to the three projections of the clusters $g1$, $g51$, and $g72$ are shown in Fig.~\ref{fig:sdmaps}. The corresponding mass profiles are displayed in Fig.~\ref{fig:mprofs}. Note that the sample of clusters considered here comprises object with different morphologies, triaxialities, and levels of substructures, although in the X-ray they all appear quite relaxed (see Fig.~\ref{fig:xraymaps}). While the cluster $g1$ has the most regular morphology, the cluster $g72$ has a massive companion ($\sim 10^{14}\;h^{-1}M_\odot$) located at $\sim 2.5\;h^{-1}$Mpc from the cluster center.
 In the projection along the $z-$axis, this secondary clump is much closer to the main halo ($\sim 300\;h^{-1}$kpc). Also, the cluster $g51$ has some substructures  within the inner $500\;h^{-1}$kpc but their masses are smaller ($\lesssim 5\times 10^{12}\;h^{-1}M_\odot$). A massive clump of mass $5\times 10^{13}\;h^{-1}M_\odot$ orbits at a distance of $3 \;h^{-1}$Mpc from the center. 

As demonstrated by the differences between the 2D-mass profiles, all three clusters are triaxial.  Their shape is prolate, with the major axis oriented nearly along the $z-$axis of the simulation box in the cases of $g51$ and $g72$, and nearly along the $x-$axis in the case of $g1$. The axis ratios, measured at $r_{200}$ by calculating the cluster inertia ellipsoids, are listed in Table~\ref{tab:mainprop}, together with some other relevant properties. We also report there the angles between the main axes of the clusters and the $x-$, $y-$, and $z-$axes of the simulation boxes. 

The clusters are well described by Navarro-Frenk-White {  (NFW)} density profiles \citep{NA97.1}, whose functional form is given by
\begin{equation}
	\rho_{\rm NFW}(r)=\frac{\rho_s}{r/r_s(1+r/r_s)^2}
	\label{eq:nfw}		
\end{equation}
where $\rho_s$ and $r_s$ are the characteristic density and the scale radius, respectively. The characteristic density is often written in terms of the concentration parameter, $c_{200}=r_{200}/r_{s}$, as
\begin{equation}
  \rho_s=\frac{200}{3}\rho_{\rm cr}\frac{c_{200}^3}{[\ln(1+c_{200})-c_{200}/(1+c_{200})]} \ .
  \label{equation:deltacpar}
\end{equation}
The last two columns of Table~\ref{tab:mainprop} report the best-fit concentrations and scale-radii obtained by fitting the 3D dark-matter density profiles of the clusters with the formula in Eq.~\ref{eq:nfw}. {  The radial fits are done between $10 h^{-1}$kpc and $r_{200}$.}

{  In the remainder of the paper}, we will refer to the projections of cluster $g$N along the simulation axis X using the abbreviation {  $g$N-X} .

\begin{table*}[htdp]
\caption{Main properties of the simulated clusters used in this work. {  Column 1: cluster name; Column 2: redshift; Column 3: $r_{200}$; Column 4: $M_{200}$; Columns 5-6: principal axes ratios: $b/a$, $c/a$, where $a>b>c$; Columns 7-9: angles between the main principal axis and the $x-$, $y-$, and $z-$axes of the simulation box; Column 10: best-fit 3D-concentration; Column 11: best-fit 3D-scale radius}}
\begin{center}
\begin{tabular}{|l||c|c|c|c|c|c|c|c|c|c|c|}
\hline
\hline
cluster & $z$ & $r_{200}$ & $M_{200}$  & $b/a$ & $c/a$ & $\theta_x$ & $\theta_y$ & $\theta_z$ & $c_{200}$ & $r_s$ \\
	& & [$h^{-1}$ Mpc] & [$h^{-1} M_\odot$] &  & & [deg] & [deg] & [deg] & & [$h^{-1}$ Mpc] \\
\hline
g1  & 0.297 & 1.54 & $1.14\times10^{15}$ & 0.64 & 0.57 & 33.3 & 57.4 & 96.1 & 4.62 & 0.310 \\ 
g51 & 0.2335 & 1.39 & $7.85\times10^{14}$ & 0.78 & 0.65 & 81.5 & 75.59 & 16.8 & 5.37 & 0.241 \\
g72 & 0.297 & 1.30 & $6.83\times10^{14}$ & 0.31 & 0.29 & 98.9 & 92.8 & 9.4 & 3.99 & 0.299 \\
\hline 
\hline
\end{tabular}
\end{center}
\label{tab:mainprop}
\end{table*}% 

\subsection{Lensing simulations}
In this section we describe the ray-tracing algorithms used to derive the lensing distortion fields of our simulated clusters. For a summary of the lensing definitions used through the rest of the paper, {  we refer the reader to Appendix~\ref{sect:lensing}. } 

In the following, we make use of the standard thin-screen approximation, i.e. we assume that the deflections occur on a plane perpendicular to the line-of-sight, passing through the cluster center. This is justified since the distances between the observer and the lens and between the lens and the background sources are much larger than the sizes of the clusters. 
 
The particles projected on each lens plane are used to calculate the deflection angles of bundles of light rays.  The light rays are traced from the observer position towards the background sources through two regular grids with different spatial resolutions. The inner $1.5 \times 1.5\,h^{-2}$Mpc$^2$ region around the cluster center is sampled with $2048 \times 2048$ light rays. This guaranties sufficient spatial resolution for reproducing accurately the positions of multiple images in the strong lensing (SL) regime \citep{2007A&A...461...25M}. For the weak lensing (WL) regime, we need to sample a much wider area, while the spatial resolution is less important. Thus, we cover the whole lens plane with a grid of $4096 \times 4096$ light rays.  The deflection angles are computed using a tree-based code, which works as follows. First, it ranks the particles based on their distances from the light ray positions, building a Barnes \& Hut oct-tree in two dimensions. The contributions to the deflection angles from nearby and distant particles are calculated separately using direct summation or higher-order Taylor expansions of the deflection potential around the light-ray positions. Precisely, given a light ray at position $\vec R$ in physical units, which corresponds to an angular position $\vec \theta=\vec R/D_{l}$, the contribution to its deflection angle by a system of mass elements, $m_a$ at positions $\vec R_a$ ($a=1,2,...,N-1,N$), with center of mass $\vec R_{CM}$, and with $|\vec R-\vec R_{CM}|\ll|\vec{ R}_a-\vec R_{CM}|$ for all the mass elements $a$, is    
\begin{eqnarray}
\alpha_{i}(\vec{ R}) &=&\frac{4GM}{c^2} \left [ F_{1}( R')\delta^{ij} +
F_{2}( R')Q^{ij} 
+ \frac{1}{2}F_{3}( R')( R'^{k}Q_{kn} R'^{n})\delta^{i,j}
\right. \nonumber 
\\
&+& \left.\frac{1}{2}F_{4}( R')P^{ij}  \right ]  R'_{j}
\label{eq_alphatree}
\end{eqnarray}
where $M$ is the total mass of the system, $\vec R'=\vec R-\vec R_{CM}$,
$\delta^{ij}$ is the Kroneker function, and the tensors $P$ and $Q$ are
defined as:
\begin{eqnarray}
Q_{ij}&=&\frac{1}{M} \sum_{a=1}^{a=N} m_{a} R'^{a}_{i} R'^{a}_{j}, \\
P_{ij}&=&\frac{1}{M} \sum_{a=1}^{a=N} m_{a}| R'^{a}|^{2}\delta_{ij}.
\end{eqnarray}
Assuming a Plummer softening to avoid that the deflection angles diverge, the $F_k( R')$ functions
are defined as:
\begin{eqnarray}
F_{1}( R')=\frac{1}{( R'^2+s^2)} \label{eq:plummer}\\
F_{2}( R')=\frac{-2}{( R'^2+s^2)^2} \\
F_{3}( R')=\frac{8}{( R'^2+s^2)^3} \\
F_{4}( R')=\frac{-2}{( R'^2+s^2)^2} \;.
\end{eqnarray}
Nearby particles are treated as point lenses and Eq.~\ref{eq_alphatree} reduces to
\begin{equation}
\alpha_i(\vec R) = \frac{4GM}{c^2} R'_iF_1(R') \;.
\label{eq:directsum}
\end{equation}
The fraction of of particles which are evaluated with Eqs.~\ref{eq_alphatree} or \ref{eq:directsum} is set by the Barnes-Hut opening criterion, $\theta_{BH}$ \citep[see e.g][]{SP05.1}, which we fix at $\theta_{BH}=0.4$. 
As shown by \cite{AU07.1}, the optimal softening length $s$ depends on the resolution of the simulation. We performed several tests to determine which values to use. Doing ray-tracing through NFW halos sampled with a similar number of particles as our simulated clusters, we verified that a softening scale of $5\,h^{-1}$kpc is appropriate for reliably reproducing the deflection angle field of the input models over the range of scales relevant for both strong and weak lensing.

Having obtained the deflection angle maps, we apply the cluster distortion fields to the images of a large number of background galaxies. While doing so, we simulate optical observations of each cluster under the different projections. For this purpose, we use the code described in \cite{2008A&A...482..403M} (quoted as {\tt SkyLens} hereafter), which has been recently further developed.  In short, the code uses a set of real galaxies decomposed into shapelets \citep{RE03.1} to model the source morphologies on a synthetic sky. In the current version of the simulator, the shapelet database contains $\sim 3000$ galaxies in the $z$-band from the GOODS/ACS  archive \citep{GIA04.1} and $\sim 10000$ galaxies in the $B,V,i,z$ bands from the {\it Hubble-Ultra-Deep-Field} (HUDF)  archive \citep{BECK06.1} . Most galaxies have spectral classifications and photometric redshifts available \citep{2000ApJ...536..571B,2006AJ....132..926C}, which are used to generate a population of sources whose luminosity and redshift distributions resemble those of the HUDF.

\begin{figure*}[t!]
  \includegraphics[width=0.5\hsize]{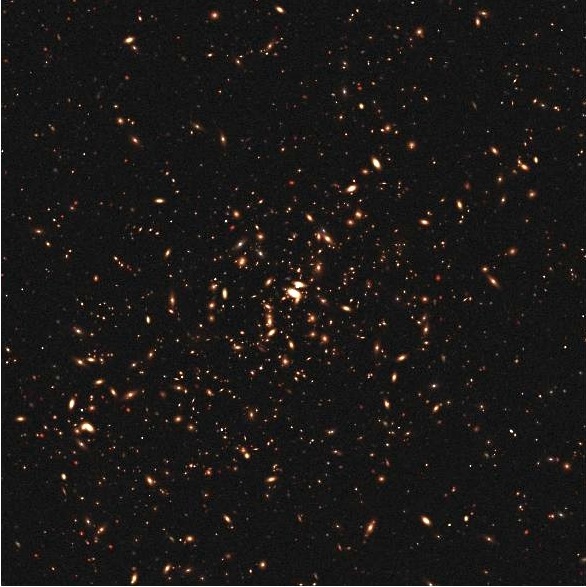}
  \includegraphics[width=0.5\hsize]{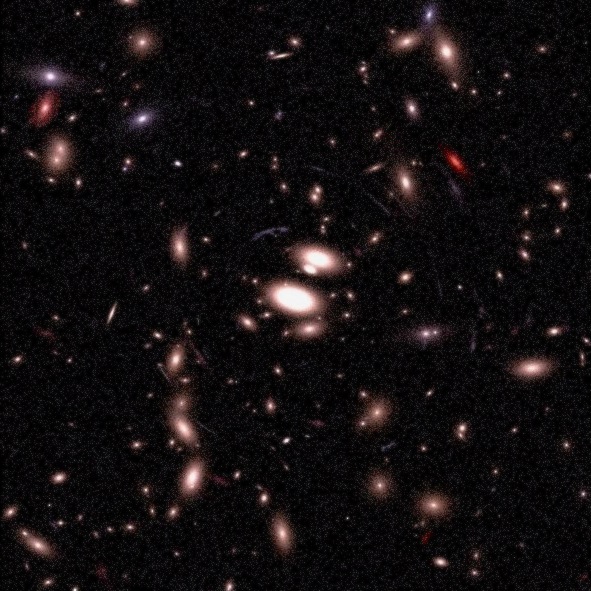}
  \caption{Left panel: Color-composite image of a simulated galaxy cluster, obtained by combining three SUBARU exposures of $2500$s each in the $B,V,I$ bands. {  The field-of-view} corresponds to $\sim 450" \times 450"$. Right panel: an HST/ACS composite image of the the central $100"\times100"$ of the same cluster. The image has been produced by combining mock observations with the filters F475W, F555W and F775W.}
\label{fig:hstg1}
\end{figure*}

{\tt SkyLens} allows to mimic observations with a variety of telescopes, both from space and from the ground. In this work we simulate wide field observations, on which we carry out a weak lensing analysis, using the SUBARU Suprime-Cam. We simulate Hubble-Space-Telescope observations of the cluster central regions using the Advanced Camera for Surveys (ACS). All simulations include realistic background and instrumental noise. As an example, in Fig.~\ref{fig:hstg1} we show two color-composite images of one simulated galaxy cluster (although in the rest of the paper we will use simulated observation in a single band). The left and the right panels show the results of simulated observations with SUBARU and with HST, respectively. Note that the galaxy colors are realistically reproduced by adopting $22$ SEDs to model the background galaxies, following the spectral classifications published by \cite{2006AJ....132..926C}. 

In the HST image, several blueish arcs and arclets are visible behind the cluster galaxies. These are originated by background spiral and irregular galaxies strongly lensed by the foreground cluster. 

{  In our analysis, we use simulated observations} with the following characteristics. For the SUBARU simulations, we assume an exposure time of $6000$s in the $I$ band, with a seeing of $0.6''$. The PSF is assumed to be isotropic and modeled using a two-dimensional Gaussian. For the HST simulations, we assume an exposure time of $7500$s with the F775W filter. {  The fields-of-view adopted for the HST and SUBARU simulations are overlaid to the surface density maps in Fig.~\ref{fig:sdmaps} (blue inner- and white outer-boxes, respectively). Note that these fields of view do not correspond to the true fields-of-view of the ACS and Suprim-CAM mounted on the HST and on the SUBARU telescope. For computational efficiency we limit the fields-of -view in the HST simulated observations to 120 arcsec, which is wide enough to contain the Einstein rings of our clusters. For the SUBARU simulations, the fields-of-view are defined such to correspond to the same comoving scale on the lens plane, i.e. $4\;h^{-1}$ Mpc, for clusters $g1$ and $g51$. For cluster $g72$, we simulate a wider field-of-view, corresponding to $5\;h^{-1}$Mpc comoving in order to include a large substructure in the  observation.} 
         
\subsection{X-ray simulations}

\begin{figure*}[t!]
 \includegraphics[width=1.0\hsize]{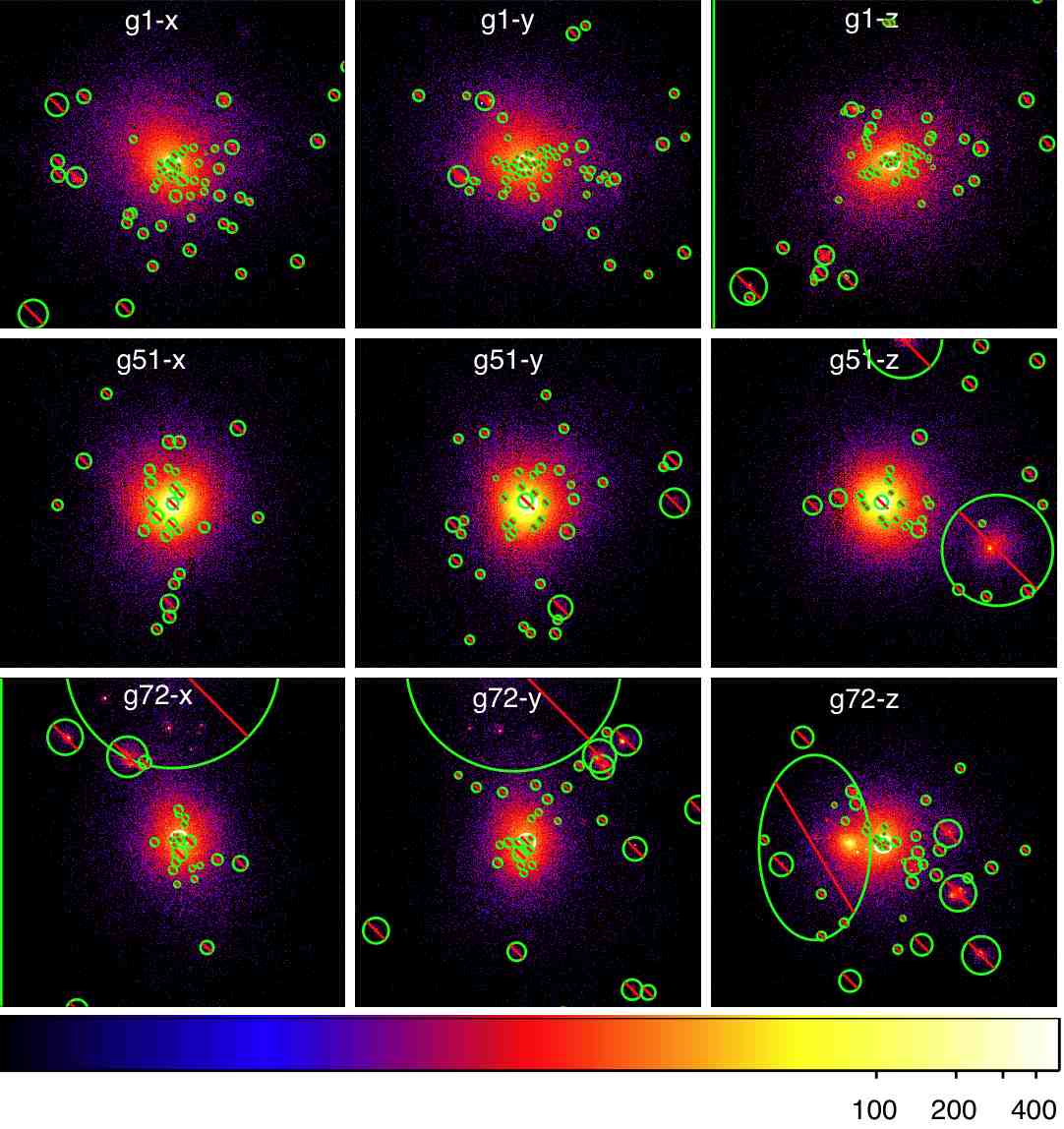}
  \caption{X-ray maps of the three projections of the clusters $g1$ (top panels), $g51$ (middle panels) and $g72$ (bottom panels) used in this work. The size of each map is $16$ arcmin, {  which corresponds to $2.5\,h^{-1}$Mpc for $g51$ and to $2.97\,h^{-1}$Mpc for $g1$ and $g72$}. In each panel, we display the masked regions. The dense cold blobs are encircled in green, while we indicate to the excluded central region in white. The color-bar on the bottom allows to convert colors into counts.}
\label{fig:xraymaps}
\end{figure*}

The X-ray images of our simulated clusters are produced with the code {\em X-ray MAp Simulator} ({\tt XMAS}). {  The software is presented in \cite{GA04.1} and \cite{2008ApJ...674..728R}, where a full description of the simulation pipeline can be found.} It generates synthetic event files which have the same format as real observations. 
 In this work, which is not intended to exploit any X-ray calibration issue, we assume a constant response over the detector. In particular, the response matrix files and ancilliary response file are those of the aimpoint of ACIS-S3 CCD on board of {\it Chandra} telescope.  For each cluster projection we produce an X-ray image. The X-ray images have a field of view of 16 arcmin on a side, which, for the considered cosmology, corresponds to $2.5\,h^{-1}$Mpc at $z=0.2335$ and to $2.97\,h^{-1}$Mpc at $z=0.297$. {  The {\it Chandra} field-of-view is overlaid to the surface density maps of the nine clusters in Fig.~\ref{fig:sdmaps} (green box).} The background is not included a priori {  since \citealt{2006MNRAS.369.2013R} showed that it does not induce any systematic on the bias}. 
The spectral model used to generate the photons considers the contributions from the different metal species present in the simulation: C, N, O, Mg, Si, and Fe. The exposure time is 500 ksec. The images for all the cluster projections analyzed in this paper are shown in Fig.~\ref{fig:xraymaps}. The color-bar on the bottom allows to convert the color levels into counts per pixel.

\section{Analyses}
\label{sect:analyses}

In this Sect. we describe the methods used for analyzing the previously outlined simulations. Firstly, we consider the mass estimates based on strong and weak lensing separately. Then we also discuss a non-parametric method which combines both the lensing regimes. Finally, we consider two methods for deriving the cluster masses from the X-ray simulated data.

\subsection{Strong lensing}
\label{sect:strlenan}
The strong lensing analysis is performed by using the  public software {\tt Lenstool}  \citep{KN93.1}. This is very well developed tool for strong lensing parametric reconstructions, which allows to fit the observed strong lensing features in a cluster field through the combination of several mass components, each of which can be characterized by a density profile and by a projected shape (ellipticity and orientation). The code uses a bayesian approach to find the best-fit lens model and to estimate the errors on the free parameters. We refer the reader to the paper by \cite{2007NJPh....9..447J} for a more detailed description of the many options available in this software.

{\tt Lenstool} allows the user to choose among several available density profiles to describe the lens components. In this work, we use their implementation of the NFW profile \citep{GO02.2} to model the main cluster halo. The functional form of this profile is given in Eq.~\ref{eq:nfw}.
Moreover, we add several sub-components, representing the contribution from the most massive galaxies in the cluster. As shown in some previous studies, it is important to include the cluster members in the model, because they can affect the positions and the magnifications of the strong lensing features \citep{ME03.2,2007A&A...461...25M}. These are modeled using Pseudo-Isothermal-Elliptical-Mass-Distributions (PIEMD) described by the following density profile:
\begin{equation}
	\rho_{\rm PIEMD}(r)=\frac{\rho_0}{(1+r^2/r_{\rm core}^2)(1+r^2/r_{\rm cut}^2)} \;.
\end{equation}
{  We choose the PIEMD model because this is widely used for modeling the lensing properties of cluster galaxies in observations \citep[see e.g.][for some recent references]{2007ApJ...668..643L,2009ApJ...693.1570R,2009MNRAS.398..438D}}.    
As shown  in the previous equation this profile is parametrized by a central density $\rho_0$, which is linked to the central velocity dispersion $\sigma_0$, and by two characteristic radii, namely the core radius $r_{\rm core}$ and the cut-off radius $r_{\rm cut}$. To incorporate the galaxy population into the global lens model we use the same approach used by \cite{2007ApJ...668..643L}, who scale the parameters as a function of the luminosity (or equivalently of the mass). The scaling relations are given by
\begin{eqnarray}
	r_{\rm core}  & = & r_{\rm core}^\star\left(\frac{L}{L^\star}\right)^{1/2} \nonumber \\ 
	r_{\rm cut} & = & r_{\rm cut}^\star\left(\frac{L}{L^\star}\right)^{1/2} \nonumber \\
	 \sigma_0 & = & \sigma_{0}^\star\left(\frac{L}{L^\star}\right)^{1/4} \;,
\end{eqnarray}
where $L^\star$ is a reference luminosity and the quantities $r_{\rm core}^\star$, $r_{\rm cut}^\star$, and $\sigma_{0}^\star$ are the corresponding core and cut radii, and central velocity dispersion, respectively.
In our analysis, we assume a constant mass-to-light ratio for all the cluster members. Such an assumption is generally adopted when studying real clusters.  

\begin{figure*}[ht!]
  \includegraphics[width=0.5\hsize]{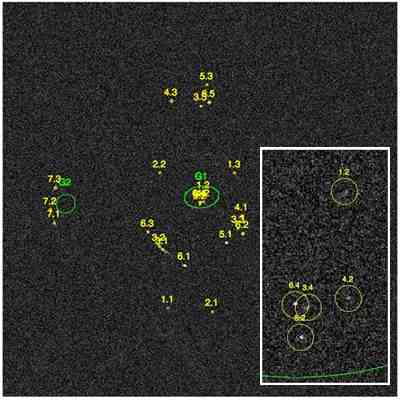}
  \includegraphics[width=0.5\hsize]{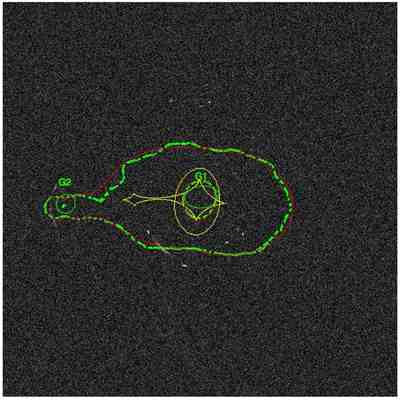}  
  \caption{Left panel: HST-like view over the central $120"\times120"$ of cluster $g1$ (projected along the $y$ axis). The light emission from the cluster galaxies has been removed to avoid confusion with the background multiple images. The identified systems are shown in yellow. The images $i$ belonging to the source $n$ are indicated with the labels $n.i$. {  The inset on the left shows a zoom over the very central region of the image, where the central images $1.2$, $3.4$, $4.2$, $5.2$, and $6.4$ are located}. We also show in green the location of the cluster members which have been included in the lens reconstruction. Right panel: the result of the reconstruction obtained by using {\em Lenstool} with the lensing constraints shown in the left panel. The critical lines for sources at redshift 1.7 are shown in red and the corresponding caustics in yellow. We also overplot the true critical lines of the cluster in green. Note that the shape of the critical lines is very well reproduced, especially where the lensing constraints are tighter. }
\label{fig:exstrlen}
\end{figure*} 

{  Observationally,
the galaxies  to be included in the model should be selected as those laying in the cluster red sequence and being brighter than a given apparent luminosity \cite[e.g.][]{2007ApJ...668..643L}. Of course what matters for lensing is not the luminosity but the mass, which is assumed to be traced by the light. Indeed, the minimal luminosity should be interpreted as a minimal mass. Working with simulations, we identify the cluster galaxies using the {\tt SUBFIND} code \citep{2001MNRAS.328..726S} and then apply a selection based directly on the stellar mass}. {\tt SUBFIND} decomposes the cluster halo into a set of disjoint substructures and then identifies each of them as a locally overdense region in the density field of the background halo. 

In our reconstructions, we include those galaxies which have stellar mass $M_{stars}\ge 10^{9}\,h^{-1}\,M_\odot$ and which are contained in a region of $500\;h^{-1}$kpc around the cluster center. This is typically more than three times the size of the Einstein rings of the clusters in our sample. The orientation and the ellipticity of each galaxy are measured from the distribution of the star particles belonging to it. Following this procedure, we typically end up with catalogs of several tenth of cluster members. 

The Brightest-Central-Galaxy (BCG) is included in the lens model by optimizing its parameters individually, rather than scaling them with the luminosity/mass. Since the BCG forms in the simulations in a strong cooling region, we assume it might have significantly different properties compared to the other cluster members. Thus, we prefer to treat it individually. Analogously, we use individual optimization with some other cluster members which lay particularly close to some multiple image systems. Indeed, their influence on the local lensing properties of the cluster requires to be carefully modeled.

The total number of free parameters in the model depends on the complexity of the lens. Usually we consider a cluster-scale mass component, a galaxy-scale component to describe the BCG, and other galaxy-scale terms to incorporate the relevant cluster members. 

We distribute the sources behind the clusters such to have $\sim 3-7$ strong lensing systems available for the optimization. For this condition to be satisfied, we randomly distribute few sources in a shell surrounding the lens caustics, enhancing the chances that they are strongly lensed. Then, we visually check if the multiple images belonging to each source are detectable in the simulation and retain those systems which are useful for the strong lensing analysis.  
The optimization is done using the bayesian method implemented in {\tt LENSTOOL} with optimization rate $\delta \lambda=0.1$. We assume the uncertainty in the lensed image positions to be $\sigma_I=0.3"$.

In Fig.~\ref{fig:exstrlen}, we illustrate the reconstruction of cluster $g1-y$. The system has a massive galaxy at the center, labeled G1, and another massive galaxy located  $\sim 45"$ west of the BCG (G2). Thus, we fit the lensing observables using three lens components, namely the main halo, the BCG and a secondary PIEMD clump coincident with the other massive galaxy. In this particular case, adding additional cluster members does not affect the reconstruction.   
Among the sources, which were distributed along the caustics of the input cluster lens, seven of them produce multiple image sytems detectable in this deep exposure (7500s) in the F775W filter. More precisely, two sources (source 3 and 6) produce five images, while the other sources are imaged into triplets. These are displayed in the left panel of Fig.~\ref{fig:exstrlen}. The bright knots in the multiple images of the same source are marked with points and identified with labels. The first digit corresponds to the source number, while the last indicates the multiplicity of the image to which the knots belong. For the sake of simplicity, the simulation is shown without including the light emission from the cluster members neither from other background sources that are not strongly lensed. The central image of source 2 cannot be detected by eye, thus it is not used in the reconstruction. The other central images of sources 1, 3, 4, 5, 6 lay behind the BCG but are detectable in the BCG subtracted frame. Using these observables, the reconstruction converges, finding a good fit to the lensing features ($\chi^2=18$ for 21 degrees of freedom). The best fit model consists of  an NFW halo with concentration $c=10.57_{-1.81}^{+2.82}$ and scale radius $r_s=29.07_{-1.85}^{+20.68}$ arcsec (corresponding to $90_{-6}^{+64}\;h^{-1}$kpc). The galaxies G1 and G2 have velocity dispersions of $340_{-30}^{+21}$ km/s and $269_{-17}^{+12}$ km/s, respectively. To illustrate the result of the modeling, we show in the right panel of Fig.~\ref{fig:exstrlen} the true and the reconstructed critical lines of the lens, assuming a source redshift of $z_{\rm s}=1.72$ (source 3). These are displayed in green and red, respectively. Both the tangential and the radial critical lines of the cluster are generally quite well reproduced by the model. The largest differences are on those portions of the critical lines along which fewer lensing constraints are present.

\begin{figure}[t!]
  \includegraphics[width=1.0\hsize]{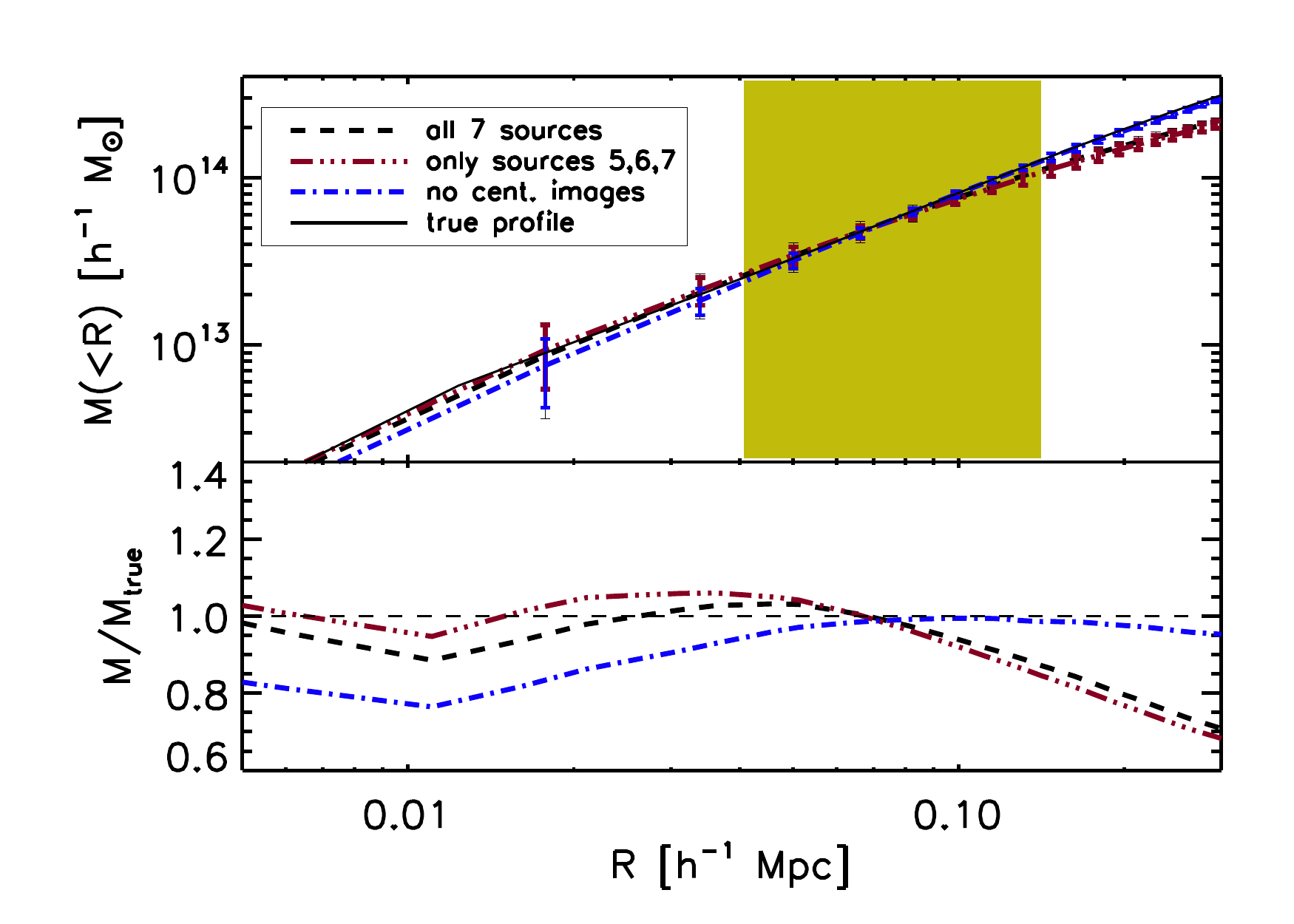}
  \caption{{\it Results of the strong lensing analysis}. The total projected mass profile of the inner region of cluster $g1-y$  as recovered from the strong lensing mass reconstruction using {\tt LENSTOOL}. The dashed line shows the result obtained by using seven multiple-image systems. The red three-dot-dashed line shows the mass profile if only three multiple-image systems are used. The blue dot-dashed line indicates the mass profile recovered by fitting all the 7 multiple-image systems, but assuming that the all the central images are not detectable. The lensing constraints are shown in the left panel of Fig.~\ref{fig:exstrlen}. Finally, the true mass profile, as drawn from the particle distribution in the input cluster, is given by the solid line. The shaded region shows the radial range of the tangential strong lensing constraints. {  The bottom panel shows the ratios between the recovered mass profiles and the true mass profile.}}
\label{fig:slmasspg1}
\end{figure}

The inner projected mass profile derived from the model, $M(<R)$, is shown in Fig.~\ref{fig:slmasspg1} (dashed line), where we also show the true profile of the cluster acting as lens in this simulation (solid line). Since the model reproduces well the lens tangential critical line, it is not surprising that the model is very reliable at estimating the mass enclosed in the strong lensing region. The shaded area in the Fig. indicates the radial range of the multiple images, excluding the central images, which are located at $R\lesssim 10\;h^{-1}$kpc. The reconstruction reproduces well the true mass profile up to $\sim 150\;h^{-1}$kpc from the center, where the deviation from the true mass profile is $\lesssim 10\%$. At larger radii, the differences become significant. Thus, extrapolating the strong lensing model to distances where no strong lensing features are observed may result in largely incorrect mass estimates. This issue will be discussed in more detail in Sect.~\ref{sect:resu}.

In order to evaluate how the reliability of the model degrades by reducing the number of constraints, we perform another reconstruction using only one system with five images (arising from the source 6) and two triplets (source 5 and 7). The final reconstruction does not differ significantly from the previous one. The projected mass profile for this new lens model is given by the red three-dot-dashed line in Fig.~\ref{fig:slmasspg1}. This result shows that reliable reconstructions can be achieved even with a limited number of lensing constraints, if they  are optimally distributed across the cluster. We also attempted a reconstruction  by neglecting the presence of the central images (and using all the seven lensed systems). This is likely to be a realistic situation, since the central images are generally de-magnified and hidden behind the BCG, and thus difficult to detect. In this case, the mass enclosed by the strong lensing region is again correctly estimated, but the reconstructed profile deviates more from the true one at small radii, as shown by the blue dot-dashed line in Fig.~\ref{fig:slmasspg1}. {  In the following, the SL models are constructed using also the central images, when they are detectable in the galaxy subtracted frames.}

\subsection{Weak lensing}

The weak lensing measurements are done using the standard KSB method, proposed by \cite{1995ApJ...449..460K} and subsequently extended by \cite{1997ApJ...475...20L} and by \cite{1998ApJ...504..636H}. Such method is now internally implemented in {\tt Skylens}.   The galaxy ellipticities are measured from the quadrupole moments of their surface brightness distributions, corrected for the PSF, and used to estimate the reduced shear under the assumption that the expectation value of the intrinsic source ellipticity vanishes (see Eq.~\ref{eq:ell}).

Selecting the galaxies with $S/N>10$, we end up with catalogs of galaxy ellipticities containing $\sim 30$ sources/sq. arcmin. The median redshift of these sources is $z_{\rm s,true}\sim 1.05$. In the following analysis we assume that all sources have the same redshift of $z_{\rm s}=1$. Furthermore, we assume that we can separate perfectly the population of background galaxies from the foreground cluster members. This is intentionally very optimistic, since we aim here at verifying the capabilities of several lensing methods to retrieve the cluster mass in the best possible conditions. The misidentification of cluster members as background galaxies leads to a dilution of the lensing signal, which causes to erroneous mass estimates \citep[see e.g.][]{2007ApJ...663..717M}. We will address in more detail this issue in a forthcoming paper. Increasing the distance from the cluster center, the probability that nearby substructures or additional mass clumps affect the mass estimates becomes higher. In this work, we do not include the effects of uncorrelated large-scale-structures (LSS) on the weak-lensing signal. The effects of the LSS on the weak lensing mass estimates have been discussed in detail in several other papers \citep{2001A&A...370..743H,2003MNRAS.339.1155H,2004APh....22...19W}. Uncorrelated LSS introduce a noise in the mass estimates, but the importance of matter along the line sight is fairly small for rich clusters at intermediate redshifts, like those in our sample, provided that the bulk of the sources 
are at high redshift compared to the cluster.           
Given that  we are taking into account all the mass in cylinders of height  $20\;h^{-1}$ Mpc in the lensing simulations, the effects of the correlated large-scale structure is  partially included.  \cite{CL04.1} have studied the weak lensing signal of our same clusters (but in a pure dark matter version) in their cosmological environment. They find that including all the matter in a cylinder of height $\sim 100 h^{-1}$ Mpc around the clusters only results in small scatter being added to the measured cluster masses, compared to simulations where a cylinder of only $\sim 10 h^{-1}$ Mpc was used. These results are in agreement with the findings of \cite{RE99.1}. 

The cluster masses are derived using the following approaches:

\paragraph{NFW fit of the tangential shear profile:} Assuming that the cluster is well described by an NFW density profile, we use the corresponding formula for the reduced shear to fit the azimutally averaged profile of the tangential component of the reduced shear. For the NFW profile, the formulas for the radial profiles of the shear and of the convergence can be found in \cite{BA96.1} and in \cite{ME03.1}. The tangential component of the reduced shear is given by 
\begin{equation}
	g_+=-{\rm Re}[g {\rm e}^{-2i\phi}] \;,
\end{equation}
where the angle $\phi$ specifies the direction from the galaxy centroid towards the center of the cluster, which we identify with the position of the BCG. The cross component of the reduced shear is given by
\begin{equation}
	g_\times=-{\rm Im}[g {\rm e}^{-2i\phi}] \;.
\end{equation}
If the distortion is caused by lensing, this component of the shear should be zero. 

\begin{figure}[t!]
  \includegraphics[width=1.0\hsize]{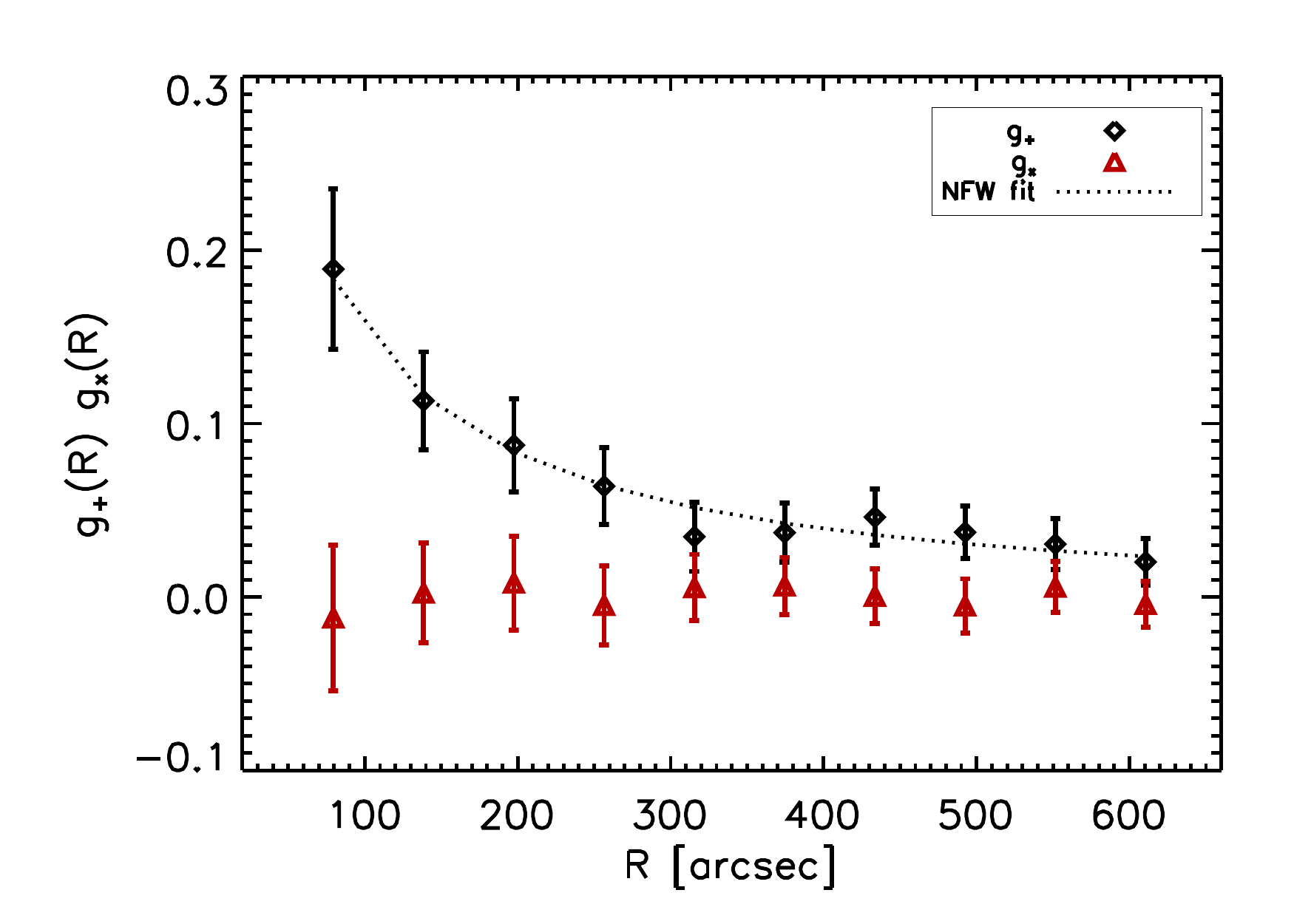}
  \caption{Radial profiles of the tangential and of the cross components of the reduced shear measured from the center of cluster $g1-y$. The dotted line shows the best fit NFW model.}
\label{fig:g1gshprof}
\end{figure}

In Fig.~\ref{fig:g1gshprof}, we show the radial profiles of both the components of the shear for the cluster $g1-y$, measured out to large radii ($\sim 3\;h^{-1}$Mpc from the cluster center). The tangential component is well fitted by an NFW profile with $c=4.82 \pm 0.64$ and $r_{s}=0.307 \pm 0.048\;h^{-1}$Mpc. As expected in absence of systematics, the cross component of the shear is consistent with zero.
  
\paragraph{Aperture Mass Densitometry:} The aperture mass densitometry \citep{1994ApJ...437...56F,1998ApJ...497L..61C} uses the fact that the shear can be related to a density contrast. More precisely, it can be shown that the following relation holds: 
\begin{eqnarray}
	\zeta(R_1) & = & \overline\kappa(<R_1)-\overline\kappa(R_2<R<R_{\rm max}) \nonumber \\
	& = & 2\int_{R_1}^{R_2} d\ln R \langle \gamma_t \rangle +\frac{2 R_{\rm max}^2}{R_{\rm max}^2-R_2^2}\int_{R_2}^{R_{\rm max}} d \ln R \langle \gamma_t \rangle \;,
\end{eqnarray} 
where $\overline\kappa(<R_1)$ is the mean convergence within a circular aperture of radius $R_1$ and   $\overline\kappa(R_2<R<R_{\rm max})$ is the mean convergence in an annulus from $R_2$ to $R_{\rm max}$. This relation shows that the mean surface density within a circle can be derived from the tangential shear profile up to a constant. This is straightforwardly converted into a mass estimate:
\begin{equation}
M(<R_1)=\pi R_1^2 \overline\kappa(<R_1) \Sigma_{\rm cr}(z_{\rm l},z_{\rm s}) \;.
\end{equation}
 
If wide field observations are avaliable, $R_2$ and $R_{\rm max}$ can be chosen to be large, so that the surface density in the annulus is negligible. Otherwise, setting the annulus term to zero, $\zeta$ allows to estimate only a lower limit to the projected mass. 

Although the $\zeta$-statistic would not require any parameterization of the lensing signal to convert the shear into a mass estimate, a complication arises from the fact we do not directly measure the shear but the reduced shear. To convert the observed signal into an estimate of the shear, it is usually necessary to make some assumption on the shape of the convergence profile. In our analysis, we follow the method of \cite{2007MNRAS.379..317H}, who uses the convergence from the best fit NFW model to the tangential shear profile. We also use this approach to estimate the mean surface density in the annulus. For our mass estimates $R_{\rm max}$ varies between $\sim640$ and $\sim800$ arcsec, depending on the cluster. We set $R_{2}=0.9\times R_{\rm max}$.

\subsection{Strong and weak lensing}
 Finally, we consider a completely non parametric, two-dimensional mass reconstruction. The method used here is not based only on weak lensing. Instead, it combines both strong and weak lensing constraints to provide a map of the lensing potential. The method is fully described in \cite{2009A&A...500..681M} \citep[see also][for another similar algorithm]{BR04.1}. To obtain the underlying lensing potential~$\psi$~of the galaxy cluster the reconstruction algorithm performs a combined $\chi^{2}$-minimisation, which consists of a weak and a strong-lensing term, in combination with an additional regularisation term:
\begin{equation}
\chi^{2}(\psi)=\chi^{2}_{\rm w}(\psi)+\chi^{2}_{\rm s}(\psi)+R(\psi).
\end{equation}
The algorithm is grid-based and expresses the derivatives of the lensing potential by finite-differencing schemes. Starting from a coarse initial grid, the resolution is steadily increased until the final reconstruction is reached. At each iteration the reconstruction is regularised on the former iteration by the regularisation function $R(\psi)$. This procedure results in a smooth reconstruction of the lensing potential and prevents the reconstruction from following noise patterns in the data.
 
The weak lensing constraints are obtained by averaging over a certain number of background-galaxy ellipticities in every reconstruction pixel. The error is given by the statistical scatter in each pixel. Afterwards the reduced shear of the cluster is fitted:
\begin{equation}
 \chi^{2}_{\rm w}(\psi)=\left(
   \left\langle\epsilon\right\rangle -\frac{Z(z)\gamma(\psi)}{1-Z(z)\kappa(\psi)}
 \right)_{i}\mathcal{C}^{-1}_{ij}\left(
   \left\langle\epsilon\right\rangle -\frac{Z(z)\gamma(\psi)}{1-Z(z)\kappa(\psi)}
 \right)_{j}\;.
\label{wchi}
\end{equation}
The sum over~$i$~and~$j$~ runs over all reconstruction pixels and the cosmological-weight function $Z(z)$ describes the redshift distribution of the sources
\begin{equation}
 Z(z)\equiv\frac{D_{\infty}D_{{\rm ls}}}{D_{{\rm l}\infty}D_{\rm s}}H(z-z_{{\rm l}}),
\end{equation}
where $D_{\infty}$ and $D_{{\rm l}\infty}$ are the angular- diameter distances between observer and infinity and between lens and infinity respectively.
Note that we have to solve the full $\chi^{2}$-function since the reconstruction 
pixels might become correlated on a high grid resolution.

The strong lensing constraints are based on the estimated positions of the critical curves of the galaxy cluster. They are given by the observed arc positions and/or multiple images bracketing the critical lines. Assuming a reconstruction pixel to be part of a critical curve, by summing over all these pixels with index $k$, we find:
\begin{equation}
\chi^{2}_{{\rm s}}(\psi)=\frac{\left(\det{\mathcal A}(\psi)\right)^{2}_{k}}{\sigma^{2}_{{\rm s}}}=
 \frac{\left((1-Z(z)\kappa(\psi))^{2}-|Z(z)\gamma(\psi)|^{2}\right)^{2}_{k}}{\sigma^{2}_{{\rm s}}}\;.
\end{equation}

The error $\sigma_{s}$ is determined by the pixelisation of the grid.\\
A final high resolution step is added, which is able to resolve the positions of the critical curve estimators in greater detail. Since there is no reliable weak lensing information on this resolution, this step is embedded in the foregoing reconstruction by regularising on the former result in the cluster centre:
\begin{equation}
 \chi^{2}_{{\rm highres}}(\psi)=\chi^{2}_{{\rm s}}(\psi)+R_{{\rm lowres}}(\psi)\;,
\end{equation}
The mass analysis is then done with a convergence map, which is calculated from the reconstructed lensing potential of the cluster using Eq.\ref{eq:k}. 

\begin{figure}[t!]
  \includegraphics[width=1.0\hsize]{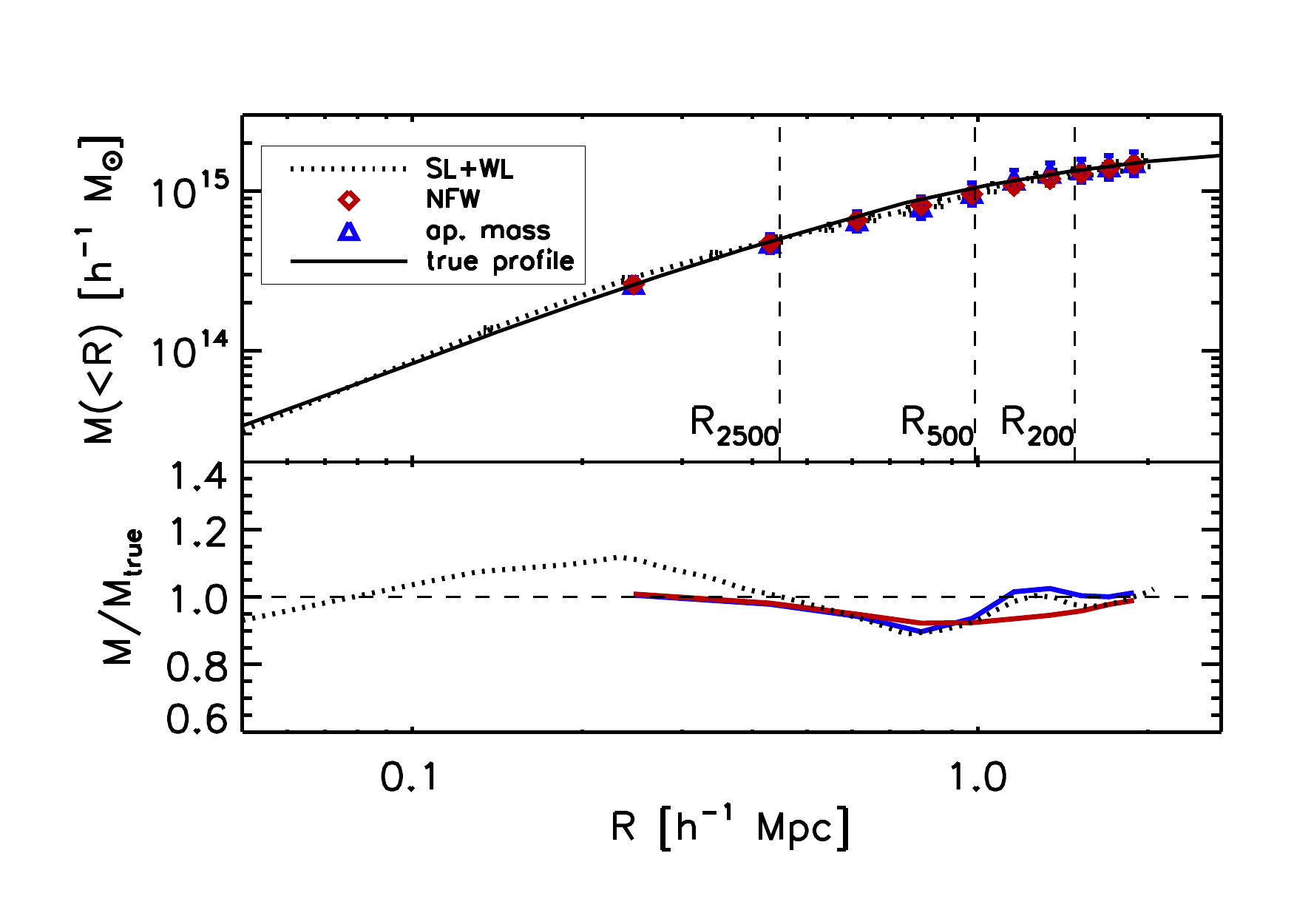}
  \caption{{\it Results of the weak lensing analysis}. Radial 2D-mass profiles of cluster $g1-y$, as obtained from the three different methods used in this work, namely the NFW fit to the shear profile (diamonds), the aperture mass densitometry (triangles), and the two-dimensional mass reconstruction combining weak and strong lensing (dotted line). The solid line shows the  true mass profile. The vertical lines indicate the positions of $R_{2500}$, $R_{500}$ and $R_{2500}$ as derived from the NFW fit. The bottom panel shows the ratios between the mass profiles recovered from the lensing analysis and the true mass profile. {  The dotted line refers again to the SL+WL method, while the red and blue solid lines indicate the results for the NFW fit and for the aperture mass, respectively. }}
\label{fig:g1wlprofs}
\end{figure}

In Fig.\ref{fig:g1wlprofs}, we show the mass profiles of cluster $g1-y$ obtained with the weak-lensing methods described above. All the methods perform well and provide consistent results for this particular cluster. The mass profiles deviate from the true one by less than $10\%$ at all radii. The vertical lines show the position of $R_{2500}$, $R_{500}$, and $R_{200}$, i.e. the estimated radii enclosing over-densities of $2500$, $500$, and $200$ times the critical density of the universe, as derived from the NFW model that best fits the tangential shear profile. In the case of the two-dimensional mass reconstruction method combining strong and weak lensing, we  estimate the errors on the masses by bootstrapping 24 galaxy catalogs and by repeating the reconstruction with each of them. This is computationally very demanding, thus we reconstruct the lensing potential a coarse grid of $32\times32$ pixels covering the whole cluster field ($1280"\times1280"$).

\subsection{X-ray}\label{sect:X-ray}
\begin{figure}[t!]
  \includegraphics[width=1.0\hsize]{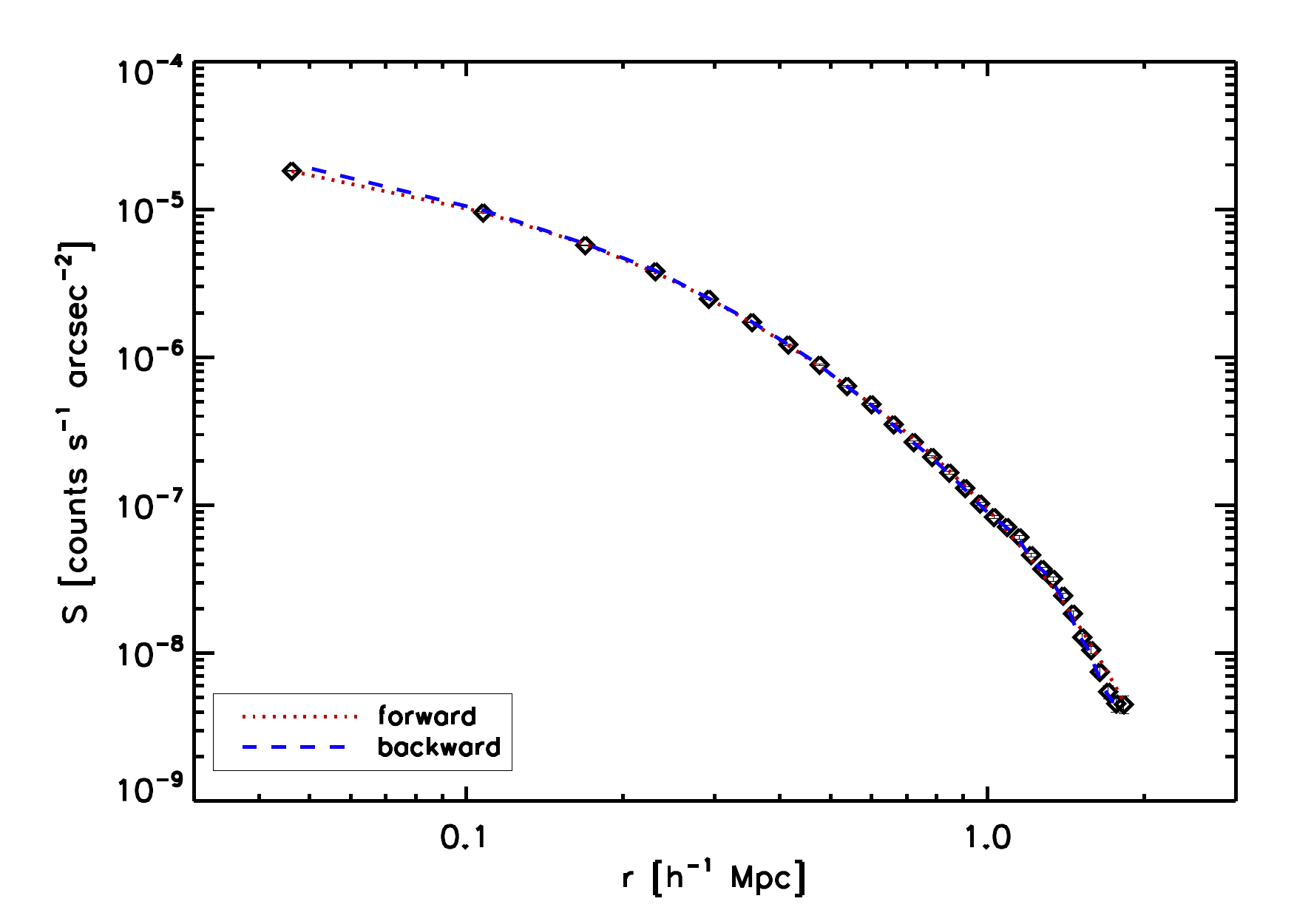}
  \caption{Radial surface brightness  profile of cluster $g1-y$, as derived from the analysis of the x-ray observation shown in upper middle panel of Fig.~\ref{fig:xraymaps} (diamonds). The dotted and the dashed lines show the fits to the data obtained with the forward and with the backward methods respectively.}
\label{fig:g1sb}
\end{figure}

\begin{figure}[t!]
  \includegraphics[width=1.0\hsize]{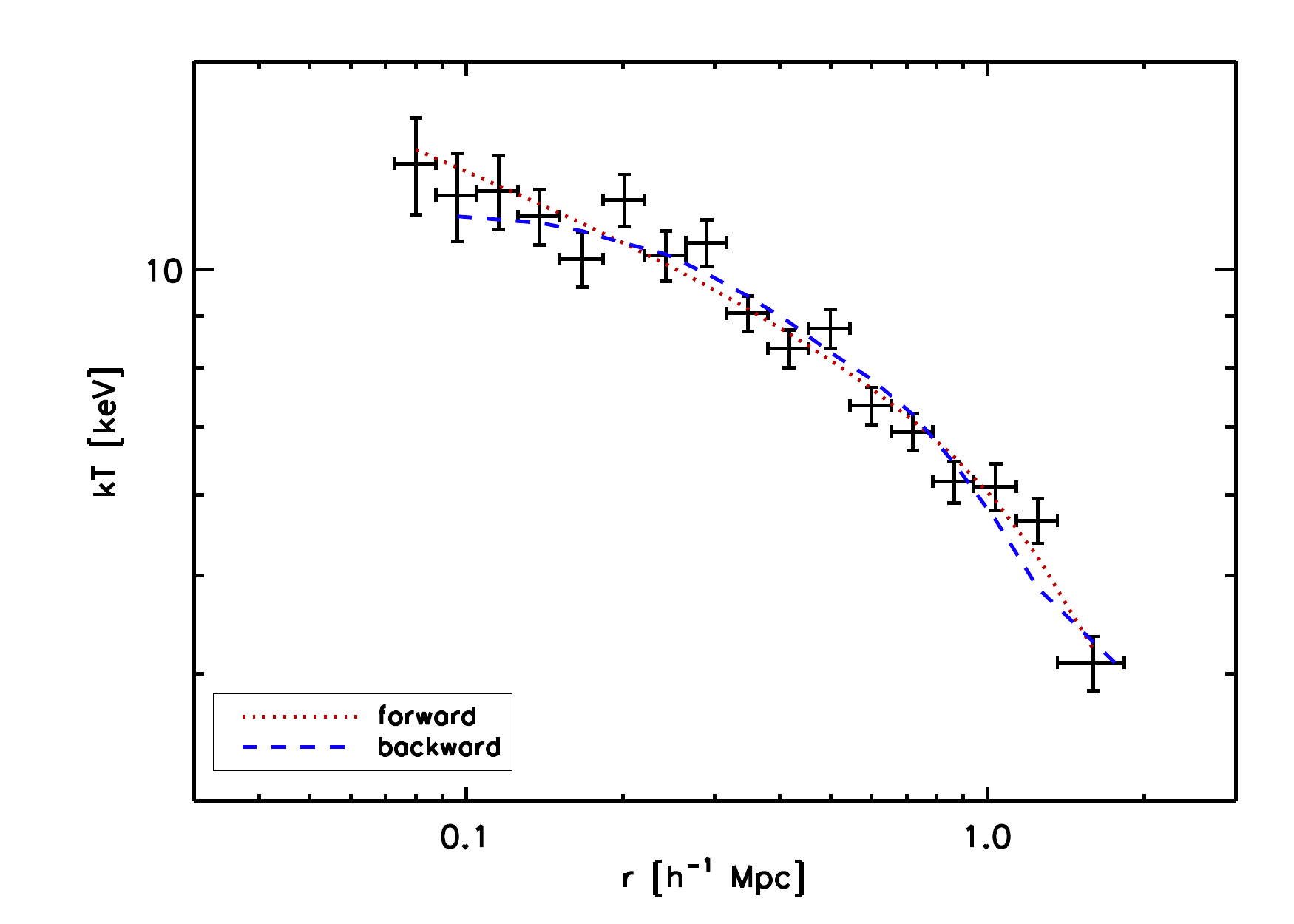}
  \caption{Radial temperature profile of cluster $g1-y$, as obtained from the spectral analysis of the X-ray observation (data points). The dotted and the dashed lines show the fits to the data obtained with the forward and with the backward methods respectively. See text for more details.}
\label{fig:g1t}
\end{figure}

Soft band [0.7-2] keV X-ray images are created from the event
files. Using them we identify the regions of dense cold blobs that
we mask in any further analysis. The masked regions are overlaid to
the X-ray images for each cluster projection in
Fig.~\ref{fig:xraymaps}. These bright point-like spots are mostly
correlated to the cores of previously merged substructures.  {  Further, we
exclude an inner region of 60 kpc $h^{-1}$ for $g51$ and 70 kpc
$h^{-1}$ for $g1$ and $g72$ (white circles in Fig.~\ref{fig:xraymaps}) to exclude
the central region affected by the overcooling problem. From the soft
band image, we finally produce the surface brightness profiles using 30
annuli spanning from 23 to 400 arcsec.}  The surface brightness profile
of the cluster $g1-y$ is given by the diamonds in
Fig.~\ref{fig:g1sb}. In the same radial range, we extract the
spectra of 10-15 annuli logarithmically spaced. We subsequently
analyze the spectra in {\tt XSPEC} \citep{1996ASPC..101...17A} using a single temperature MEKAL
model \citep{1985A&AS...62..197M,1995ApJ...438L.115L} to obtain the temperature and the normalization. The values of
redshift and hydrogen column density are set accordingly to the input
simulation. The temperature profile of $g1-y$ is shown in
Fig.~\ref{fig:g1t}.


The X-ray gas and total masses are recontructed using two different methods, here
labelled {\it backward} and {\it forward} method.
In brief, starting from the observed X-ray surface brightness and the radially resolved
spectroscopic temperature measurements and under the assumption of spherical geometry
and hydrostatic equilibrium between the ICM and the underlying dark matter potential,  
the {  {\it backward}} puts constraints on a parametric functional form of the total mass
by deprojecting the observed quantities, while the {  {\it forward}} makes direct use 
of the 3-D model of the gas density and temperature profiles estimating
the model parameters by projecting the profiles and fitting them to the observed ones.

We describe here in detail the two methods.

\paragraph{Backward method.}
A functional form of the total mass has to be assumed. From this, and moving
{\it backward}, the profiles of the observed quantities are recovered. 
In this work, the NFW model in Eq.~\ref{eq:nfw} is used to describe the density profile of the clusters, so that 
\begin{equation}
M_{\rm tot}(<r) = 4 \pi \ r_{\rm s}^3 \ \rho_{\rm s} \ f(x) \;, 
\label{eq:mass_nfw}
\end{equation}
where
\begin{equation}
f(x) =  \ln (1+x) - \frac{x}{1+x} \;. 
\end{equation}
and $x = r/r_{\rm s}$.

The two free parameters $(r_{\rm s}, c_{200})$ are constrained by minimizing
a $\chi^2$ statistic defined as 
\begin{equation}
\chi^2 = \sum_i\frac{ \left(T_{\rm data, i} - T_{\rm model, i} \right)^2}{\epsilon_{T, i}^2}
\label{eq:chi}
\end{equation}
where $T_{\rm data}$ are the either deprojected or observed
temperature measurements obtained in the spectral analysis; $T_{\rm
model}$ are the either 3-D or projected values recovered from the
inversion of the hydrostatic equilibrium equation for a given gas
density and total mass profiles; $\epsilon_T$ is the error on the
spectral measurements.  The gas density profile, $n_{\rm gas}$, is
estimated from the geometrical deprojection (Fabian et al. 1981, Kriss
et al. 1983, McLaughlin 1999, Buote 2000, Ettori et al. 2002) of the
measured X-ray surface brightness and, in the present analysis, {  is
used to project $T_{\rm 3D}$ accordingly to the recipe in Mazzotta et
al. (2004) to recover $T_{\rm model}$. The latter values are then
compared to the results of the {\tt XSPEC} analysis using the $\chi^2$
statistics (Eq.~\ref{eq:chi}). The values of $T_{\rm 3D}$ are obtained
from}
%The values of $T_{\rm model}$ are then obtained from
\begin{eqnarray}
-G \mu m_{\rm p} \frac{n_{\rm gas} M_{\rm tot}(<r)}{r^2} =
\frac{d\left(n_{\rm gas} \times T_{\rm 3D} \right)}{dr},  
\label{eq:mtot}
\end{eqnarray}
where $G$ is the gravitational constant, $m_{\rm p}$ is the proton mass and
$\mu$=0.59 is the mean molecular weight in a.m.u. as adopted in the present simulations.
Further detail on variations and applications of the technique here described 
are presented in Ettori et al. (2002) and Morandi et al. (2007). The best fit surface brightness and temperature profiles obtained with this method for the cluster $g1-y$ are given by the dashed lines in Figs.~\ref{fig:g1sb} and \ref{fig:g1t}, respectively.

\paragraph{Forward method.} 
By this technique, the total mass profile is recovered through a direct ({\it forward}) 
application of the hydrostatic equilibrium equation (see Eq.~\ref{eq:mtot}) 
once a parametric form of the gas density and temperature profiles are estimated.
In the following analysis, we adopt the approach by Vikhlinin et al. (2006) 
that follows similar techniques presented in, e.g., Lewis et al. (2003) and Pratt \& Arnaud
(2003), and extends the number of parameters to increase the modeling freedom.
In particular, the formula describing the gas density is a $\beta-$model modified
to accomodate a power-law behaviour in the center and the observed steepening of the
surface brightness in the outskirts (Vikhlinin et al. 1999, Neumann 2005, Ettori \& Balestra 2009)
with the addition of a second $\beta-$model to better reproduce the core profile:
\begin{eqnarray}
\rho(r)&=&\frac{N_1}{(r/r_{c1})^{-\alpha}[1 +(r/r_{c1})^2]^{(3\beta_1-\alpha/2)}} \frac{1}{[1 +(r/r_s)^\gamma]^{(\epsilon/\gamma)}}\\
        &+& \frac{N_2}{[1 +(r/r_{c2})^2]^{3 \beta_2}} \nonumber
\end{eqnarray}
$\alpha,\beta_1, \beta_2, r_ s, \epsilon, \gamma, r_{c1}$, and $r_{c2}$ are all free parameters in the fit.
The temperature profile is modeled by a 5 parameter $- T_0, r_t,  a, b, c -$ function
\begin{equation}
T=T_0 \frac{(r/r_t)^{-a}}{[1+(r/r_t)^b]^{c/b}}T_{Cool}.
\end{equation}

{  For these simulations, where the cool cores are well inside the excluded inner regions (see Fig.~\ref{fig:xraymaps}), we consider $N_2=0$ and $T_{cool}=1$.}
These profiles are then projected along the line of sight {  considering Mazzotta et al. (2004) prescription} and the best-fit parameters determined
from a $\chi^2$ minimization technique by comparing the projected quantities with the
observed ones. The best fit surface brightness and temperature profiles for cluster $g1-y$ are given by the dotted lines in Figs.~\ref{fig:g1sb} and \ref{fig:g1t}. 

To summarize, by using the observed X-ray surface brightness and temperature profiles
these two methods provide: ({\it backward} method) starting from a given gas density profile, 
the gas mass $M_{\rm gas}$ and the best-fit parameters $(r_{\rm s}, c_{200})$ from which a 
total mass profile and a 3D temperature profiles are recovered; 
({\it forward} method) for some given parametric forms of the gas and temperature profiles, their best-fit parameters 
from which $M_{\rm gas}$ and $M_{\rm tot}$ are estimated.
In the following analysis, the results obtained from these two methods are compared 
to assess systematics in the X-ray analysis of the matter distribution.

\begin{figure}[t!]
  \includegraphics[width=1.0\hsize]{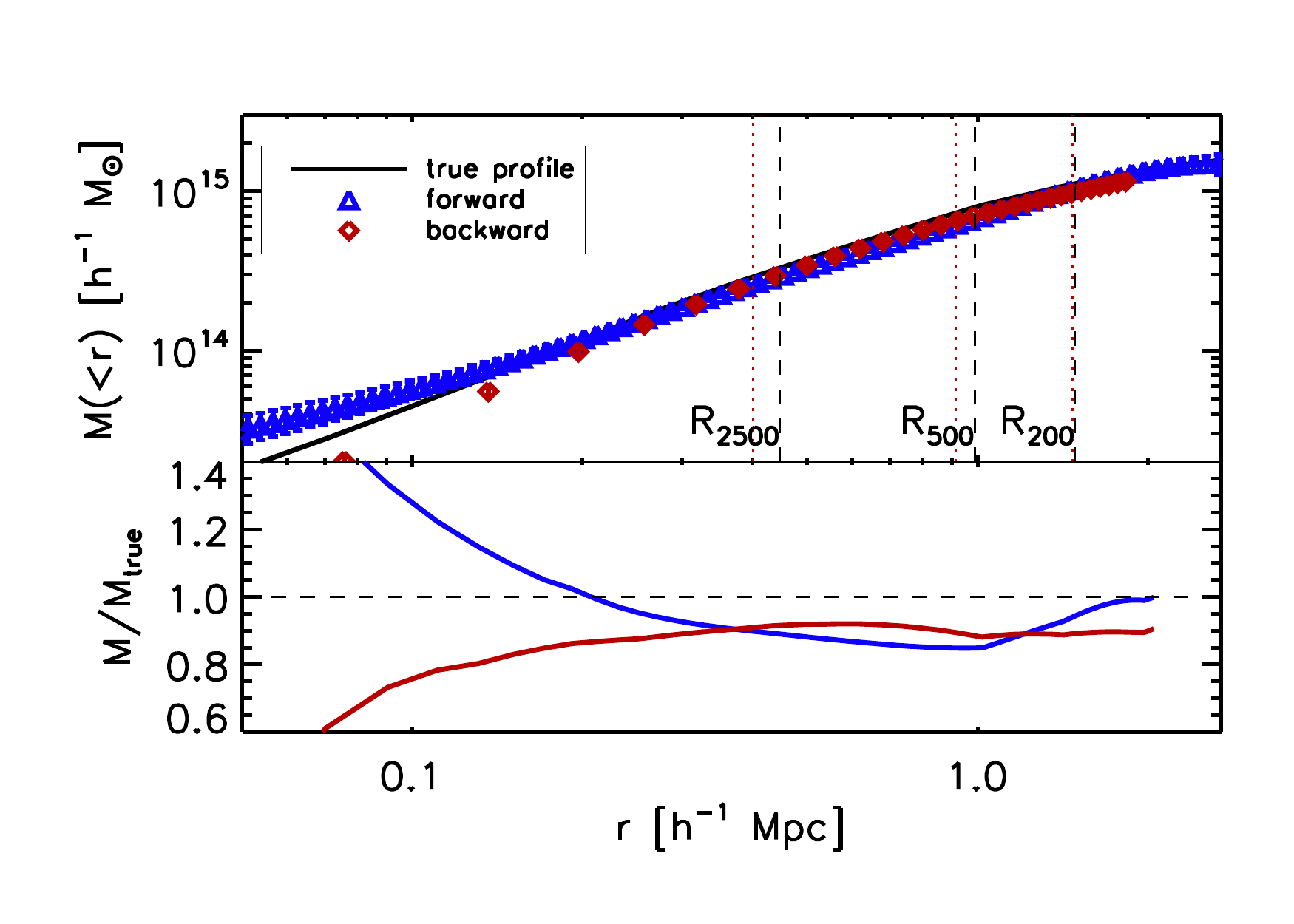}
  \caption{{\it Results of the X-ray analysis}. Radial 3D-mass profiles of cluster $g1-y$, as obtained from the forward (triangles) and from the backward methods (diamonds) used in this work. The solid line shows the  true mass profile. The vertical lines indicate the positions of $R_{2500}$, $R_{500}$ and $R_{2500}$ as derived from the lensing (see Fig.~\ref{fig:g1wlprofs}, dashed lines) and from the X-ray {  backward} analyses (dotted lines).}
\label{fig:g1xrayprofs}
\end{figure}  

\begin{figure}[t!]
  \includegraphics[width=1.0\hsize]{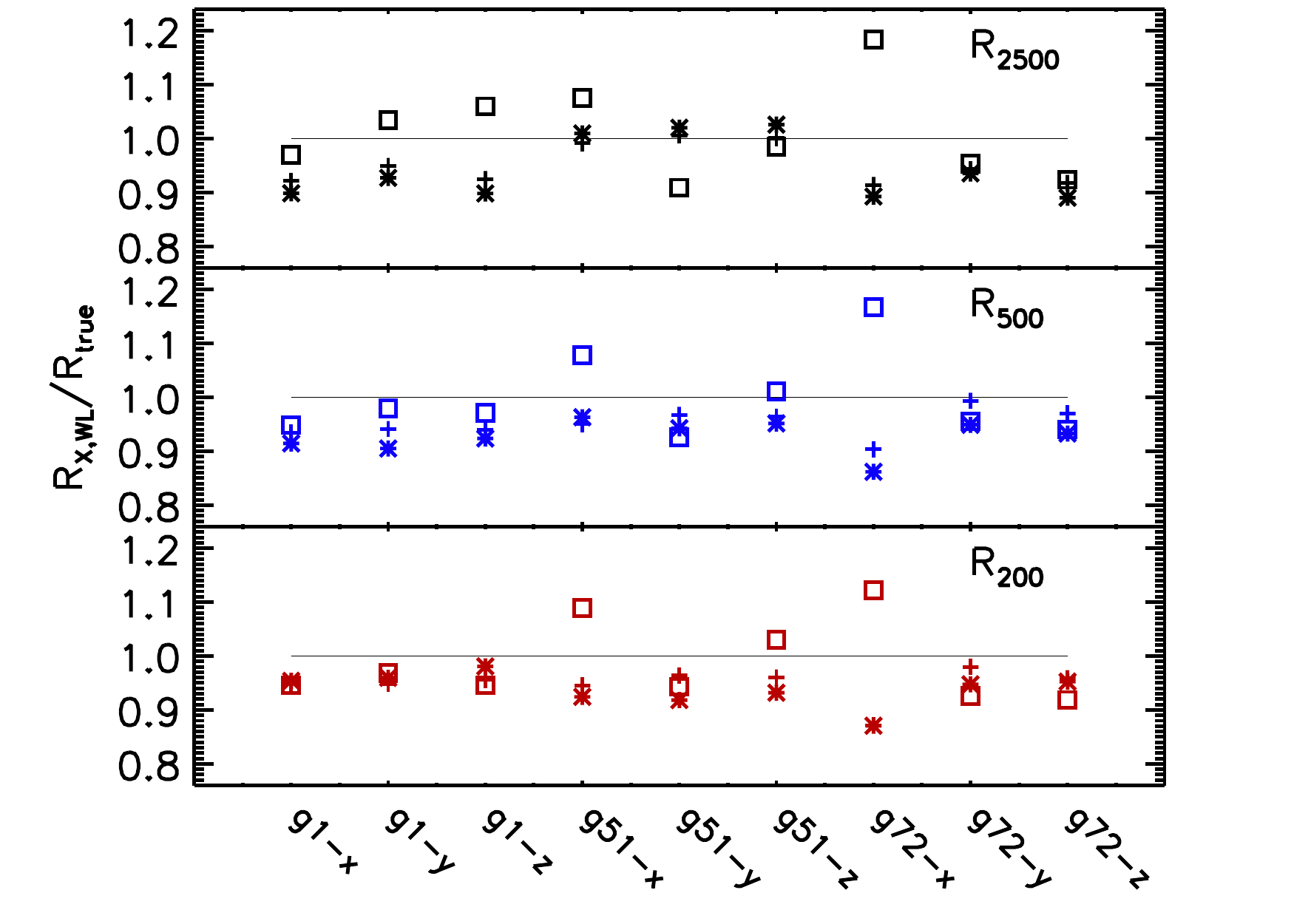}
  \caption{Ratio of the estimated to the true characteristic radii calculated for each cluster projection. Squares, Stars, and Crosses refer to size derived
using the weak lensing, the X-ray forward,  and the X-ray backward method, respectively. The continuous line indicate where the true and the estimated size are equal.  
The Top, Middle, and Lower panel refer to   $R_{2500}$, $R_{500}$, and $R_{200}$, respectively.}
\label{fig:radii_ratio}
\end{figure}

In Fig.~\ref{fig:g1xrayprofs} we show the 3D-mass profiles of cluster $g1-y$, obtained with both the forward (triangles) and the backward (diamonds) methods. In the region between $R_{2500}$ and $R_{200}$ the two methods are consistent with each other and they under-estimate the true mass profile, given by the solid line, by $\sim 10-20\%$. Similar results are found for the other clusters in the sample, and will be discussed in detail in Sect.~\ref{sect:xresu}. The vertical lines in the figure indicate the estimated sizes of $R_{2500}$, $R_{500}$, and $R_{200}$. In particular, the dashed and the dotted lines refer to the estimates based on the weak-lensing and on the X-ray {\it forward} analyses, respectively. 
Note that, due to the mass under-estimate in the X-ray case, the radii measured through lensing are typically larger than those measure from the X-ray methods. 
This is clearly shown in Fig.~\ref{fig:radii_ratio} where we report the ratio of the extimated to the true characteristic radii calculated for each cluster projection
using the weak lensing signal or the X-ray analysis.
It is worth noting even if the  X-ray method always underestimate the true radii, this bias seem to have a quite small scatter.
On the contrary while the weak lensing is a  less biased estimator, we find for it a much larger scatter.
Just for conveniece, in the following analysis we compare the X-ray and the lensing masses at the same physical radii, which we choose to be $R_{2500}$, $R_{500}$, $R_{200}$ as derived from lensing. 
 
\section{Results}
\label{sect:resu}
{  In this Sect. we discuss the results of the analyses outlined in the previous sections. We start with a discussion of the two-dimensional mass estimates obtained with the lensing techniques. Then, we consider the deprojection of the lensing profiles and the X-ray 3D-mass estimates. Finally, we compare lensing and X-ray mass profiles and discuss how our results match the observations.}

\subsection{Lensing 2D mass profiles}

\begin{figure}[t!]
  \includegraphics[width=1.0\hsize]{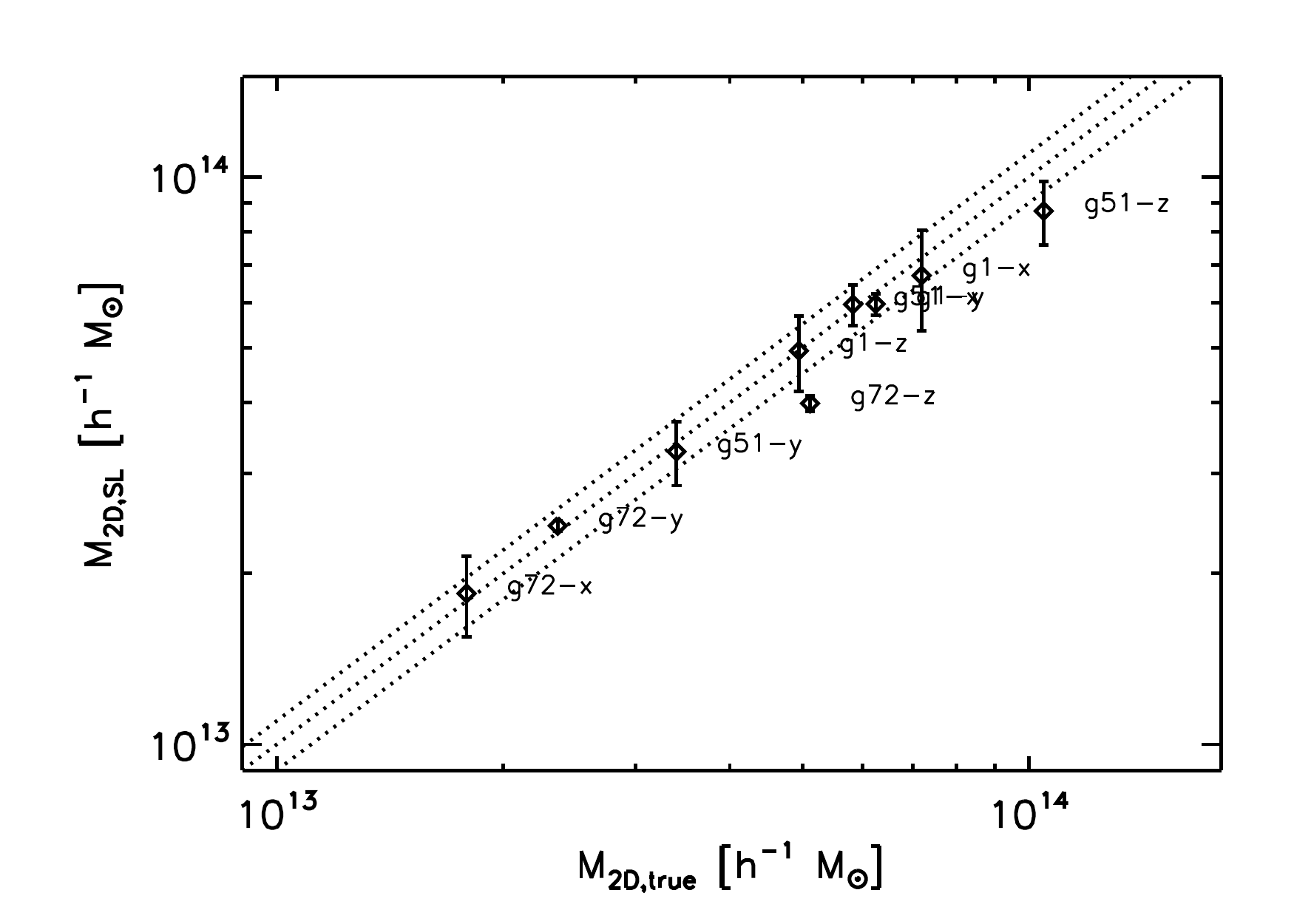}
  \caption{The projected masses estimated through the strong lensing analysis vs  the corresponding true masses of the lenses. {  The dotted lines correspond to $M_{\rm 2D,SL}=M_{\rm true}$ and to $M_{\rm 2D,SL}=M_{\rm true}\pm 10\%$}. The masses are measured within a circle centered on the BCG and having a radius equal to the mean distance of the lensing constraints from the cluster center.}
\label{fig:mslmt}
\end{figure}  

\begin{figure}[t!]
  \includegraphics[width=1.0\hsize]{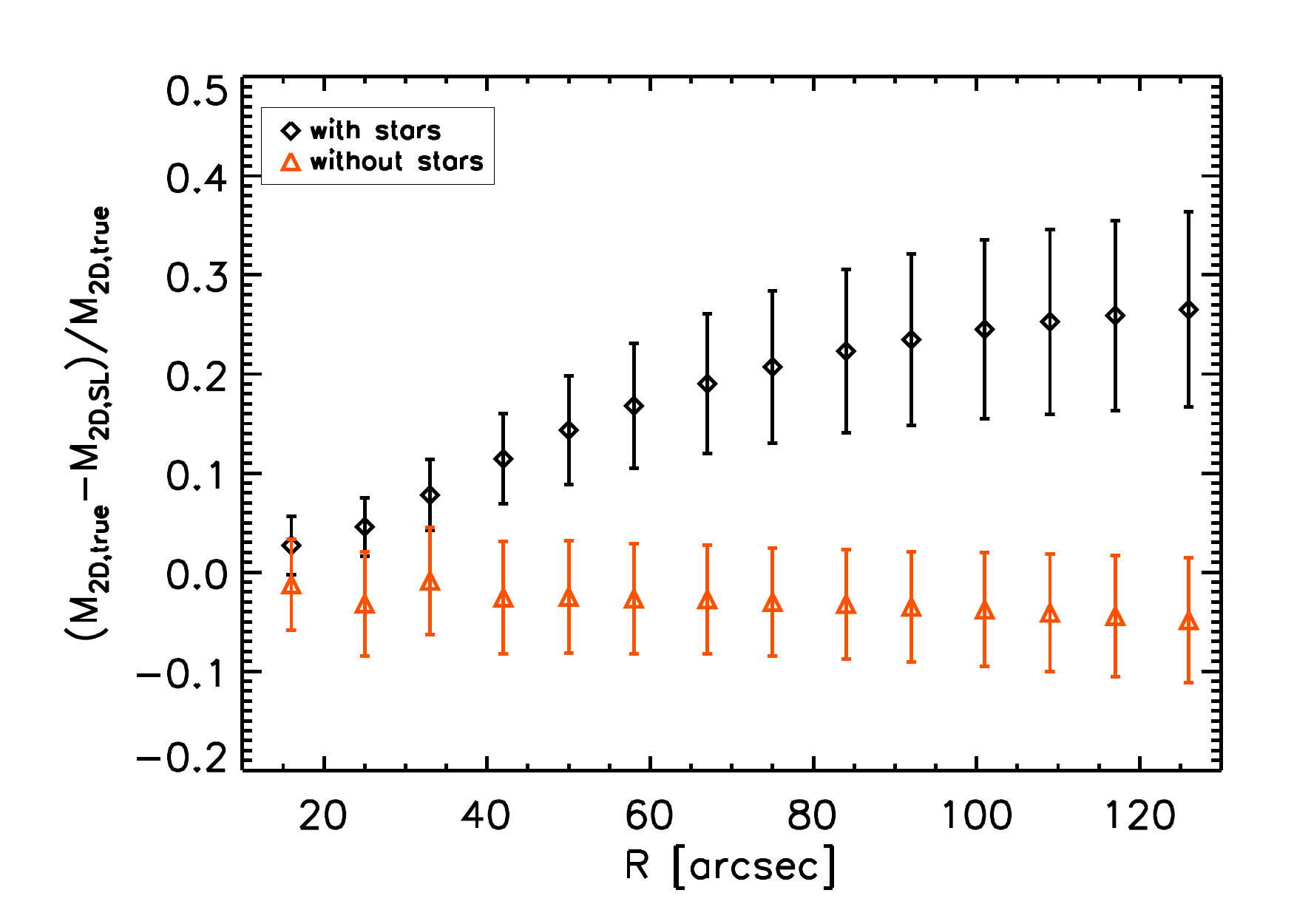}
  \caption{The relative difference between true and estimated projected masses as a function of the distance from the cluster centers. The results are obtained by averaging over all the clusters in the sample. The diamonds and the triangles refer to the simulations including and excluding the contribution of the BCG to the lensing signal (see text for more details). The errorbars show the scatter among all the reconstructions. The regions probed by SL are typically smaller than $\sim 30-40$ arcsec, thus the mass estimates at larger radii are extrapolations of the SL mass model to distances which are un-constrained by the data.}
\label{fig:slmassbias}
\end{figure}

\subsubsection{Strong lensing masses} 
\label{sect:sl}
In Fig.~\ref{fig:mslmt}, we compare the true and the estimated strong-lensing masses of all clusters in our sample. The two-dimensional masses are measured at the limits of the strong lensing regions, given by the mean distance of the tangential images from the cluster center. The agreement between the estimated and true masses is remarkable, showing that, in most cases, strong lensing methods based on parametric modeling are accurate at the level of few percent at predicting the projected inner mass. The worst results are obtained for clusters $g51-z$ and $g72-z$, for which the offsets between the true and the estimated mass are $\sim 15\%$ and $\sim 20\%$, respectively. These two clusters have complex morphologies, {  as shown in Fig.~\ref{fig:sdmaps}}, being characterized by multiple mass components and double cores. For $g51-z$ the mass model combines 12 mass components, three of which are optimized individually. The cluster $g72-z$ is modeled with 78 components. In this projection the cluster appears as a double cluster, with two main mass clumps separated by $\sim 50"$. Each of these clumps is modeled with an NFW profile and has a massive star concentration at the center, which is modeled as a PIEMD. These four components are optimized individually. Given the complexity of the systems and the large number of free parameters (24 for $g72-z$), it is reasonable to expect less accurate mass estimates for these two clusters. 

Although the mass measurement is very precise at the position of the tangential critical lines, {  as anticipated in Sect.~\ref{sect:strlenan}}, the extrapolation of strong lensing models to larger radii such as $R_{2500}$, $R_{500}$, or $R_{200}$ could lead to wrong mass estimates. In particular, we find that the results are very sensitive to the parameterization chosen to model the BCG, when we include the central images in the optimization. Using the wrong model leads to biased mass estimates. In our simulations we adopted a PIEMD which is a broadly used profile for describing the central galaxies in clusters. Such model, however, is not adequate to describe the profile of the BCGs in our simulations. While the PIEMD model is isothermal, i.e. the surface density profile decreases with the distance from the center as $\Sigma_{\rm PIEMD}(R)\propto R^{-1}$, the true surface density  distribution of the stars in our numerical simulations is steeper, i.e. $\Sigma_{\rm BCG} \propto R^{-1.7}$. This implies that, by fitting the lensing constraints as we do, we  impose  that the BCG is more spatially extended,  and an additional contribution to the central mass has to be provided by the dark matter halo.  This causes to systematically over-estimate of the halo concentration and under-estimate of the scale radius, as shown in Table~\ref{tab:bfc}.
For this reason, the masses extrapolated to large radii are systematically under-estimated. Such a dependence on the BCG mass has been reported recently by \cite{ 2009MNRAS.398..438D} modeling the cluster MS2137 \citep[see also][]{CO05.1}. 
The diamonds in Fig.~\ref{fig:slmassbias} show the mean relative differences between true and estimated projected masses of the clusters at different distances from the centers. As the distance from the cluster center grows, the amplitude of the bias in the mass estimates increases, being of the order of $\sim 30\%$ at a distance of two arcmins. The size of the errorbars reflects the scatter among the reconstructions. Note that also the scatter grows as a function of the distance from the center.

In order to better illustrate how the results shown above depend on the stellar  component of the lenses,  we ran a new set of lensing simulations using only the dark matter halos of the clusters as deflectors. Then, we repeated the fitting procedure and we derived the mass profiles as done before, but without modeling the BCGs. In this case, even extrapolating the profiles to radii much larger than the size of the strong lensing region, we find a much smaller disagreement between true and estimated masses. In this case, at large radii the strong lensing masses tend to be on average only slightly larger than the true masses ($<5\%$). This is shown by the triangles in Fig.~\ref{fig:slmassbias}.     

\subsubsection{Weak Lensing masses}
\label{sect:wlmasses}
{  We discuss now the results obtained from the weak-lensing analysis of the clusters in our sample.}

The accuracy of the mass estimates depends on the morphology of the lenses and on their substructures.
No big substructures are contained into the cylinders in the case of cluster $g1$. As shown in Fig.~\ref{fig:sdmaps}, the cluster appears regular in all the three projections. Instead, clusters $g51$ and $g72$ have more complex morphologies. In particular, as explained above, $g72$ has a massive companion ($\sim 1.4 \times 10^{14}\;h^{-1}$Mpc) located at $\sim 2.5\;h^{-1}$Mpc from the center. When projected along the $z$-axis of the simulation box the substructure is very close to the cluster center ($\sim 300\,h^{-1}$kpc). For this reason, this mass clump has been included in the strong lensing model of $g72-z$. Although it lays outside $R_{200}$ in the projections $g72-x$ and $g72-y$, this substructure produces a significant shear which complicates the weak lensing mass measurements of the main cluster clump, as described below.  

\begin{figure*}[t!]
  \includegraphics[width=0.33\hsize]{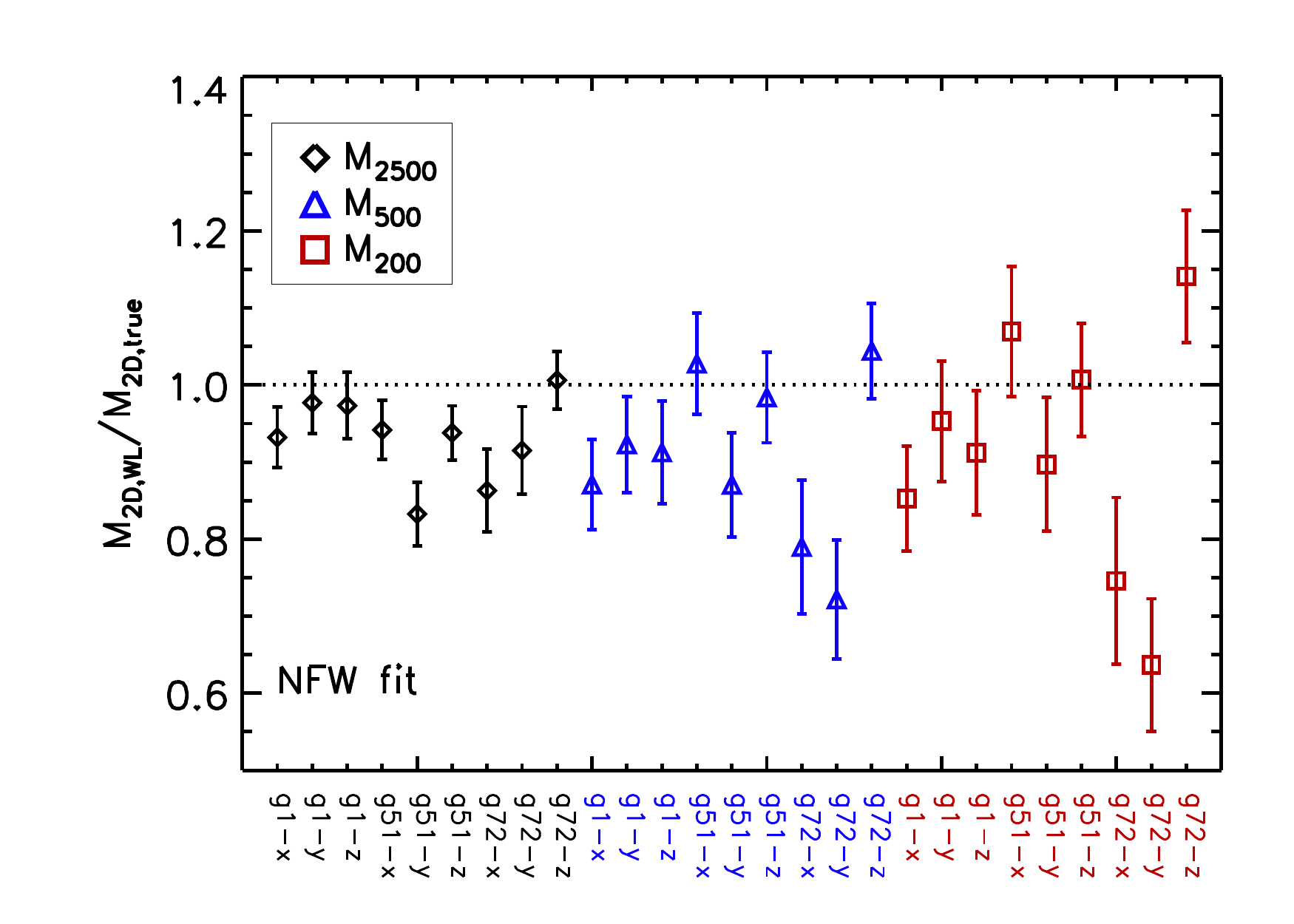}
  \includegraphics[width=0.33\hsize]{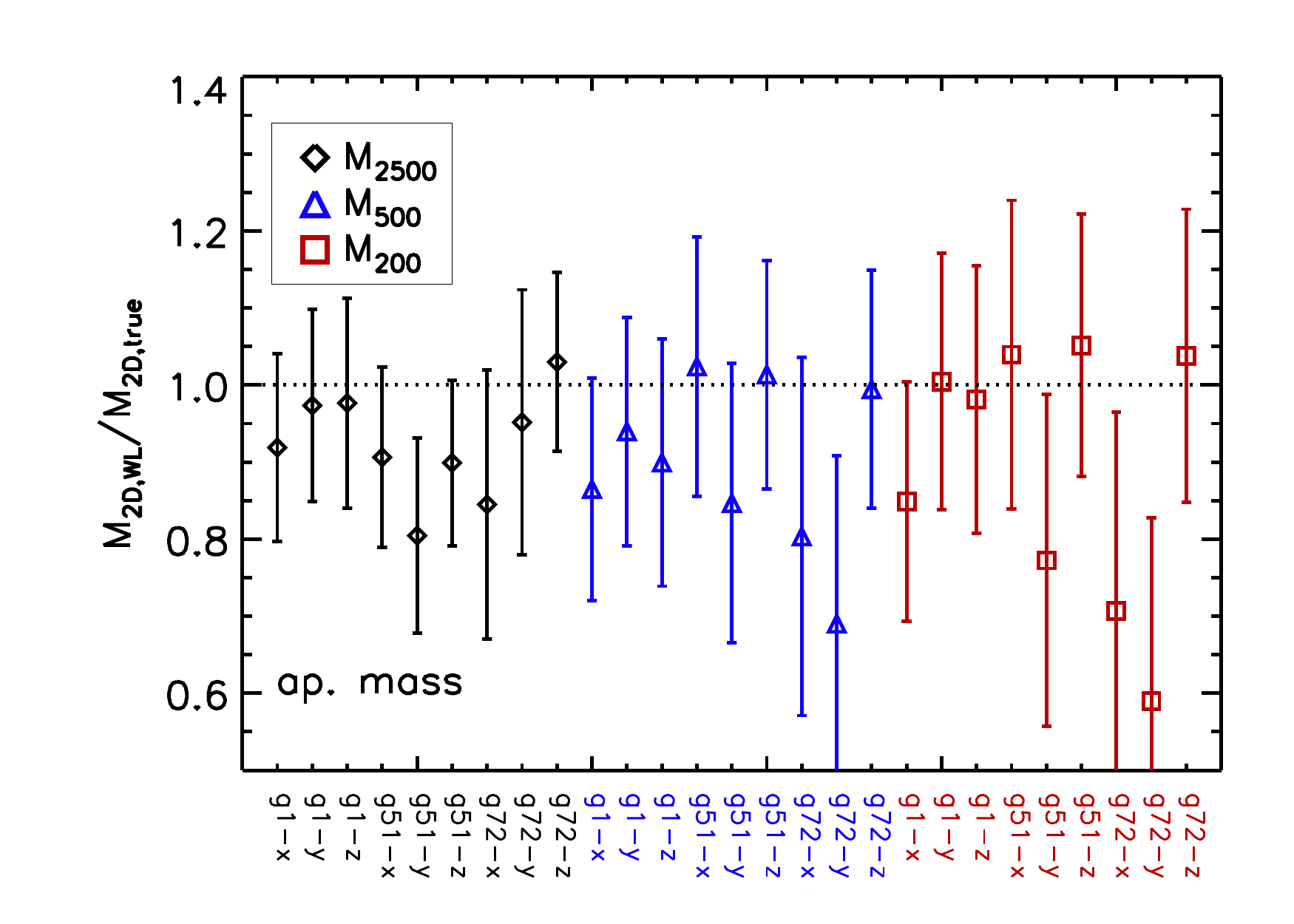}
  \includegraphics[width=0.33\hsize]{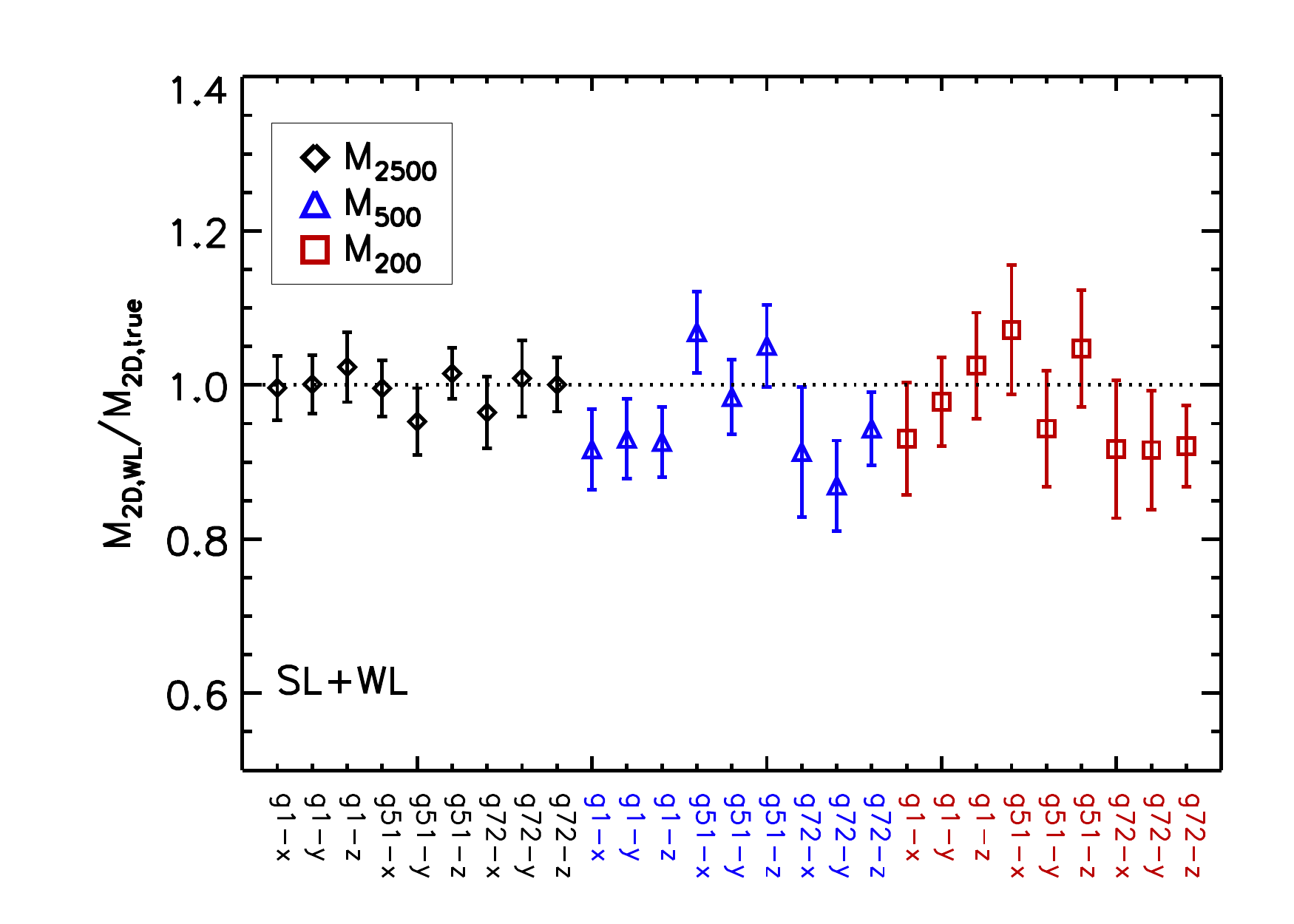}
  \caption{Comparison between the weak-lensing and the true 2D-masses of all the simulated clusters. From left to right, the panels refer to the methods based on the NFW fit of the shear profile, on the aperture mass densitometry, and on the non-parametric SL+WL reconstruction of the lensing potential, respectively. Shown are the ratios between the estimated and the true masses measured at three characteristic radii, namely $R_{2500}$ (diamonds), $R_{500}$ (triangles), and $R_{200}$ (squares), {  versus the cluster names.}}
\label{fig:wlmasses}
\end{figure*}  

In Fig.~\ref{fig:wlmasses} we show the ratios of the estimated and of the true 2D-masses of the clusters in the sample. The measurements have been made at $R_{2500}$ (diamonds), $R_{500}$ (triangles), and $R_{200}$ (squares). The different panels refer to the three methods implemented in this work, namely the NFW fit of the tangential shear profile (left panel), the aperture mass densitometry (central panel), and the SL+WL method (right panel). 
 The reliability of the methods depends strongly on the morphology of the lenses. 

Two methods assume that the lensing signal is tangential to the center of the lens, which is assumed to coincide with the BCG. This implies that all shear is assumed to be produced by a main cluster clump. These are the NFW fit and the aperture mass densitometry methods.  For some lenses, like the three projections of $g1$, such approximation is rather well met. Indeed, this cluster is free of large substructures and the shear is dominated by the main cluster clump. For some other lenses, like the three projections of $g72$, this is not the case. In all three projections of this cluster the shear produced by the secondary mass clump contributes significantly to the total shear. As a result, in the projection along the $x-$ and the $y-$axes, the tangential shear with respect to cluster center is smaller than it should be in absence of the secondary mass component. When fitting the shear profile with a single NFW model or converting the shear into a mass estimate using the aperture mass densitometry, this causes to under-estimate the mass. The amplitude of this effect grows with increasing distance from the cluster center, because the secondary mass clump is approached. In the case of $g72-x$ and $g72-y$, using the NFW fit method, the mass is underestimated by $\sim 10-15\%$ at $R_{2500}$ and by $\sim 30-40\%$ at $R_{200}$. Similar results are found with the aperture mass densitometry. In the case of $g72-z$, where the secondary clump is located near the cluster center, the shear produced by the main cluster and by the substructure sum up. The resulting tangential shear mimics that of a lens with a larger mass and with an  extended core, because of the offset between the two mass clump centers. As a result, we find that the estimated $M_{200}$ exceeds the true value by $\sim 15-20\%$.      

In order to validate our interpretation of these results, we perform the following tests. First, we place two NFW halos of mass $M_1=6.8\times 10^{14}\;h^{-1}M_\odot$ and $M_2=1.4\times10^{14}$ at a distance of $2.5\;h^{-1}$ Mpc, and we calculate the shear field produced by these two mass clumps. Then, we sample the shear field at random galaxy positions and add the noise due to the intrinsic galaxy ellipticities, following the method outlined in \cite{MA05.2}. Finally we measure the tangential shear profile with respect to the center of the most massive halo and we fit it using a single NFW profile, over a radial range similar to that used in our numerical simulations. We repeat the experiment after removing the least massive halo. We find that, if the shear produced by the second halo is included in the simulation, the masses at $R_{2500}$, $R_{500}$, and $R_{200}$ are under-estimated, with respect to the simulation including only the most massive halo, by $4\%$, $21\%$, and $30\%$, respectively. This is qualitatively in agreement with the results for $g72-x$ and $g72-y$. Note that in both cases, the tangential shear profile is well fitted by an NFW profile, i.e. with reduced $\chi^2\lesssim1$. A a second test, we place the least massive halo at a distance of $300\;h^{-1}$kpc from the main halo. The resulting tangential shear profile is compared with that produced by a single mass component of total mass $M=M_1+M_2$. By fitting with NFW models, we find that, in the two-halo case, the masses at $R_{2500}$, $R_{500}$, and $R_{200}$ are over-estimated, by $3\%$, $17\%$, and $25\%$, respectively, with the respect to the single halo case. This is also in good agreement with our results for $g72-z$.  

For the remaining clusters, the mass estimates obtained with these two methods are more accurate. Typically, the estimated masses deviate by $\lesssim 15\%$ from the true masses. Even in the case of $g51$, despite the presence of few substructures in the cluster surroundings, the mass estimates are in good agreement with the input masses. The reason is that the above mentioned substructure are less massive than in the case of $g72$ and the shear is dominated by the cluster halo.

The non-parametric SL+WL method does not require any assumption on the symmetry of the lensing signal. Substructures are included in the mass model by construction. For this reason, the SL+WL method can recover the input mass with good precision even in the case of morphologically disturbed clusters. We find that the deviations between estimated and true masses are typically below the $10\%$ level. 

Note that, for all the three methods the scatter between estimated and true masses grows as a function of the radius, in agreement with \cite{2009arXiv0903.1103O}.

\subsection{Lensing 3D masses}
\label{sect:l3d}
As stated several times above, lensing measures the total mass contained in cylinders and projected on the sky. To convert the 2D-mass into 3D-mass estimates, deprojection needs to be implemented. This requires to make some assumptions on the shape of the clusters and on their three-dimensional density profile. We assume here that clusters are spherical and that their density profile is well described by the NFW model.

\begin{figure*}[t!]
  \includegraphics[width=0.33\hsize]{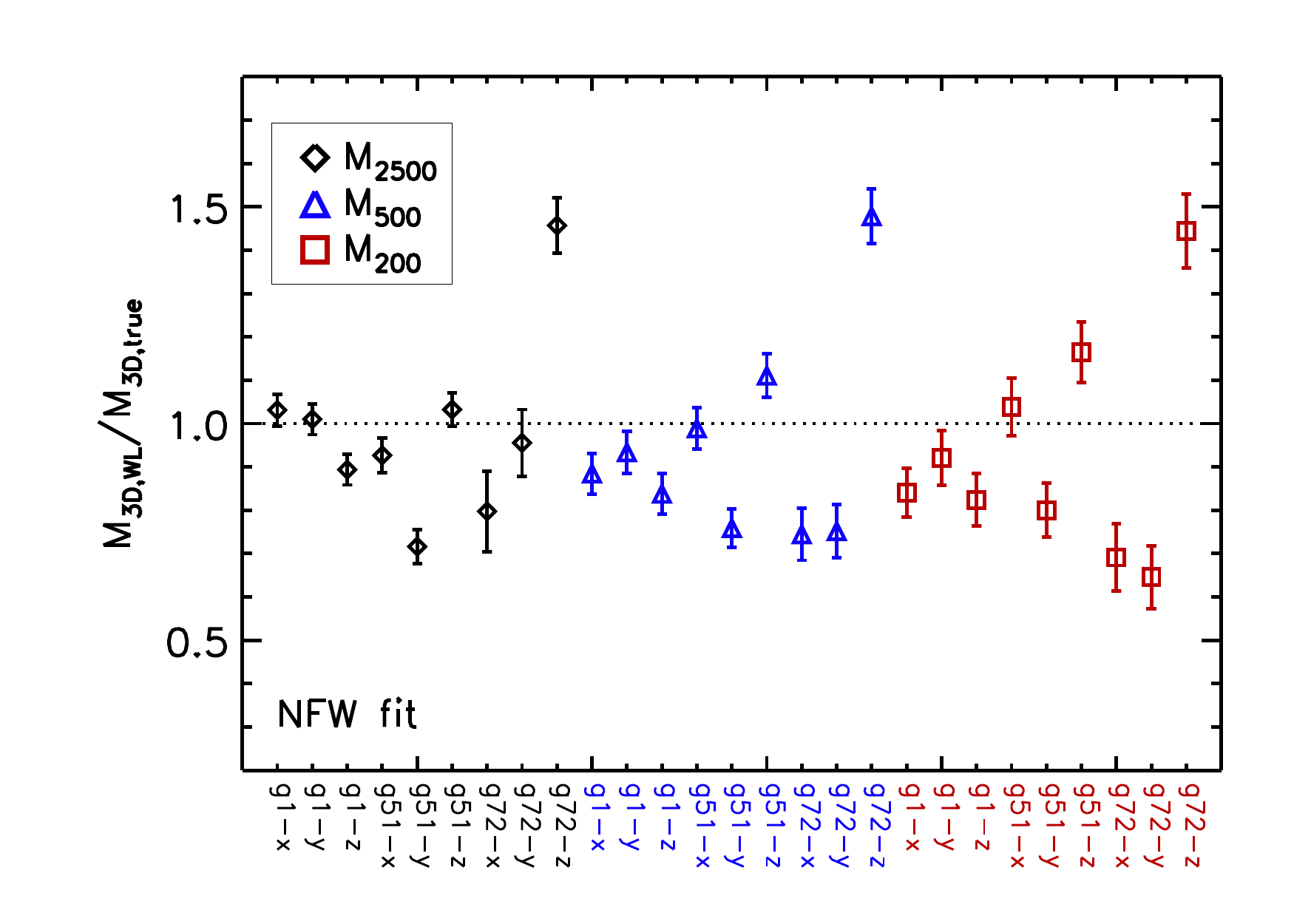}
  \includegraphics[width=0.33\hsize]{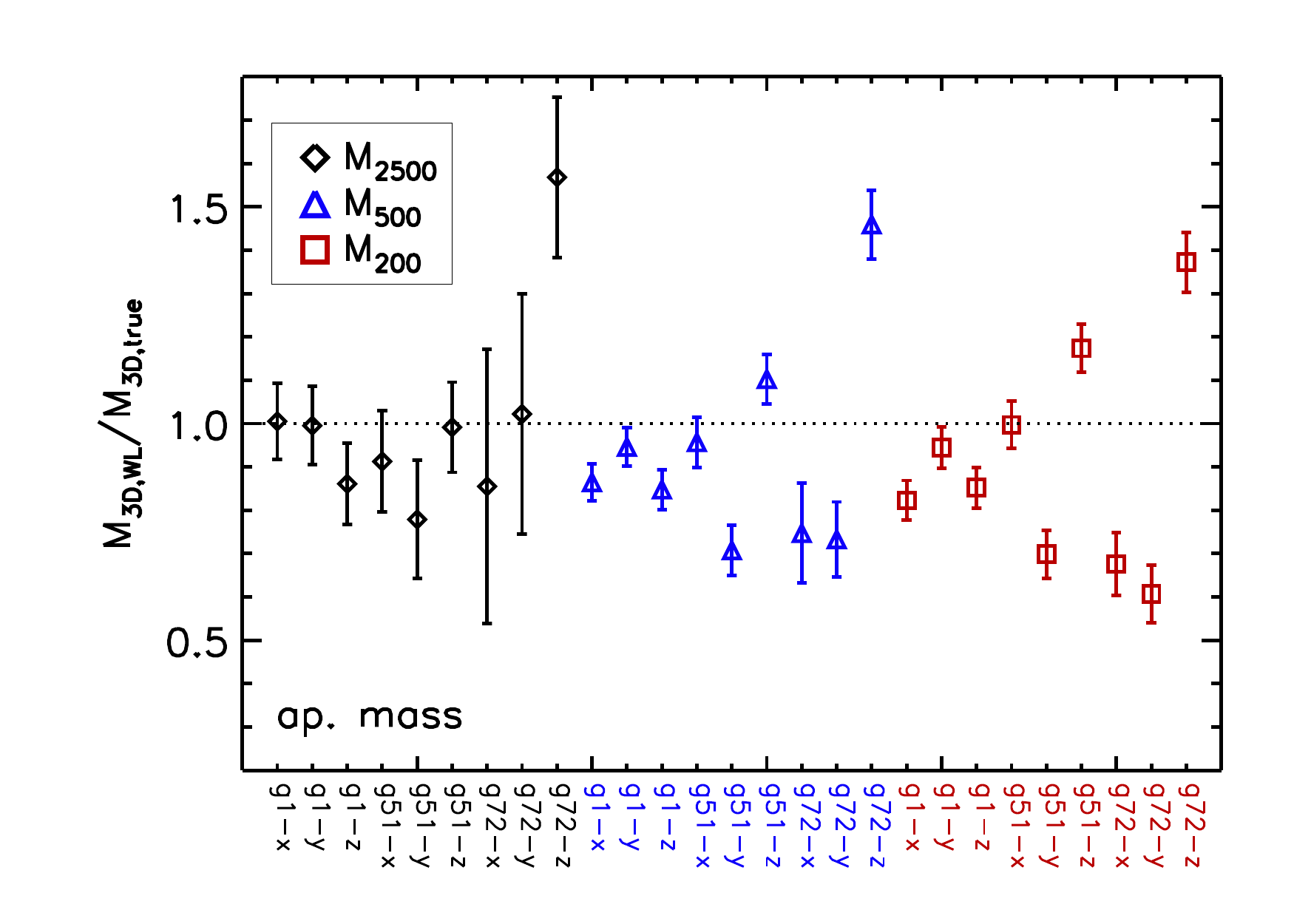}
  \includegraphics[width=0.33\hsize]{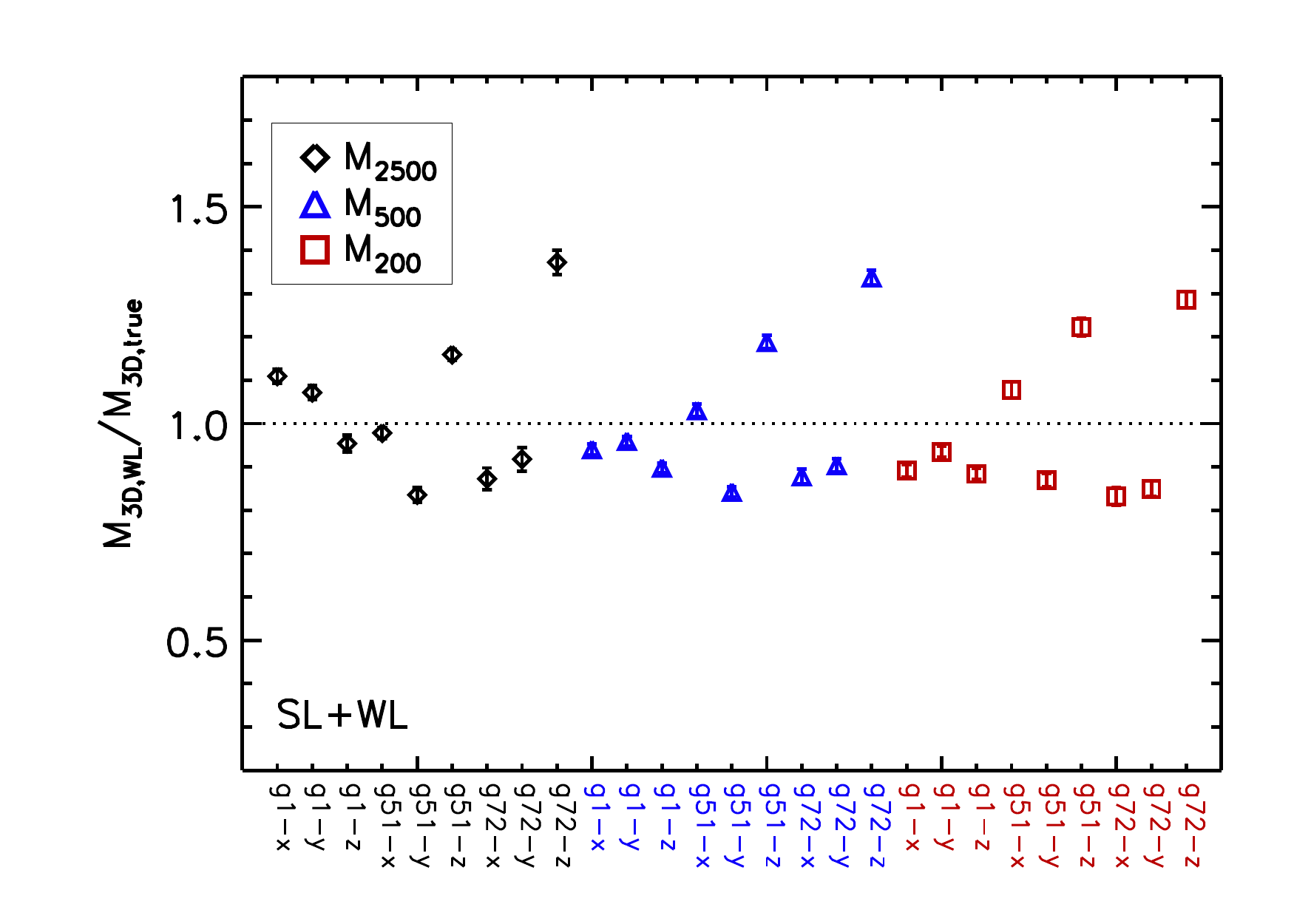}
  \caption{As in Fig.~\ref{fig:wlmasses}, but comparing the estimated and true 3D-masses.}
\label{fig:wl3dmasses}
\end{figure*} 

De-projection is done differently for the three methods investigated here. 
For the NFW fit method, we use the NFW parameters obtained from the fit of the tangential shear profile to calculate the 3D-mass profile of the lenses. For the aperture mass densitometry and for the SL+WL methods, we fit the 2D-mass profiles with projected NFW models and we use the best fit parameters to derive the 3D-mass profiles.  {  These 3D-profiles are shown in Fig.~\ref{fig:indmasses} for the individual cluster projections.}

\begin{figure}[t!]
  \includegraphics[width=1.0\hsize]{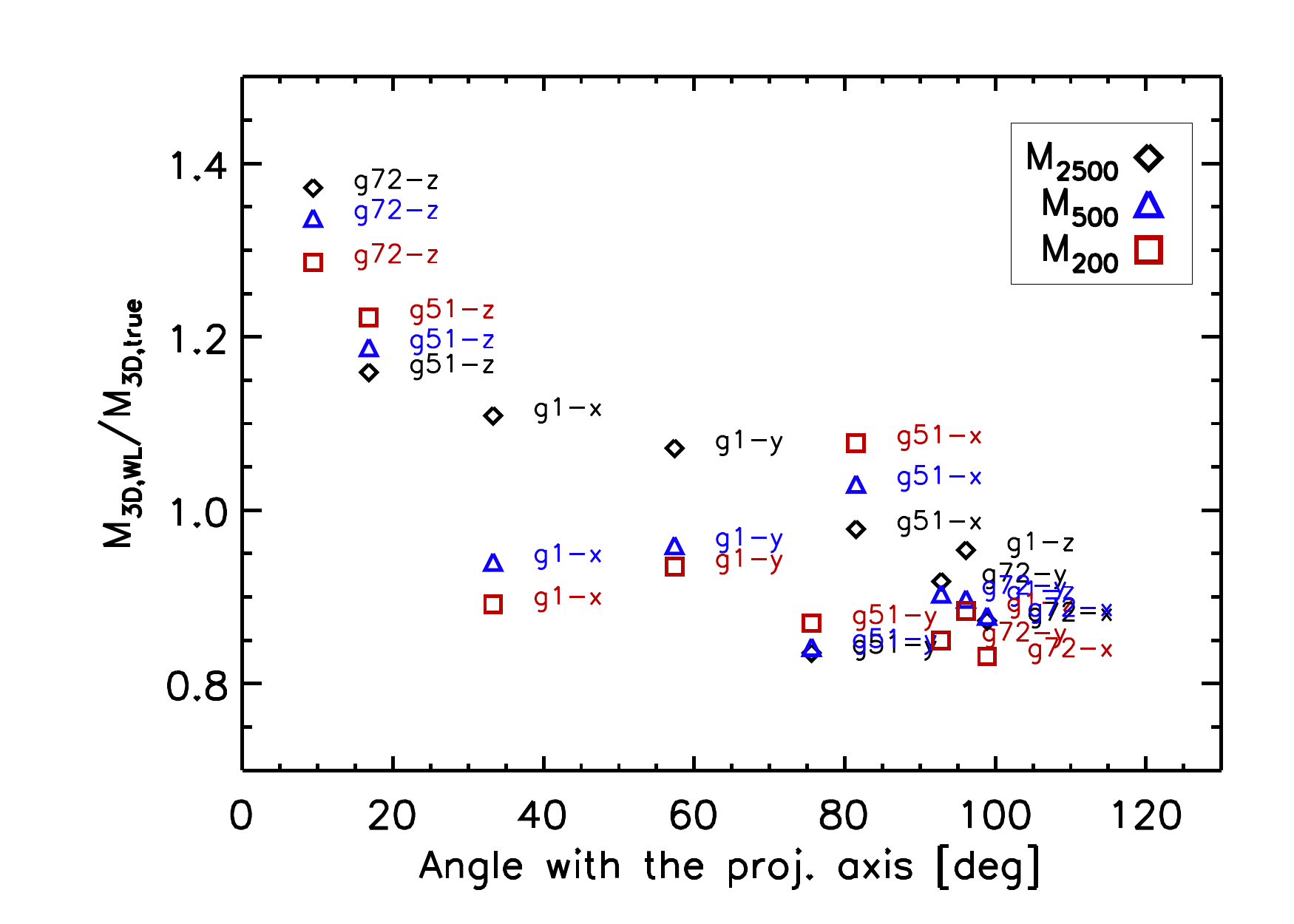}
  \caption{Ratio between estimated and true lensing masses as a function of the angle between the major axis of the cluster inertia ellipsoid and the axis along which the mass distribution is projected. The results are shown for the lensing masses obtained with the SL+WL method. Squares, triangles and diamonds indicate the mass measurements at $R_{200}$, $R_{500}$, and $R_{2500}$, respectively.}
\label{fig:3do}
\end{figure}

Similarly to Fig.~\ref{fig:wlmasses}, we show in Fig.~\ref{fig:wl3dmasses} the ratios between estimated and true 3D-masses at three over-density radii. It is clear that in 3D the scatter between the estimates and the input masses is significantly larger than in 2D. This is due to two main reasons. First, lenses are triaxial, while we are assuming spherical symmetry during the de-projection. Depending on the degree of triaxiality and on the orientation of the clusters with respect to the line of sight, 3D-masses may result to be over- or under-estimated. In particular, we find that, in the cases of good alignment (i.e. small angles) between the major axis of the cluster and the projection axis, the lensing masses tend to be systematically larger than the true masses, while the opposite occurs in those cases where the major axis is nearly perpendicular to the line of sight. This is shown in Fig.~\ref{fig:3do}, where the lensing masses are derived with the SL+WL method, which even in 3D seems to provide the most accurate mass estimates. Given that the masses of ellipsoids and spheres with the same azimuthal density profile tend to converge at large distances from their centers, the effect is strongest for $M_{2500}$ and for $M_{500}$, and mildest for $M_{200}$. However, even at $R_{200}$, the analysis of our sample shows that the scatter due to triaxiality is of the order of $\sim 20\%$. Similar results are found by \cite{2007MNRAS.380..149C}. 

The second factor which makes the 3D lensing mass estimates so noisy is the presence of substructures along the line of sight. Since their distance from the lens plane is unknown, the 3D-mass estimates can be severely affected by these mass clumps, especially if they are located close to the cluster core in projection. The high ratio between the estimated and the true mass of $g72-z$ is in large part due to the presence of the massive sub-clump previously mentioned in the paper. This accounts for $\sim 15\%$ of the total cluster mass, but its erroneous inclusion in the central $300\,h^{-1}$kpc significantly affects the mass estimates at small radii. 

Apart from this particular cluster projection, in the cases of systems without large substructures along the line of sight, 3D-lensing masses are affected by an intrinsic uncertainty due to triaxiality, which we estimate to be of the order of $20\%$ at $R_{200}$. Unfortunately, at smaller radii where the lensing measurements of the 2D masses would be more robust, the scatter becomes larger, being of the order of $50\%$.  

Recently \cite{2009arXiv0903.1103O}, by studying the weak lensing signal of 30 galaxy clusters observed with the SUBARU telescope, found that the mean ratios between 3D- and 2D-masses at $R_{\rm vir}$ and $R_{500}$ are $1.34\pm 0.17$ and $1.40 \pm 0.10$, respectively, where $R_{\rm vir}$ is the virial radius. They derive the 2D-masses using the aperture mass densitometry method, while the 3D-masses are obtained from the NFW fits of the shear profiles. Their sample spans a range of masses which is much wider than that covered by our sample. From their Fig.~8, we can estimate that, limiting the analysis to masses $M_{\rm vir} \ge 8\times10^{14}\;h^{-1}\;M_\odot$ and $M_{500}\ge 4\times 10^{14}\;h^{-1}\;M_\odot$, the ratios are smaller ($\sim 1.06$ and $\sim 1.37$ at $R_{\rm vir}$ and at $R_{500}$, respectively). Averaging over our sample, we find $M^{\rm ap}_{\rm vir,2D}/M^{\rm NFW}_{\rm vir,3D}=1.14\pm 0.09$ and  $M^{\rm ap}_{\rm 500,2D}/M^{\rm NFW}_{\rm 500,3D}=1.29\pm 0.08$. 

These are quite in a good agreement with the the ratios between the true 2D- and 3D-masses, which are $M^{\rm true}_{\rm vir,2D}/M^{\rm true}_{\rm vir,3D}=1.13$ and 
$M^{\rm true}_{\rm 500,2D}/M^{\rm true}_{\rm 500,3D}=1.34$. From these results, we can deduce that a) on average, the ratios between 2D- and 3D-masses are well recovered, despite triaxiality and substructures affecting individual mass estimates, and b) the agreement between simulations and observations is an indication that the density profiles of real and simulated clusters are, within the errors, compatible with each other. 

\subsection{X-ray masses}
\label{sect:xresu}

\begin{figure*}[t!]
  \includegraphics[width=0.5\hsize]{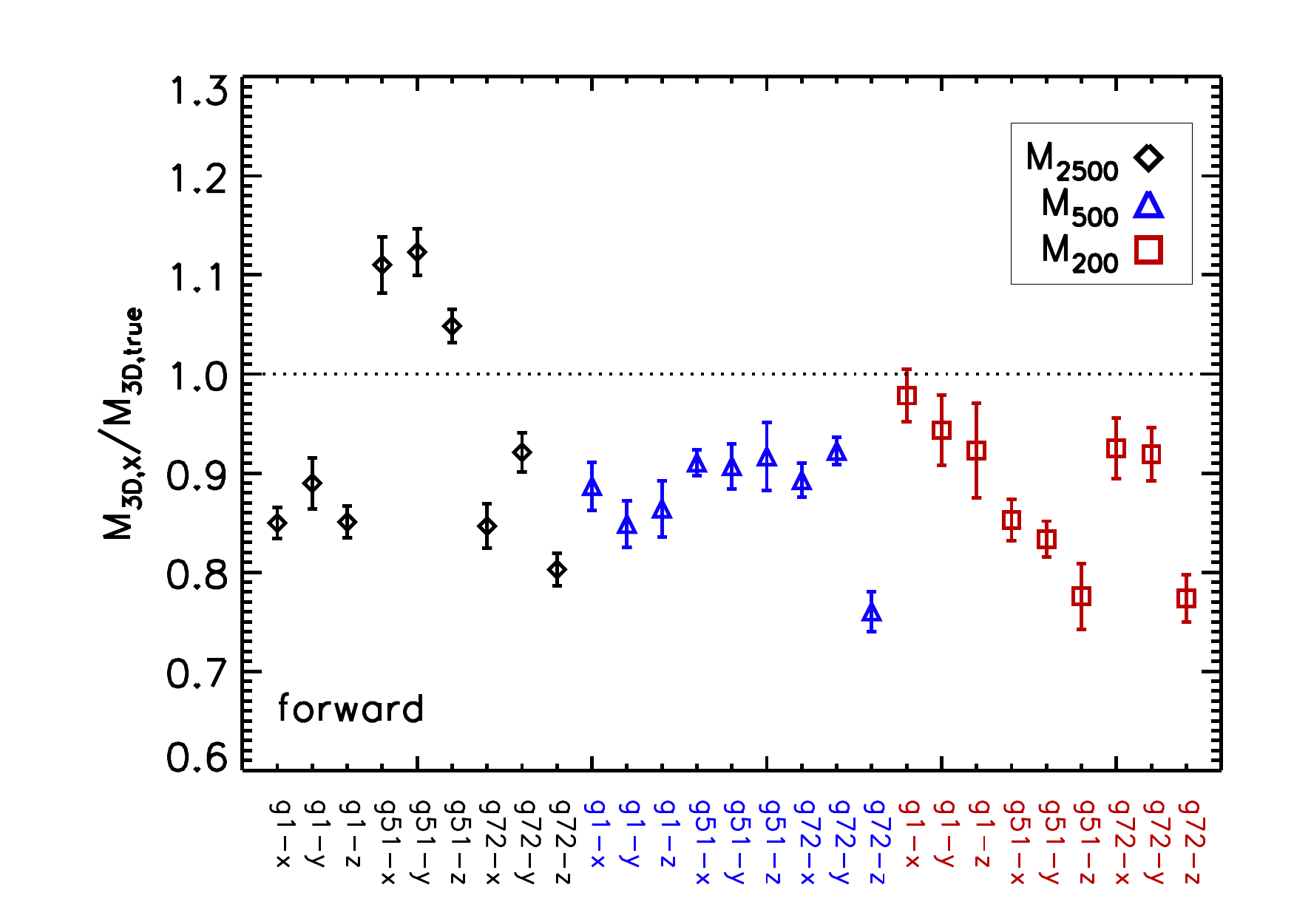}
  \includegraphics[width=0.5\hsize]{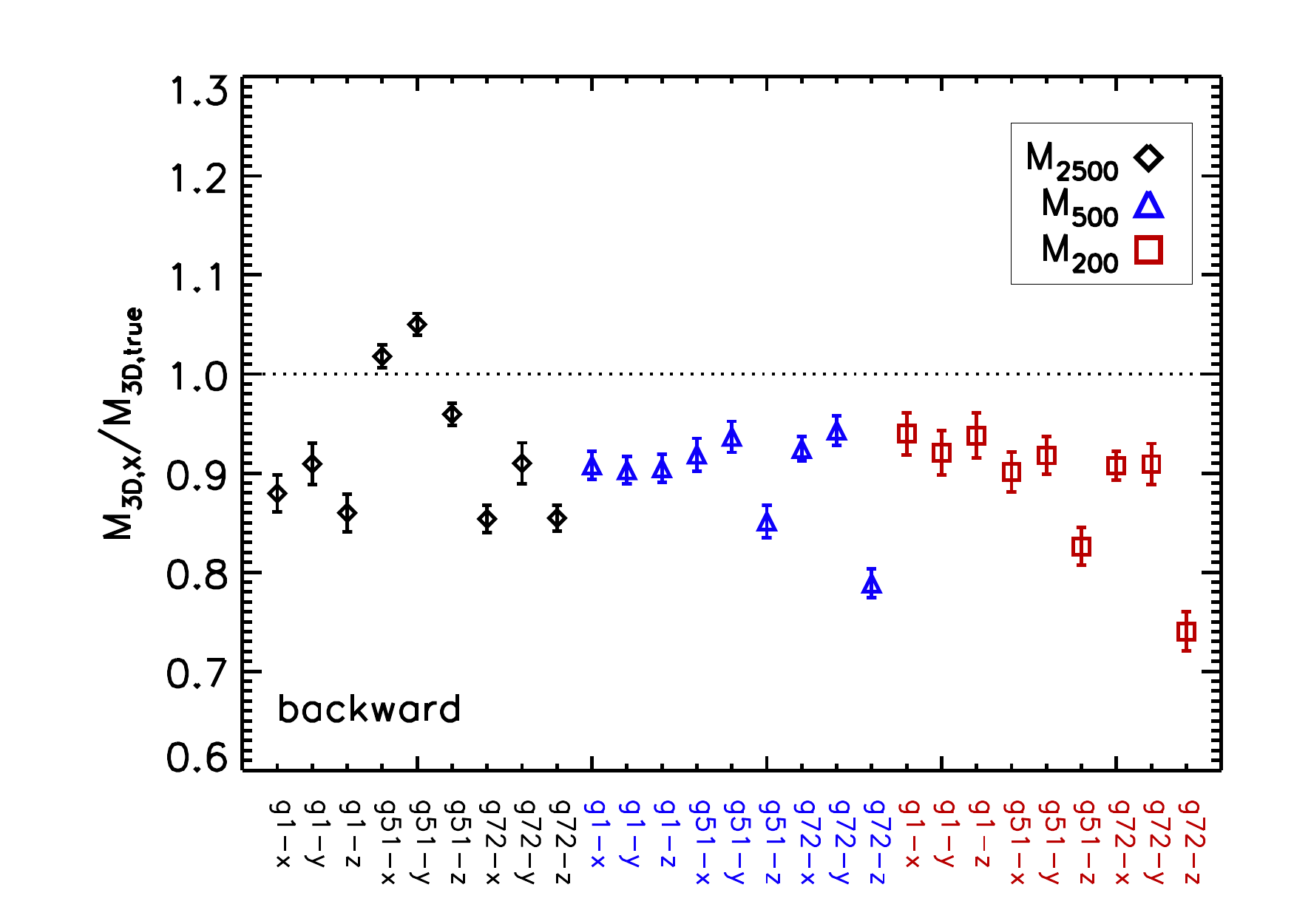}
    \includegraphics[width=0.5\hsize]{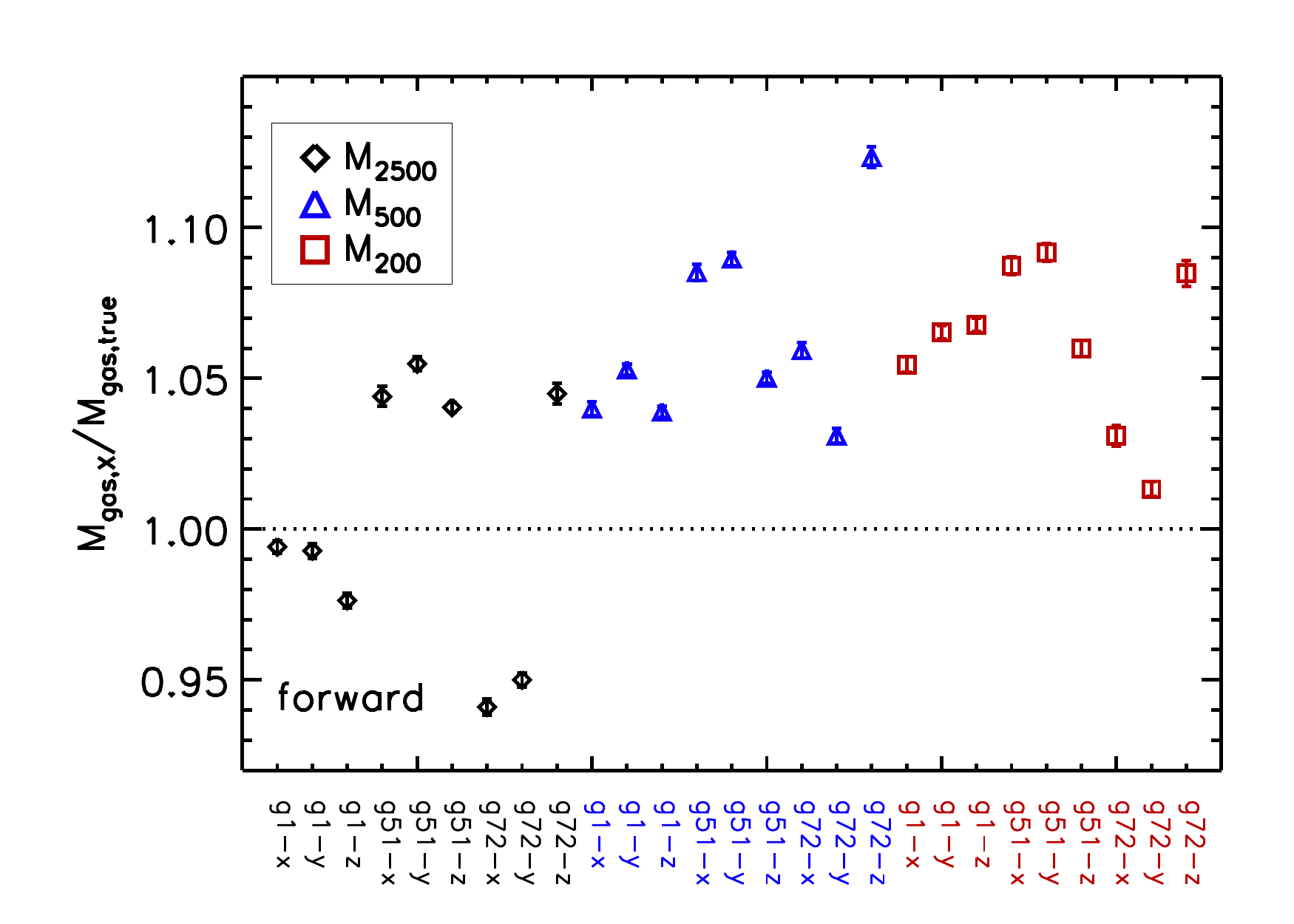}
  \includegraphics[width=0.5\hsize]{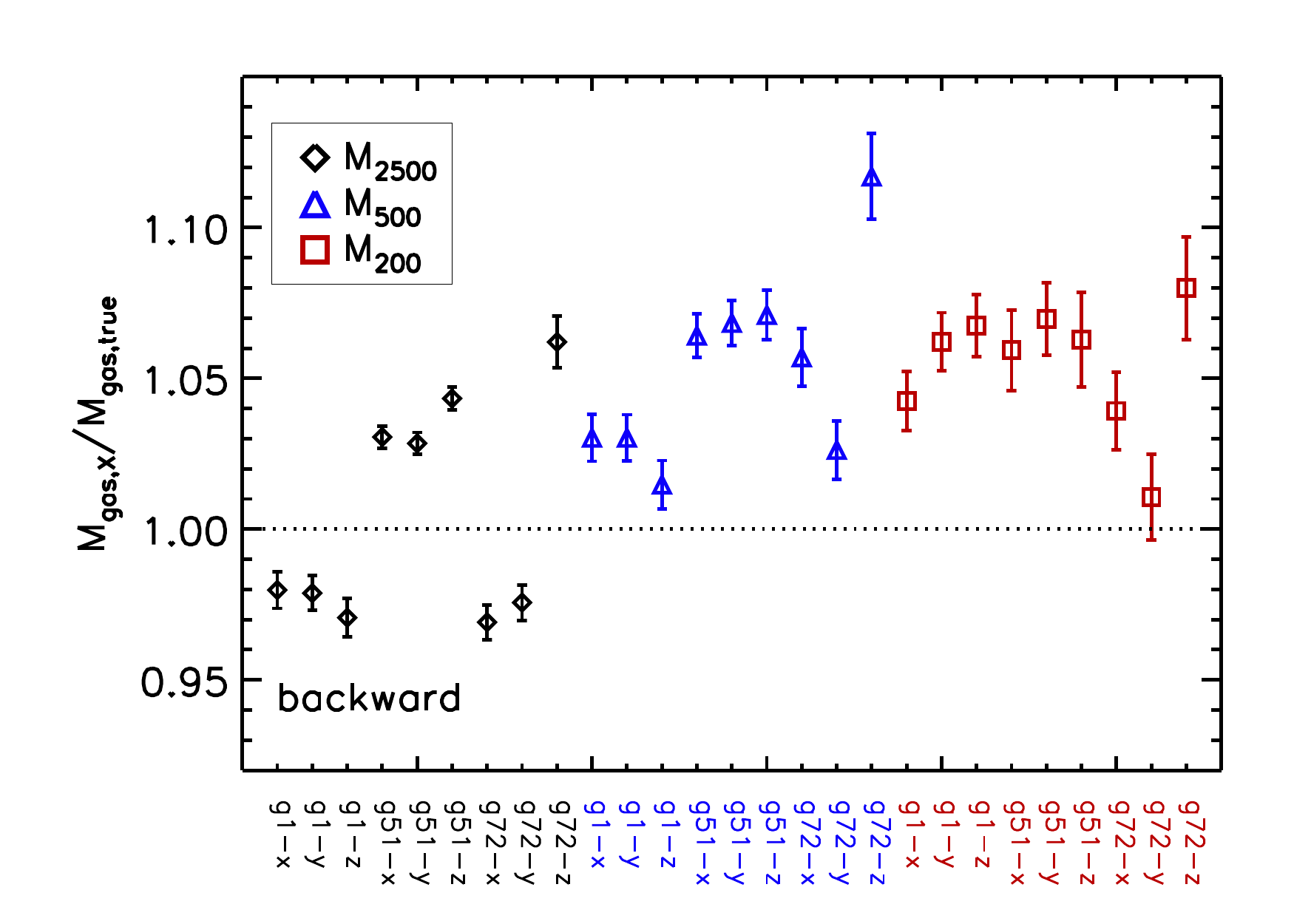}
  \caption{Top panels: comparison between the X-ray and the true 3D-masses of all the simulated clusters. The left and the right panels refer to the forward and to the backward methods, respectively. Shown are the ratios between the estimated and the true masses measured at three characteristic radii, namely $R_{2500}$ (diamonds), $R_{500}$ (triangles), and $R_{200}$ (squares), as a function of  the cluster name. 
  Bottom panels: as for the upper panels, but comparing the gas masses derived from the X-ray analysis to the true gas masses of the clusters.}
\label{fig:xraymasses}
\end{figure*}

In Fig~\ref{fig:xraymasses} we compare the true and estimated total and gas masses for all the cluster in our sample. The left and the right panels refer to the {\it forward} and to the {\it backward} methods described in Sect.~\ref{sect:X-ray}.
{\it The top panels} show  the ratios between the estimated and the true masses measured at three characteristic radii, namely $R_{2500}$ (diamonds), $R_{500}$ (triangles), and $R_{200}$ (squares), as a function of  the cluster name. {  A comparison of the mass profiles at any other radius is shown in Fig.~\ref{fig:indmasses}.} The forward and backward methods give results which, at the three different characteristic radii, are consistent within $2\sigma$ for all the clusters. {  The reconstruction of the X-ray masses through the two methods can differ because, while the backward method assumes a functional form of the total density profile imposing the gas temperature profile for given gas density profile, the forward method reproduces the gas density and temparture with very flexible models that, in particular in the outer regions, can induce not stable extrapolations. For the forward method, the confidence intervals on mass profiles
have been estimated by minimising the distance between
the projected models and a set of 100 random realisations of observed
profiles. These profiles have been obtained assuming
Gaussian statistics around observed values of the surface brightness and temperature profiles
with the constraint of rejecting realisations leading to 
non-physical solution of non-monotonically increasing mass
profiles. For the backward method, errors on $M_{\rm gas,X}$ are estimated by propagation of the errors on the gas density profile as obtained from the direct deprojection of the surface brightness, reconstructed 100 times by considering the Gaussian error in each radial bin. The error on $M_{\rm 3D,X}$ is obtained from the propagation of the errors on the concentration parameter and scale radius estimated through Eq.~25.}
Similarly  to what shown in other work \citep[see e.g.][and references therein]{2006MNRAS.369.2013R,2008ApJ...674..728R, 2007ApJ...655...98N} we find that, overall the X-ray mass estimates are biased low by 5\%-20\%. More quantitatively we find that, on average, both methods underestimate the true mass
by 10\% with a scatter of 6\% almost independently of  the  characteristic radius considered.
The most deviant mass estimate comes only from the $z$ projection of cluster g72. This is due again to the fact that this cluster is composed by two structures that line-up along the line of sight (see Fig~1).
When we fit the surface brightness profile this appears to be steeper in the center and shallower at the characteristic radius.

{\it In  the bottom panels} of Fig.~\ref{fig:xraymasses} we compare the gas masses derived from the X-ray analysis to the true gas masses of the clusters.
Again, we find that the two methods give results consistent within $2\sigma$ errors.
Furthermore, in agreement with other work, we find that the total gas mass is well reconstructed \citep{2007ApJ...655...98N}. More quantitatively we find
that, on average, the gas mass is recovered to better than 1\% with a 3\% scatter  at $R_{2500}$ and    7\% with a 3\% scatter
at $R_{500}$ and $R_{200}$. It is important to say that the trend to slightly overestimate the gas mass at large radii is due to the fact that both X-ray methods 
tend to slightly overestimate the cluster density profiles in the outskirts due to a non perfect azimuthal symmetry of the cluster surface brightness at larger radii.
The most deviant gas mass estimates is given by the $z$ projection of cluster g72. Again this is due to  the same effect explained above for the mass. 

\subsection{NFW fit parameters}
\label{sect:nfwfits}

\begin{table*}[htdp]
\caption{Radial ranges used for fitting NFW profiles to the lensing and X-ray data. All radii are expressed in units of $r_{200}$ reported in Table 1.}
\centering
\begin{tabular}{lcccccccc}
\hline
\hline
cluster & min-max SL &  min-max WL & min-max SL+WL & min-max X-ray\\
\hline
g1 - x  & 0.002-0.08 & 0.16-1.00 & 0.03-1.00 & 0.45-0.90\\
g1 - y  & 0.002-0.09 & 0.16-1.00 & 0.03-1.00 & 0.45-0.90\\
g1 - z  & 0.002-0.06 & 0.16-1.00 & 0.03-1.00 & 0.45-0.90\\
g51 - x  & 0.003-0.07 & 0.20-1.00 & 0.05-1.00 & 0.50-1.00\\
g51 - y  & 0.002-0.05 & 0.20-1.00 & 0.05-1.00 & 0.50-1.00\\
g51 - z  & 0.002-0.12 & 0.20-1.00 & 0.05-1.00 & 0.50-1.00\\
g72 - x  & 0.006-0.05 & 0.20-1.00 & 0.05-1.00 & 0.54-1.08\\
g72 - y & 0.006-0.05 & 0.20-1.00 & 0.05-1.00 & 0.54-1.08\\
g72 - z & 0.002-0.05 & 0.20-1.00 & 0.05-1.00 & 0.54-1.08\\
\hline 
\hline
\end{tabular}
\label{tab:fitranges}
\end{table*}%

The best-fit NFW parameters for all the clusters in the sample, obtained both through the lensing and X-ray analyses {  in the radial ranges listed in Table~\ref{tab:fitranges}}, are summarized in Table~\ref{tab:bfc}. In column 4 of the table we report the best fit concentrations and scale radii obtained for the main cluster halo using the strong-lensing constraints. These results are obtained by including also the central images in the strong lensing modeling. As discussed in Sect.~\ref{sect:sl}, this leads to systematically overestimate the concentration and underestimate the scale radius, being the PIEMD model used in the SL modeling inadequate to describe the distribution of the stars in the simulations. In particular, we note that, in some cases, the concentrations are off by a factor of $\sim 3$ with respect to the true values reported in the second column. {  On average, the concentrations derived from SL alone are almost $85\%$ larger than the true concentrations of the DM-only profiles. Conversely, the scale radii are almost $\sim 50\%$ smaller. We remind that the SL fits refer to the cluster halos, while the stars are modeled apart.}. Such bias is not present in the simulations without stars, where the average ratio between estimated and true concentrations is around unity. 

The weak lensing best fit parameters, obtained from the fit of the shear profiles, are given in column 5. The concentrations tend to be slightly larger than those of the dark-matter distributions in the input models ($\langle c^{\rm WL}/c_{\rm true}^{\rm DM only} \rangle=1.17$) and the scale radii are on average $\sim 10\%$ smaller that the input values ($\langle r_s^{\rm WL}/r_{\rm s,true}^{\rm DM only} \rangle=0.93$). Similar results are found combining SL and WL (columns 6). In this case the average ratio between estimated and true concentrations is $\langle c^{\rm SL+WL}/c_{\rm true}^{\rm DM only} \rangle=1.15$ (for the scale radius we find $\langle r_s^{\rm SL+WL}/r_{\rm s,true}^{\rm DM only} \rangle=0.93$). The tendency to over-estimate more the concentration when adding the SL constraints is caused by the large contribution of the stellar and gas masses within the inner $100\;h^{-1}$kpc, as shown in Fig.~\ref{fig:mprofs}. We remind that in the cases of the WL and WL+SL methods the fits are done over the total projected mass profiles, thus without distinguishing the dark-matter component from the stellar and gas masses. To support this interpretation, we note that the largest discrepancies with the true concentrations arise for the cluster $g51$, which is characterized by an extended strong over-cooling region in the center, where the total density profile steepens compared to the DM only profile. This mimics a larger concentration, being the mass profile still compatible with an NFW model. Fitting the total density profiles of the inputs clusters indeed leads to higher concentrations and smaller scale radii, as reported in column 3, which are in a much better agreement with the results of  the lensing fits. 
As highlighted in the previous sections, the WL fits of some systems, like the three projections of $g72$,  can be strongly biased assuming a single mass component. Indeed, their substructures need to be properly modeled when deriving the cluster mass from the shear signal. The SL+WL method provides a better chance to measure the density profile, at least in some cases, like $g72-x$ and $g72-y$, where the large substructure laying at the edge of the cluster virial region is by construction properly modeled. For $g72-z$, where the substructure is near the cluster core, a correct fit of the projected mass distribution would require multiple mass components also for the SL+WL method. 

Comparable estimates of $c$ and $r_{\rm s}$ are obtained with the two X--ray techniques. Even the X-ray analysis probes the total mass distribution, including the gas and the stars. We note that, using  the {\it Backward method}, the concentration and the scale  radius are obtained
as best-fit parameters that minimize the reconstructed temperature profile with respect to the
observed values, whereas using the {\it Forward method} we have estimated them with a--posteriori
fit with a NFW functional form performed in the radial range $0.7-1.4\,h^{-1}$ Mpc on the mass model obtained
by appling the hydrostatic equilibrium equation on the gas density and temperature models described in Eqs.~(27) and (28), respectively.
Overall, the deviations from the true estimates go in the same direction: in $g1$, $c$ is underestimated 
consistently by a factor $0.7-0.8$, with a corresponding overestimate of $r_{\rm s}$ up to a factor $1.45$, if compared with the true concentrations obtained by fitting the input total mass distributions of the clusters;
the same considerations apply to $g72$, apart from $g72-z$, where the concentration (scale radius) 
is estimated higher (lower) than $c_{\rm true}^{\rm total}$ ($r_{\rm s, true}^{\rm total}$) by about 15 per cent as consequence of the alignment
along the line of sight of the two main clumps; more critical is the case of $g51$, where both the X--ray
methods provide a measure of the concentration that is twice $c_{\rm true}^{\rm total}$, due again to the large contribution
of the cool substructures in the central regions, as also shown from the overestimate of the total mass obtained
through X-ray analysis within $R_{2500}$ (see Fig.~18). Although we attempted to mask it in the X-ray analysis, the over-cooling region in this cluster is very extended, thus it is still affecting the mass reconstruction (see Fig.~\ref{fig:indmasses} in Appendix \ref{sect:indmasses}). {  On average the X-ray concentrations are $\sim 10\%$ and $\sim 20\%$ smaller than $c_{\rm true}^{\rm DMonly}$ and $c_{\rm true}^{\rm total}$, respectively, if the three projections of $g51$ are not included.}

\begin{table*}[htdp]
\caption{The NFW concentrations and scale radii (upper and lower part of the Table, respectively) resulting from the strong-lensing (column 4), weak-lensing (column 5), strong+weak lensing (column 6), and X-ray analyses (column 7-8 for the forward and for the backward methods) of the clusters in our sample. The weak-lensing estimates are obtained by fitting the shear profile with an NFW model. In columns 2 and 3, we quote the true concentrations obtained by fitting the DM-only and the total density profiles of the three clusters in the radial range between $10\,h^{-1}$ kpc and $r_{200}$. {  Below each column, we report the mean ratios of the estimated and the true parameters. These are calculated both including and neglecting the three projections of $g51$, which strongly bias the mean concentrations and scale radii derived from the X-ray analyses. The numbers in parentheses are the corresponding r.m.s. values.}}
\centering
\begin{tabular}{lccccccc}
\hline
\hline
cluster & $c_{\rm true}^{\rm DM only}$ &  $c_{\rm true}^{\rm total}$ & $c^{\rm SL}$  & $c^{\rm WL}$ & $c^{\rm SL+WL}$  & $c^{\rm X,forw}$ & $c^{\rm X,back}$\\
\hline
g1 - x & $ 4.62 $ & $ 5.38 $ & $11.31_{-0.59}^{ 0.19}$ & $ 6.21\pm 0.87$ & $ 6.5
9\pm 0.32$ & $ 3.58_{-0.06}^{ 0.06}$ & $ 3.59_{-0.28}^{ 0.32}$ \\
g1 - y & $ 4.62 $ & $ 5.38 $ & $10.57_{-1.81}^{ 2.82}$ & $ 4.82\pm 0.64$ & $ 5.3
9\pm 0.23$ & $ 4.51_{-0.08}^{ 0.08}$ & $ 4.05_{-0.25}^{ 0.40}$ \\
g1 - z & $ 4.62 $ & $ 5.38 $ & $ 6.92_{-1.12}^{ 2.25}$ & $ 4.44\pm 0.62$ & $ 4.5
4\pm 0.23$ & $ 4.06_{-0.07}^{ 0.07}$ & $ 3.62_{-0.31}^{ 0.22}$ \\
g51 - x & $ 5.37 $ & $ 7.20 $ & $ 3.36_{-0.11}^{ 0.46}$ & $ 5.83\pm 0.96$ & $ 6.
10\pm 0.29$ & $11.15_{-0.20}^{ 0.19}$ & $10.76_{-0.65}^{ 0.47}$ \\
g51 - y & $ 5.37 $ & $ 7.20 $ & $ 9.72_{-0.61}^{ 0.68}$ & $ 5.13\pm 1.05$ & $ 6.
35\pm 0.42$ & $12.39_{-0.22}^{ 0.22}$ & $10.80_{-0.56}^{ 0.56}$ \\
g51 - z & $ 5.37 $ & $ 7.20 $ & $ 8.78_{-0.46}^{ 0.19}$ & $ 6.06\pm 0.93$ & $ 7.
41\pm 0.32$ & $11.04_{-0.22}^{ 0.21}$ & $11.30_{-0.62}^{ 0.65}$ \\
g72 - x & $ 3.99 $ & $ 4.22 $ & $ 6.46_{-2.54}^{ 0.99}$ & $ 4.17\pm 1.54$ & $ 3.
88\pm 0.30$ & $ 3.31_{-0.05}^{ 0.05}$ & $ 3.22_{-0.02}^{ 0.06}$ \\
g72 - y & $ 3.99 $ & $ 4.22 $ & $ 7.26_{-0.14}^{ 1.47}$ & $ 7.91\pm 2.86$ & $ 4.
17\pm 0.31$ & $ 3.48_{-0.05}^{ 0.05}$ & $ 3.29_{-0.02}^{ 0.06}$ \\
g72 - z & $ 3.99 $ & $ 4.22 $ & $11.39_{-0.85}^{ 0.42}$ & $ 4.19\pm 0.62$ & $ 4.
50\pm 0.24$ & $ 4.92_{-0.08}^{ 0.08}$ & $ 4.49_{-0.29}^{ 0.38}$ \\
\hline
$c/c_{\rm true}^{\rm DM only}$ & & & 1.84 (0.60) & 1.17 (0.30) & 1.15 (0.15) & 1.33 (0.59) & 1.25 (0.56) \\  
$c/c_{\rm true}^{\rm total}$ & & & 1.59 (0.59) & 1.01 (0.33) & 0.98 (0.11) & 1.09 (0.38) & 1.03 (0.36) \\  
\hline
$c/c_{\rm true}^{\rm DM only}$ & no g51 & & 2.09 (0.48) & 1.23 (0.35) & 1.12 (0.15) & 0.92 (0.15) & 0.86 (0.12) \\  
$c/c_{\rm true}^{\rm total}$ & no g51 & & 1.88 (0.45) & 1.12 (0.35) & 1.00 (0.12) & 0.83 (0.15) & 0.78 (0.13) \\  
\hline 
\hline
 & $r_{\rm s,true}^{\rm DM only}$ & $r_{\rm s,true}^{\rm total}$ & $r_{\rm s}^{\rm SL}$  & $r_{\rm s}^{\rm WL}$ & $r_{\rm s}^{\rm SL+WL}$  & $r_{\rm s}^{\rm X,forw}$ & $r_{\rm s}^{\rm X,back}$\\
\hline
g1 - x & $  0.310 $ & $  0.278 $ & $  0.097_{ -0.002}^{  0.007}$ & $   0.229\pm   0.037$ & $   0.221\pm   0.012$ & $  0.408_{ -0.007}^{  0.007}$ & $  0.409_{ -0.037}^{  0.038}$ \\
g1 - y & $  0.310 $ & $  0.278 $ & $  0.090_{ -0.006}^{  0.064}$ & $   0.307\pm   0.048$ & $   0.276\pm   0.013$ & $  0.315_{ -0.006}^{  0.006}$ & $  0.360_{ -0.036}^{  0.026}$ \\
g1 - z & $  0.310 $ & $  0.278 $ & $  0.152_{ -0.031}^{  0.029}$ & $   0.317\pm   0.053$ & $   0.320\pm   0.017$ & $  0.351_{ -0.007}^{  0.007}$ & $  0.403_{ -0.027}^{  0.043}$ \\
g51 - x & $  0.241 $ & $  0.189 $ & $  0.385_{ -0.053}^{  0.072}$ & $   0.242\pm   0.045$ & $   0.235\pm   0.012$ & $  0.120_{ -0.002}^{  0.002}$ & $  0.124_{ -0.006}^{  0.009}$ \\
g51 - y & $  0.241 $ & $  0.189 $ & $  0.099_{ -0.012}^{  0.008}$ & $   0.246\pm   0.057$ & $   0.206\pm   0.014$ & $  0.107_{ -0.002}^{  0.002}$ & $  0.124_{ -0.007}^{  0.008}$ \\
g51 - z & $  0.241 $ & $  0.189 $ & $  0.141_{ -0.003}^{  0.002}$ & $   0.244\pm   0.043$ & $   0.204\pm   0.010$ & $  0.120_{ -0.003}^{  0.003}$ & $  0.116_{ -0.007}^{  0.008}$ \\
g72 - x & $  0.299 $ & $  0.299 $ & $  0.087_{ -0.009}^{  0.032}$ & $   0.262\pm   0.105$ & $   0.306\pm   0.025$ & $  0.366_{ -0.006}^{  0.006}$ & $  0.386_{ -0.007}^{  0.001}$ \\
g72 - y & $  0.299 $ & $  0.299 $ & $  0.094_{ -0.027}^{  0.029}$ & $   0.135\pm   0.053$ & $   0.287\pm   0.023$ & $  0.355_{ -0.006}^{  0.005}$ & $  0.386_{ -0.008}^{  0.002}$ \\
g72 - z & $  0.299 $ & $  0.299 $ & $  0.059_{ -0.004}^{  0.008}$ & $   0.367\pm   0.062$ & $   0.324\pm   0.018$ & $  0.223_{ -0.004}^{  0.004}$ & $  0.251_{ -0.022}^{  0.019}$ \\
\hline
$r_{\rm s}/r_{\rm s,true}^{\rm DM only}$ & & & 0.49 (0.40) & 0.93 (0.20) & 0.93 (0.11) & 0.89 (0.33) & 0.97 (0.36) \\  
$r_{\rm s}/r_{\rm s,true}^{\rm total}$ & & & 0.59 (0.53) & 1.05 (0.27) & 1.04 (0.12) & 0.98 (0.32) & 1.06 (0.34) \\  
\hline
$r_{\rm s}/r_{\rm s, true}^{\rm DM only}$ & no g51 & & 0.31 (0.09) & 0.88 (0.24) & 0.95 (0.12) & 1.10 (0.18) & 1.20 (0.17) \\  
$r_{\rm s}/r_{\rm s, true}^{\rm total}$ & no g51 & & 0.34 (0.11) & 0.94 (0.26) & 1.001 (0.11) & 1.17 (0.22) & 1.27 (0.21) \\  
\hline 
\hline

\end{tabular}
\label{tab:bfc}
\end{table*}%

\subsection{Lensing vs. X-ray 3D mass profiles}

Finally, we attempt a comparison between the lensing and the X-ray mass estimates. 

\begin{figure}[t!]
  \includegraphics[width=1.0\hsize]{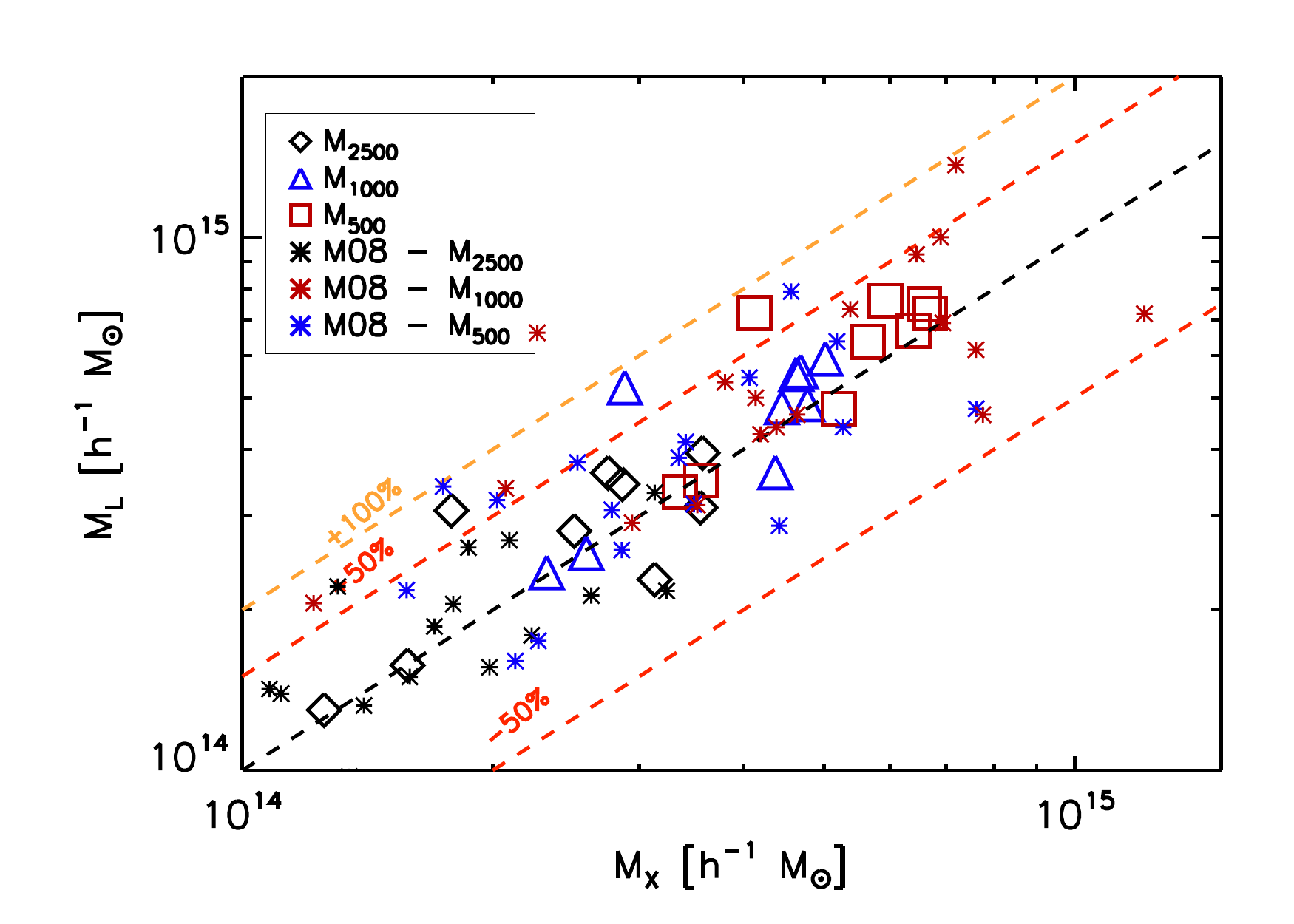}
  \caption{Lensing vs X-ray masses within $R_{2500}$ (diamonds), $R_{500}$ (triangles), and $R_{200}$ (squares). {  The asterisks} show the mass estimates  published by M08 for a sample of 18 galaxy clusters. The dashed lines correspond to $M_L=[0.5,1.0,1.5,2.0]\times M_X$.}
\label{fig:XL}
\end{figure}

First of all, given the results discussed above, it is not surprising that lensing and X-ray mass estimates in individual galaxy clusters can differ by up to $100\%$. This is clear in Fig.\ref{fig:XL}, where the lensing masses are shown as a function of their X-ray equivalents. {  The lensing masses are obtained with the SL+WL method.} The X-ray masses are obtained with the forward method. The discrepancies in the simulated sample are consistent with those observed in several galaxy clusters. As an example, we over-plot with {  asterisks} the mass measurements of a sample of 18 galaxy clusters reported by \cite{2008MNRAS.384.1567M} (M08 in the following discussion).   

Several recent studies seem to agree on the fact that the lensing masses are on average larger than the X-ray masses. For example, analyzing a sample of $19$ clusters observed with SUBARU and {\it XMM-Newton}, \cite{2008A&A...482..451Z} find that $M_{500,{\rm WL}}/M_{500,{\rm X}}=1.09 \pm 0.08$. M08 report similar results. In particular, they find a trend in the ratio between lensing and X-ray masses as a function of the overdensity radius. While X-ray and the lensing masses are consistent with each other at $R_{2500}$ ($M_{2500,{\rm X}}/M_{2500,{\rm WL}}=1.03\pm0.07$), their mean ratio becomes $0.78\pm0.09$ at $R_{500}$. Correcting for excess correlated structure outside the virial radius, they find $M_{2500,{\rm X}}/M_{2500,WL}=1.06\pm0.07$ and $M_{500,{\rm X}}/M_{500,{\rm WL}}=0.85\pm0.10$.  {  Very recently, \cite{zhang09} (Z09 in the following), studying a sample of 12 clusters divided into relaxed and unrelaxed on the basis of their X-ray morphology, also find that the weak lensing masses exceed the X-ray masses at large radii (or low over-densities), while they are comparable at small radii (or high over-densities). However, they measure a shallower but still significant radial evolution of $M_{\rm X}/M_{\rm L}$ compared to M08. For the sub-sample of relaxed clusters, they find  $M_{2500,{\rm X}}/M_{2500,{\rm WL}}=1.04\pm0.08$ and $M_{500,{\rm X}}/M_{500,{\rm WL}}=0.91\pm0.06$ on average, while for the whole sample of objects they find $M_{2500,{\rm X}}/M_{2500,{\rm WL}}=0.97\pm0.07$ and $M_{500,{\rm X}}/M_{500,{\rm WL}}=0.94\pm0.05$.} Such trend is interpreted as an indication of lack of hydrostatic equilibrium in galaxy clusters \citep{2008MNRAS.388.1062C}.
     
\begin{figure}[t!]
  \includegraphics[width=1.0\hsize]{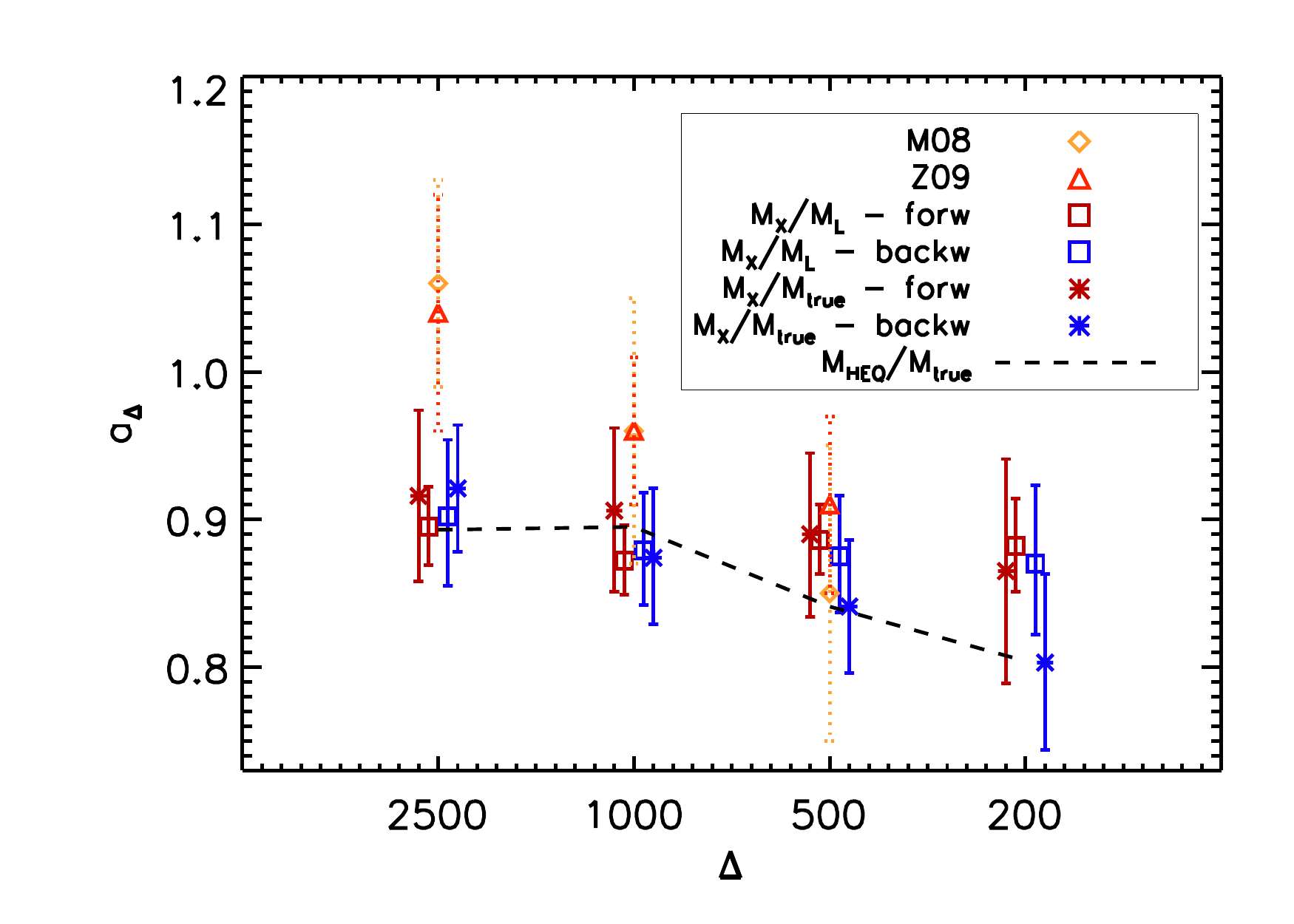}
  \caption{{  Ratio between X-ray and lensing masses as a function of the overdensity $\Delta$ (squares). The results are shown for the X-ray masses obtained with the forward (red) and with the backward (blue) methods and for the lensing masses obtained with the SL+WL method. For comparison, we also show the ratios between the X-ray masses and the true masses of the clusters (asterisks) and the ratios between the masses determined via the hydrostatic equilibrium equation using the true gas density and temperature profiles and the true masses (dashed line). The diamonds and the triangles show the results published by M08 and Z09, based on the analysis of a sample of 18 and 12 galaxy clusters, respectively. The data-points are slightly shifted along the $\Delta$-axis at each over-density, in order to avoid overlapping and facilitating the comparisons.}}
\label{fig:xlens}
\end{figure}

We repeat here the analysis of M08 and Z09 using our simulated clusters. Following the terminology of M08, we indicate with $a_\Delta$ the ratio between the X-ray and 3D-lensing masses for a particular overdensity $\Delta$. We measure $a_{\Delta}$ for $\Delta=2500$, {  1000}, 500, and 200. {  Consistently with M08 and Z09, we estimate $a_{\Delta}$ by minimizing a $\chi^2$ statistic defined as
\begin{equation}
	\chi^2=\sum\frac{(M_{\Delta,X}-a_{\Delta}M_{\Delta,L})^2}{\sigma_{\Delta,X}^2+a_{\Delta}^2\sigma_{\Delta,L}^2} \;,
\end{equation}
where $\sigma_{\Delta,X}$ and $\sigma_{\Delta,L}$ are the errors on the X-ray and lensing masses corresponding to the overdensity $\Delta$. The errors on $a_{\Delta}$ are estimated by Æ by 
locating the values at which $\chi^2-\chi_{\rm min} = 1$, which correspond 
to the 68\% conÞdence interval.}
 For this analysis, we use the lensing masses obtained from the SL+WL method and the X-ray masses obtained from both the forward and the backward methods. {  Note that the lensing masses in M08 and Z09 are obtained with the NFW fit method. We opt for the SL+WL method because our sample is limited and a significant fraction of our clusters has significant substructures. Under these conditions the efficiency of the NFW fit method to obtain reliable mass estimates is limited, as we discussed in Sect.~\ref{sect:wlmasses}.} The results are shown by the squares in Fig.~\ref{fig:xlens}, where we plot $a_\Delta$ as a function of $\Delta$. For comparison, we over-plot the data-points taken from Fig.~4 of M08. {  We also plot the results obtained by Z09 for their sub-sample of relaxed clusters.}. 
We find that the ratios between X-ray and lensing masses are below unity at all over-density radii. For $\Delta=2500$, we find {  $a_{\Delta}=0.9^{+0.05}_{-0.03}$ and $a_{\Delta}\sim 0.91^{+0.05}_{0.04}$} using the masses derived from the forward and from the backward methods, respectively. For $\Delta=500$ such ratio becomes {  $a_{\Delta}=0.88^{+0.03}_{-0.02}$} using the forward method to measure the X-ray mass,  and using the backward method it {  is $a_{\Delta}=0.87^{+0.04}_{-0.04}$}. {  We do not detect a strong radial trend as reported in M08 and Z09. On the contrary, the ratios between X-ray and lensing masses decline very gently as a function of the overdensity radius. }.

The important question which arises now is if the ratio between lensing and X-ray masses is a robust indicator of the lack of hydrostatic equilibrium in the simulated clusters. This can be easily checked by measuring the mass profile of the simulated clusters using the hydrostatic equilibrium approximation. For doing this, we plug the gas density and the temperature profiles measured from the particle distributions into the hydrostatic equilibrium equation and we derive the mass $M_{\rm HEQ}$. This mass should be recovered by the X-ray analysis, i.e. $M_{\rm X}$ is a measurement of $M_{\rm HEQ}$. This mass is compared to the true mass $M_{\rm true}$ as in \cite{RA04.1}. By considering the three clusters, we obtain the median profile of the ratio between hydrostatic equilibrium and total mass, shown by the dashed line in Fig.~\ref{fig:xlens}. For comparison, we indicate with asterisks the ratios between X-ray and true masses at each over-density radius. 
{   In these simulations,  the median $M_{\rm HEQ}/M_{\rm true}$  is $\sim 0.9$ between $R_{2500}$ and $R_{1000}$ and then it decreases to 0.84 and 0.8 at $R_{500}$ and $R_{200}$. The X-ray masses are compatible (within the $68\%$ confidence limit) with $M_{\rm HEQ}$. The mass derived with the backward method declines more steeply than that obtained with the forward  method as $\Delta$ decreases, and seems to follow more closely the behaviour of $M_{\rm HEQ}$, although the differences are not significant given the error-bars. $M_{\rm X}/M_{\rm L}$ is also compatible with $M_{\rm HEQ}/M_{\rm true}$. This is true only at the $2\sigma$ significance level for $\Delta < 500$. Thus, we conclude that lensing and X-ray observations and their comparison to the simulations can provide important information on the physics of the gas in galaxy clusters. Note that, although we have considered a small sample of clusters, averaging over several objects significantly reduces the impact of triaxiality. Indeed, at all overdiensities, $M_{\rm X}/M_{\rm L}$ is very similar to $M_{\rm X}/M_{\rm true}$, indicating that  $M_{\rm L}\sim M_{\rm true}$ on average.}

In Table~\ref{tab:summary} we report the mean, the r.m.s., the median,
the first and the third quartiles, and the minimal and maximal values
of the distributions of several mass ratios. Not only we compare 3D vs
3D masses, but we also compare 3D vs 2D mass estimates and
vice-versa. We note that triaxiality and substructure affect the 2D
X-ray mass estimates consistently to the 3D lensing mass
estimates. For example the r.m.s. of the ratio $M_{\rm 2D,X}/M_{\rm
2D,L}$ is of order $\sim17-18\%$, similar to the r.m.s. of the ratio
$M_{\rm 3D,X}/M_{\rm 3D,L}$. Not surprisingly, the mass ratio which
has the smaller scatter is $M_{\rm 3D,X}/M_{\rm 2D,L}$, whose
r.m.s. is $\sim 13\%$. However, the radial dependence of this mass
ratio is stronger than for $M_{\rm 3D,X}/M_{\rm 3D,L}$ or $M_{\rm
2D,X}/M_{\rm 2D,L}$ as indicated by the variation of the mean and of
the median as a function of the over-density $\Delta$. Obviously, the
mass ratio which is affected by the largest scatter is  $M_{\rm
2D,X}/M_{\rm 3D,L}$ ($\sim 25\%$).

\begin{table*}[htdp]
\caption{Summary of the comparison between lensing, X-ray, and true masses. We report the mean, the r.m.s., the median, the first and the third quartiles, and the minimal and maximal values of the ratios between 3D and 2D masses obtained via lensing and X-ray analyses and from the input simulations. The lensing masses are measured with the SL+WL method. For the X-ray masses we use both the forward and the backward methods.}
{\tiny
\begin{center}
\begin{tabular}{lcccccccc}
\hline
\hline
ratio & $\Delta$ & mean  & r.m.s. & median & 1st quart. & 3rd quart. & min & max \\
\hline
$M_{\rm 3D,X}^{\rm forw}/M_{\rm 3D,true}$ & 2500 &  0.9241 & 0.1147 & 0.8854 & 0.8451 & 0.9730 & 0.8007 & 1.1143 \\
$M_{\rm 3D,X}^{\rm forw}/M_{\rm 3D,true}$ & 500  &  0.8811 & 0.0538 & 0.8950 & 0.8594 & 0.9059 & 0.7640 & 0.9493 \\
$M_{\rm 3D,X}^{\rm forw}/M_{\rm 3D,true}$ & 200  &  0.8947 & 0.0625 & 0.9162 & 0.8534 & 0.9293 & 0.7749 & 0.9759 \\
\hline
$M_{\rm 3D,X}^{\rm back}/M_{\rm 3D,true}$ & 2500 & 0.9205 & 0.0690 & 0.9018 & 0.8669 & 0.9669 & 0.8512 & 1.0407 \\
$M_{\rm 3D,X}^{\rm back}/M_{\rm 3D,true}$ & 500  & 0.8961 & 0.0582 & 0.8826 & 0.8817 & 0.9288 & 0.7753 & 0.9853 \\
$M_{\rm 3D,X}^{\rm back}/M_{\rm 3D,true}$ & 200 & 0.8832 & 0.0664 & 0.9019 & 0.8878 & 0.9126 & 0.7192 & 0.9559 \\
\hline  
$M_{\rm 3D,L}/M_{\rm 3D,true}$ & 2500 & 1.0233 & 0.1686 & 0.9683 & 0.9093 & 1.1033 & 0.8290 & 1.3687 \\
$M_{\rm 3D,L}/M_{\rm 3D,true}$ & 500  &  0.9963 & 0.1661 &  0.9374 & 0.8930 & 1.0251 & 0.8362 & 1.3438 \\
$M_{\rm 3D,L}/M_{\rm 3D,true}$ & 200 &  0.9807 & 0.1734 & 0.8891 & 0.8682 & 1.0783 & 0.8234 & 1.2879 \\
\hline 
$M_{\rm 3D,X}^{\rm forw}/M_{\rm 3D,L}$ & 2500 & 0.9303 & 0.2193 & 0.8917 & 0.8303 & 1.0034 & 0.5850 & 1.3441 \\
$M_{\rm 3D,X}^{\rm forw}/M_{\rm 3D,L}$ & 500  & 0.9062 & 0.1521 & 0.9429 & 0.8837 & 1.0167 & 0.5686 & 1.0768 \\
$M_{\rm 3D,X}^{\rm forw}/M_{\rm 3D,L}$ & 200  & 0.9395 & 0.1802 & 1.0091 & 0.7915 & 1.0815 & 0.6017 & 1.1127 \\
\hline
$M_{\rm 3D,X}^{\rm back}/M_{\rm 3D,L}$ & 2500 & 0.9230 & 0.1776 & 0.9109 & 0.8416 & 0.9934 & 0.6219 & 1.2553 \\
$M_{\rm 3D,X}^{\rm back}/M_{\rm 3D,L}$ & 500  & 0.9244 & 0.1731 & 0.9415 & 0.8926 & 1.0690 & 0.5770 & 1.1108 \\
$M_{\rm 3D,X}^{\rm back}/M_{\rm 3D,L}$ & 200  & 0.9307 & 0.1937 & 1.0088 & 0.8374 & 1.0541 & 0.5584 & 1.1353 \\
\hline  
$M_{\rm 2D,L}/M_{\rm 3D,true}$ & 2500 & 1.4898 & 0.2485  & 1.4317 & 1.3671 & 1.5457 & 1.1611 & 2.0612 \\
$M_{\rm 2D,L}/M_{\rm 3D,true}$ & 500  & 1.2924 & 0.2151  & 1.2127 & 1.1838 & 1.3059 & 1.0607 & 1.7788 \\
$M_{\rm 2D,L}/M_{\rm 3D,true}$ & 200  & 1.2171 & 0.2092  & 1.1109 & 1.0700 & 1.3212 & 1.0545 & 1.6244 \\
\hline
$M_{\rm 2D,X}^{\rm forw}/M_{\rm 3D,true}$ & 2500 &  1.2850 & 0.0757  & 1.2744 & 1.2627 & 1.3288  & 1.1276 & 1.3839 \\
$M_{\rm 2D,X}^{\rm forw}/M_{\rm 3D,true}$ & 500 &   1.0477 & 0.0852  & 1.0376 & 0.9853 & 1.0810  & 0.9179 & 1.1760 \\
$M_{\rm 2D,X}^{\rm forw}/M_{\rm 3D,true}$ & 200 &   0.8961 & 0.0835  & 0.8839 & 0.8446 & 0.9382  & 0.7449 & 1.0162 \\
\hline 
$M_{\rm 2D,X}^{\rm back}/M_{\rm 3D,true}$ & 2500 &  1.3211 & 0.0754 & 1.3317 & 1.2903 & 1.3520 & 1.2052 & 1.4415 \\
$M_{\rm 2D,X}^{\rm back}/M_{\rm 3D,true}$ & 500 &   1.1164 & 0.1121 & 1.1106 & 1.0643 & 1.1270 & 0.9497 & 1.3130 \\
$M_{\rm 2D,X}^{\rm back}/M_{\rm 3D,true}$ & 200 &   1.0124 & 0.0990 & 1.0200 & 0.9919 & 1.0436 & 0.8108 & 1.1585 \\
\hline 
$M_{\rm 2D,X}^{\rm forw}/M_{\rm 2D,L}$ & 2500 & 0.8875 & 0.1733 & 0.8901 & 0.8221 & 0.9779 & 0.5471 & 1.1702 \\ 
$M_{\rm 2D,X}^{\rm forw}/M_{\rm 2D,L}$ & 500  & 0.8314 & 0.1503 & 0.8790 & 0.7545 & 0.9272 & 0.5160 & 0.9834 \\
$M_{\rm 2D,X}^{\rm forw}/M_{\rm 2D,L}$ & 200  & 0.7596 & 0.1629 & 0.7949 & 0.6388 & 0.8664 & 0.4586 & 0.9636 \\
\hline 
$M_{\rm 2D,X}^{\rm back}/M_{\rm 2D,L}$ & 2500 & 0.9103 & 0.1657 & 0.9301 & 0.8634 & 1.0259 & 0.5847 & 1.1584 \\ 
$M_{\rm 2D,X}^{\rm back}/M_{\rm 2D,L}$ & 500  & 0.8889 & 0.1825 & 0.9475 & 0.8150 & 1.0299 & 0.5339 & 1.0827 \\
$M_{\rm 2D,X}^{\rm back}/M_{\rm 2D,L}$ & 200  & 0.8609 & 0.1951 & 0.9259 & 0.7507 & 0.9638 & 0.4991 & 1.0827 \\
\hline 
$M_{\rm 2D,X}^{\rm forw}/M_{\rm 3D,L}$ & 2500 &  1.2929 & 0.2579 & 1.3392 & 1.1445 & 1.5218 & 0.8239 & 1.6389 \\
$M_{\rm 2D,X}^{\rm forw}/M_{\rm 3D,L}$ & 500 &   1.0801 & 0.2064 & 1.1532 & 0.9612 & 1.1762 & 0.6830 & 1.3297 \\
$M_{\rm 2D,X}^{\rm forw}/M_{\rm 3D,L}$ & 200 &   0.9450 & 0.2123 & 0.9728 & 0.7828 & 1.0552 & 0.5783 & 1.2341 \\
\hline 
$M_{\rm 2D,X}^{\rm back}/M_{\rm 3D,L}$ & 2500 &  1.3274 & 0.2560 & 1.3326 & 1.2020 & 1.5853 & 0.8806 & 1.6223 \\
$M_{\rm 2D,X}^{\rm back}/M_{\rm 3D,L}$ & 500 &   1.1553 & 0.2502 & 1.2023 & 1.0383 & 1.3064 & 0.7067 & 1.4481 \\
$M_{\rm 2D,X}^{\rm back}/M_{\rm 3D,L}$ & 200 &   1.0711 & 0.2524 & 1.1663 & 0.9198 & 1.1748 & 0.6295 & 1.3760 \\
\hline 
$M_{\rm 3D,X}^{\rm forw}/M_{\rm 2D,L}$ & 2500 &  0.6407 & 0.1612 & 0.6212 & 0.5728 & 0.6556 & 0.3885 & 0.9597 \\
$M_{\rm 3D,X}^{\rm forw}/M_{\rm 2D,L}$ & 500  &  0.6985 & 0.1166 & 0.7281 & 0.6838 & 0.7519 & 0.4295 & 0.8489 \\
$M_{\rm 3D,X}^{\rm forw}/M_{\rm 2D,L}$ & 200  &  0.7557 & 0.1407 & 0.8147 & 0.6459 & 0.8509 & 0.4771 & 0.9013 \\
\hline 
$M_{\rm 3D,X}^{\rm back}/M_{\rm 2D,L}$ & 2500 &  0.6352 & 0.1302 & 0.6173 & 0.5895 & 0.6479 & 0.4129 & 0.8963 \\
$M_{\rm 3D,X}^{\rm back}/M_{\rm 2D,L}$ & 500  &  0.7120 & 0.1299 & 0.7391 & 0.7006 & 0.7906 & 0.4359 & 0.8757 \\
$M_{\rm 3D,X}^{\rm back}/M_{\rm 2D,L}$ & 200  &  0.7485 & 0.1510 & 0.8009 & 0.6834 & 0.8633 & 0.4428 & 0.8933 \\
\hline\hline\end{tabular}
\end{center}
}
\label{tab:summary}
\end{table*}%

\section{Summary and conclusions}
\label{sect:summary}

In this paper, we have tested several methods to derive the mass of galaxy clusters via lensing and X-ray observations. To do so, we have used two pipelines for simulating observations of galaxy clusters obtained from N-body and hydrodynamical simulations. These are the code {\tt SkyLens}, which produces mock observations with a variety of telescopes in the optical bands, including lensing effects by matter along the line of sight, and the code {\tt XMAS}, which allows to mimic X-ray observations of galaxy clusters with the {\it Chandra} and the {\it XMM-Newton} telescopes. In our analysis, we have used three massive galaxy clusters at redshifts between $z=0.2335$ and $z=0.297$. Each cluster has been projected along three orthogonal axes and each projection has been analyzed individually. 
For each cluster projection, we have carried out both the lensing and the X-ray analysis using standard techniques to derive the mass profiles.

Our results can be summarized as follows:
\begin{itemize}
\item strong lensing parametric mass reconstructions provide very accurate estimates of the projected mass, but only in the regions probed by the strong lensing features. Our simulations show that within the Einstein rings, the input masses are recovered with an accuracy of  $\lesssim 20\%$. For most of the cases, the measured masses differ from the input ones by only a few percent. The precision depends on the complexity of the lens and on the number of lensing constraints available. For example, using central images improves the mass estimates in the very inner regions. On the other hand, extrapolating the mass models to radii larger than those probed by strong lensing may easily result in mass estimates  wrong by up to $40-50\%$. In particular, the assumptions made for modeling the star distribution in the BCG affect the mass profiles at large radii. 
Although the very central regions of numerically simulated clusters may be not well representative of those of real clusters, given the uncertainties on the treatment of small-scale physics of the intra-cluster gas, our results warn against potential biases which may affect the strong lensing mass estimates;    
\item weak lensing methods for measuring the projected mass perform well with galaxy clusters characterized by regular mass distributions, i.e. without massive substructures which produce a significant shear. If present, such substructures need to be properly modeled. If not, the cluster masses can be over- or under-estimated by up to $50\%$, when assuming that the shear signal is tangential to a unique mass center. Non-parametric methods combining weak and strong lensing, which by construction do not assume any symmetry in the lensing signal, perform well even in the case of clusters disturbed by massive substructures. The mass estimates at $R_{2500}$, $R_{500}$, and $R_{200}$ are typically recovered within a $20\%$ accuracy, being more accurate at the smallest radii;
\item deprojection of the lensing masses introduces a significant scatter in the ratio between estimated and true masses. Cluster triaxiality causes that the accuracy of 3D-mass estimates depends on the orientation of the lens with respect to the line of sight. For clusters whose major axis points towards the observer the 3D-mass is over-estimated, while the opposite happens for clusters oriented perpedicularly to the line of sight. Cluster substructures also cause errors during the de-projection, as the distance of the substructures from the cluster center along the line of sight is unknown;
\item X-ray masses are typically biased low by $5-20\%$ due to the lack of hydrostatic equilibrium in the simulated clusters. The average bias between $R_{2500}$ and $R_{200}$ is $\sim 10\%$. In fact, the accretion of material from the surroundings during the cluster formation process causes bulk motions which contribute to the pressure support of the gas. As shown in previous works, this effect is radial dependent, i.e. it is larger in the outer regions.  We investigated whether the ratio of X-ray-to-lensing masses $a_{\Delta}$ can be used as a tracer of hydrostatic equilibrium. By averaging over the sample, we find that such ratio is indeed reproducing the lack of hydrostatic equilibrium in the input simulations, {  although the number of clusters available is limited and despite the effects of triaxiality and substructures which affect the cluster 3D-mass estimates.} 
\item the gas mass is well reconstructed within the region from which we can extract a surface brightness profile.
Almost indipendently of the dynamical state of the cluster the gas mass con be recovered, on average, with average deviation of $1\pm 3\%$ at $R_{2500}$ and $7\pm 4\%$ between $R_{500}$ and $R_{200}$.
\end{itemize}   

These results indicate that the comparison between lensing and X-ray mass estimates in individual systems can be very misleading, due to the fact that these two observables probe different quantities. While lensing probes the matter projected along the line of sight to the clusters, the X-ray emission from the intra-cluster gas probes the three-dimensional gravitational potential. Thus, a direct comparison between lensing and X-ray masses is possible only in those systems which are characterized by large degrees of spherical symmetry.  
Additional complications are caused by substructures, which largely affect the weak lensing measurements, at least using some methods of converting the shear into a mass estimate. Combining strong and weak lensing, we obtain on average a very small bias in the mass estimates ($\sim 2\%$) but the scatter is pretty high ($\sim 17-23\%$). On the contrary, X-ray masses are affected by a more substantial bias ($\sim 10\%$) but the scatter is smaller by almost a factor of three.

Finally, strong lensing alone provides mass estimates which are reliable only in a very small region around the cluster core, which cannot be probed by X-ray. Extrapolating the strong lensing mass models to the radii probed by X-ray (and vice-versa) may result in significantly large mismatches between mass estimates. Note however that the difficulty to fit the strong lensing data in simulations with cooling and star formation using the approximations which are commonly used in observations is very likely to be due to the unrealistic core structure of the simulated clusters.  

Nevertheless, it is remarkable that,
by averaging over the whole sample, we have been able to reproduce the
ratio between true and hydrostatic equilibrium masses seen in the
simulations, under the assumption that the former is given by lensing
and the latter by X-ray. This suggests that, with a sufficiently large
sample of clusters with wide-field imaging and X-ray observations, the
impact of several noises affecting the mass estimates can be
reduced. These are positive news and suggest that weak lensing masses
may be used as reliable calibrators for the scaling relations
involving the X-ray observables, like the $M-L_X$, the $M-T_X$, and
the $M-Y_X$ relations. Moreover, the scatter due to triaxiality in the
calibration can potentially be reduced.
{  In fact,} although its radial dependence is stronger, the ratio between 3D X-ray masses and 2D lensing masses is affected by a smaller scatter than the ratio between 3D or 2D X-ray and lensing masses.

\acknowledgements{We acknoweldge
  financial contribution from contracts ASI-INAF I/023/05/0, ASI-INAF
  I/088/06/0 and INFN PD51. Support for this work was provided also by NASA through {\it Chandra} Postdoctoral Fellowship grant number PF5-70042 awarded by the {\it Chandra} X-ray Center, which is operated by the Smithsonian Astrophysical Observatory for NASA under the contract NAS8-03060. ER acknoweldges the Michigan Society of Fellow. We are grateful to L. Moscardini for having carefully read the manuscript and provided useful comments. MM thanks Yu-Ying Zhang and collaborators for providing a draft of their paper. MM thanks the Institut f\"ur Theoretische Astrophysik of the University of Heidelberg for hospitality during the preparation of this manuscript. The work has been performed under the HPC-EUROPA2 project (project number: 228398) with the support of the European Commission - Capacities Area - Research Infrastructures. We are grateful to Y. Suto and S. Miyazaki for their help implementing the Subaru interface in the SkyLens simulator. We thank the anonymous referee for his useful comments. MM thanks Paolo for his cooperation during the preparation of the manuscript.}

\bibliography{./TeXMacro/master}

\begin{thebibliography}{99}
\expandafter\ifx\csname natexlab\endcsname\relax\def\natexlab#1{#1}\fi

\bibitem[{Allen(1998)}]{AL98.1}
Allen, S. 1998, MNRAS, 296, 392

\bibitem[{{Allen} {et~al.}(2001){Allen}, {Ettori}, \&
  {Fabian}}]{2001MNRAS.324..877A}
{Allen}, S.~W., {Ettori}, S., \& {Fabian}, A.~C. 2001, \mnras, 324, 877

\bibitem[{{Allen} {et~al.}(2008){Allen}, {Rapetti}, {Schmidt}, {Ebeling},
  {Morris}, \& {Fabian}}]{2008MNRAS.383..879A}
{Allen}, S.~W., {Rapetti}, D.~A., {Schmidt}, R.~W., {et~al.} 2008, \mnras, 383,
  879

\bibitem[{{Allen} {et~al.}(2004){Allen}, {Schmidt}, {Ebeling}, {Fabian}, \&
  {van Speybroeck}}]{2004MNRAS.353..457A}
{Allen}, S.~W., {Schmidt}, R.~W., {Ebeling}, H., {Fabian}, A.~C., \& {van
  Speybroeck}, L. 2004, \mnras, 353, 457

\bibitem[{{Ameglio} {et~al.}(2009){Ameglio}, {Borgani}, {Pierpaoli}, {Dolag},
  {Ettori}, \& {Morandi}}]{2009MNRAS.394..479A}
{Ameglio}, S., {Borgani}, S., {Pierpaoli}, E., {et~al.} 2009, \mnras, 394, 479

\bibitem[{{Arnaud}(1996)}]{1996ASPC..101...17A}
{Arnaud}, K.~A. 1996, in Astronomical Society of the Pacific Conference Series,
  Vol. 101, Astronomical Data Analysis Software and Systems V, ed.
  {G.~H.~Jacoby \& J.~Barnes}, 17--+

\bibitem[{{Aubert} {et~al.}(2007){Aubert}, {Amara}, \& {Metcalf}}]{AU07.1}
{Aubert}, D., {Amara}, A., \& {Metcalf}, R.~B. 2007, \mnras, 376, 113

\bibitem[{Bartelmann(1995)}]{BA95.2}
Bartelmann, M. 1995, A\&A, 299, 11

\bibitem[{Bartelmann(1996)}]{BA96.1}
Bartelmann, M. 1996, A\&A, 313, 697

\bibitem[{Bartelmann \& Schneider(2001)}]{BA01.1}
Bartelmann, M. \& Schneider, P. 2001, Physics Reports, 340, 291

\bibitem[{Bartelmann \& Steinmetz(1996)}]{BA96.2}
Bartelmann, M. \& Steinmetz, M. 1996, MNRAS, 283, 431

\bibitem[{{Beckwith} {et~al.}(2006){Beckwith}, {Stiavelli}, {Koekemoer},
  {Caldwell}, {Ferguson}, {Hook}, {Lucas}, {Bergeron}, {Corbin}, {Jogee},
  {Panagia}, {Robberto}, {Royle}, {Somerville}, \& {Sosey}}]{BECK06.1}
{Beckwith}, S.~V.~W., {Stiavelli}, M., {Koekemoer}, A.~M., {et~al.} 2006, \aj,
  132, 1729

\bibitem[{{Ben{\'{\i}}tez}(2000)}]{2000ApJ...536..571B}
{Ben{\'{\i}}tez}, N. 2000, \apj, 536, 571

\bibitem[{{Brada{\v c}} {et~al.}(2005){Brada{\v c}}, {Erben}, {Schneider},
  {Hildebrandt}, {Lombardi}, {Schirmer}, {Miralles}, {Clowe}, \&
  {Schindler}}]{BR04.1}
{Brada{\v c}}, M., {Erben}, T., {Schneider}, P., {et~al.} 2005, \aap, 437, 49

\bibitem[{{Buote} {et~al.}(2007){Buote}, {Gastaldello}, {Humphrey},
  {Zappacosta}, {Bullock}, {Brighenti}, \& {Mathews}}]{2007ApJ...664..123B}
{Buote}, D.~A., {Gastaldello}, F., {Humphrey}, P.~J., {et~al.} 2007, \apj, 664,
  123

\bibitem[{{Churazov} {et~al.}(2008){Churazov}, {Forman}, {Vikhlinin},
  {Tremaine}, {Gerhard}, \& {Jones}}]{2008MNRAS.388.1062C}
{Churazov}, E., {Forman}, W., {Vikhlinin}, A., {et~al.} 2008, \mnras, 388, 1062

\bibitem[{{Clowe} {et~al.}(2004){Clowe}, {Gonzalez}, \& {Markevitch}}]{CL04.1}
{Clowe}, D., {Gonzalez}, A., \& {Markevitch}, M. 2004, \apj, 604, 596

\bibitem[{{Clowe} {et~al.}(1998){Clowe}, {Luppino}, {Kaiser}, {Henry}, \&
  {Gioia}}]{1998ApJ...497L..61C}
{Clowe}, D., {Luppino}, G.~A., {Kaiser}, N., {Henry}, J.~P., \& {Gioia}, I.~M.
  1998, \apjl, 497, L61+

\bibitem[{{Coe} {et~al.}(2006){Coe}, {Ben{\'{\i}}tez}, {S{\'a}nchez}, {Jee},
  {Bouwens}, \& {Ford}}]{2006AJ....132..926C}
{Coe}, D., {Ben{\'{\i}}tez}, N., {S{\'a}nchez}, S.~F., {et~al.} 2006, \aj, 132,
  926

\bibitem[{{Comerford} {et~al.}(2006){Comerford}, {Meneghetti}, {Bartelmann}, \&
  {Schirmer}}]{CO05.1}
{Comerford}, J.~M., {Meneghetti}, M., {Bartelmann}, M., \& {Schirmer}, M. 2006,
  \apj, 642, 39

\bibitem[{{Comerford} \& {Natarajan}(2007)}]{2007MNRAS.379..190C}
{Comerford}, J.~M. \& {Natarajan}, P. 2007, \mnras, 379, 190

\bibitem[{{Corless} \& {King}(2007)}]{2007MNRAS.380..149C}
{Corless}, V.~L. \& {King}, L.~J. 2007, \mnras, 380, 149

\bibitem[{{Dolag} {et~al.}(2005){Dolag}, {Vazza}, {Brunetti}, \&
  {Tormen}}]{DO05.1}
{Dolag}, K., {Vazza}, F., {Brunetti}, G., \& {Tormen}, G. 2005, \mnras, 364,
  753

\bibitem[{{Donnarumma} {et~al.}(2009){Donnarumma}, {Ettori}, {Meneghetti}, \&
  {Moscardini}}]{2009MNRAS.398..438D}
{Donnarumma}, A., {Ettori}, S., {Meneghetti}, M., \& {Moscardini}, L. 2009,
  \mnras, 398, 438

\bibitem[{{Eisenstein} {et~al.}(2005){Eisenstein}, {Zehavi}, {Hogg},
  {Scoccimarro}, {Blanton}, {Nichol}, {Scranton}, {Seo}, {Tegmark}, {Zheng},
  {Anderson}, {Annis}, {Bahcall}, {Brinkmann}, {Burles}, {Castander},
  {Connolly}, {Csabai}, {Doi}, {Fukugita}, {Frieman}, {Glazebrook}, {Gunn},
  {Hendry}, {Hennessy}, {Ivezi{\'c}}, {Kent}, {Knapp}, {Lin}, {Loh}, {Lupton},
  {Margon}, {McKay}, {Meiksin}, {Munn}, {Pope}, {Richmond}, {Schlegel},
  {Schneider}, {Shimasaku}, {Stoughton}, {Strauss}, {SubbaRao}, {Szalay},
  {Szapudi}, {Tucker}, {Yanny}, \& {York}}]{2005ApJ...633..560E}
{Eisenstein}, D.~J., {Zehavi}, I., {Hogg}, D.~W., {et~al.} 2005, \apj, 633, 560

\bibitem[{{Ettori} {et~al.}(2002){Ettori}, {De Grandi}, \&
  {Molendi}}]{2002A&A...391..841E}
{Ettori}, S., {De Grandi}, S., \& {Molendi}, S. 2002, \aap, 391, 841

\bibitem[{{Ettori} \& {Lombardi}(2003)}]{2003A&A...398L...5E}
{Ettori}, S. \& {Lombardi}, M. 2003, \aap, 398, L5

\bibitem[{{Ettori} {et~al.}(2003){Ettori}, {Tozzi}, \&
  {Rosati}}]{2003A&A...398..879E}
{Ettori}, S., {Tozzi}, P., \& {Rosati}, P. 2003, \aap, 398, 879

\bibitem[{{Evrard} {et~al.}(1996){Evrard}, {Metzler}, \&
  {Navarro}}]{1996ApJ...469..494E}
{Evrard}, A.~E., {Metzler}, C.~A., \& {Navarro}, J.~F. 1996, \apj, 469, 494

\bibitem[{{Fahlman} {et~al.}(1994){Fahlman}, {Kaiser}, {Squires}, \&
  {Woods}}]{1994ApJ...437...56F}
{Fahlman}, G., {Kaiser}, N., {Squires}, G., \& {Woods}, D. 1994, \apj, 437, 56

\bibitem[{Gardini {et~al.}(2004)Gardini, Rasia, Mazzotta, Tormen, {De Grandi},
  \& Moscardini}]{GA04.1}
Gardini, A., Rasia, E., Mazzotta, P., {et~al.} 2004, MNRAS, 351, 505

\bibitem[{{Giavalisco} {et~al.}(2004){Giavalisco}, {Ferguson}, {Koekemoer},
  {Dickinson}, {Alexander}, {Bauer}, {Bergeron}, {Biagetti}, {Brandt},
  {Casertano}, {Cesarsky}, {Chatzichristou}, {Conselice}, {Cristiani}, {Da
  Costa}, {Dahlen}, {de Mello}, {Eisenhardt}, {Erben}, {Fall}, {Fassnacht},
  {Fosbury}, {Fruchter}, {Gardner}, {Grogin}, {Hook}, {Hornschemeier}, {Idzi},
  {Jogee}, {Kretchmer}, {Laidler}, {Lee}, {Livio}, {Lucas}, {Madau},
  {Mobasher}, {Moustakas}, {Nonino}, {Padovani}, {Papovich}, {Park},
  {Ravindranath}, {Renzini}, {Richardson}, {Riess}, {Rosati}, {Schirmer},
  {Schreier}, {Somerville}, {Spinrad}, {Stern}, {Stiavelli}, {Strolger},
  {Urry}, {Vandame}, {Williams}, \& {Wolf}}]{GIA04.1}
{Giavalisco}, M., {Ferguson}, H.~C., {Koekemoer}, A.~M., {et~al.} 2004, \apjl,
  600, L93

\bibitem[{Golse {et~al.}(2002)Golse, Kneib, \& Soucail}]{GO02.2}
Golse, G., Kneib, J.-P., \& Soucail, G. 2002, A\&A, 387, 788

\bibitem[{{Henriksen} \& {Mushotzky}(1986)}]{1986ApJ...302..287H}
{Henriksen}, M.~J. \& {Mushotzky}, R.~F. 1986, \apj, 302, 287

\bibitem[{{Hoekstra}(2001)}]{2001A&A...370..743H}
{Hoekstra}, H. 2001, \aap, 370, 743

\bibitem[{{Hoekstra}(2003)}]{2003MNRAS.339.1155H}
{Hoekstra}, H. 2003, \mnras, 339, 1155

\bibitem[{{Hoekstra}(2007)}]{2007MNRAS.379..317H}
{Hoekstra}, H. 2007, \mnras, 379, 317

\bibitem[{{Hoekstra} {et~al.}(1998){Hoekstra}, {Franx}, {Kuijken}, \&
  {Squires}}]{1998ApJ...504..636H}
{Hoekstra}, H., {Franx}, M., {Kuijken}, K., \& {Squires}, G. 1998, \apj, 504,
  636

\bibitem[{Jenkins {et~al.}(2001)Jenkins, Frenk, White, Colberg, Cole, Evrard,
  Couchman, \& Yoshida}]{JE01.1}
Jenkins, A., Frenk, C., White, S., {et~al.} 2001, MNRAS, 321, 372

\bibitem[{{Jullo} {et~al.}(2007){Jullo}, {Kneib}, {Limousin},
  {El{\'{\i}}asd{\'o}ttir}, {Marshall}, \& {Verdugo}}]{2007NJPh....9..447J}
{Jullo}, E., {Kneib}, J.-P., {Limousin}, M., {et~al.} 2007, New Journal of
  Physics, 9, 447

\bibitem[{{Kaiser} {et~al.}(1995){Kaiser}, {Squires}, \&
  {Broadhurst}}]{1995ApJ...449..460K}
{Kaiser}, N., {Squires}, G., \& {Broadhurst}, T. 1995, \apj, 449, 460

\bibitem[{Kneib {et~al.}(1993)Kneib, Mellier, Fort, \& Mathez}]{KN93.1}
Kneib, J., Mellier, Y., Fort, B., \& Mathez, G. 1993, A\&A, 273, 367

\bibitem[{{Komatsu} {et~al.}(2009){Komatsu}, {Dunkley}, {Nolta}, {Bennett},
  {Gold}, {Hinshaw}, {Jarosik}, {Larson}, {Limon}, {Page}, {Spergel},
  {Halpern}, {Hill}, {Kogut}, {Meyer}, {Tucker}, {Weiland}, {Wollack}, \&
  {Wright}}]{2009ApJS..180..330K}
{Komatsu}, E., {Dunkley}, J., {Nolta}, M.~R., {et~al.} 2009, \apjs, 180, 330

\bibitem[{{Kravtsov} {et~al.}(2006){Kravtsov}, {Vikhlinin}, \&
  {Nagai}}]{2006ApJ...650..128K}
{Kravtsov}, A.~V., {Vikhlinin}, A., \& {Nagai}, D. 2006, \apj, 650, 128

\bibitem[{{Lau} {et~al.}(2009){Lau}, {Kravtsov}, \&
  {Nagai}}]{2009ApJ...705.1129L}
{Lau}, E.~T., {Kravtsov}, A.~V., \& {Nagai}, D. 2009, \apj, 705, 1129

\bibitem[{{Liedahl} {et~al.}(1995){Liedahl}, {Osterheld}, \&
  {Goldstein}}]{1995ApJ...438L.115L}
{Liedahl}, D.~A., {Osterheld}, A.~L., \& {Goldstein}, W.~H. 1995, \apjl, 438,
  L115

\bibitem[{{Limousin} {et~al.}(2007){Limousin}, {Richard}, {Jullo}, {Kneib},
  {Fort}, {Soucail}, {El{\'{\i}}asd{\'o}ttir}, {Natarajan}, {Ellis}, {Smail},
  {Czoske}, {Smith}, {Hudelot}, {Bardeau}, {Ebeling}, {Egami}, \&
  {Knudsen}}]{2007ApJ...668..643L}
{Limousin}, M., {Richard}, J., {Jullo}, E., {et~al.} 2007, \apj, 668, 643

\bibitem[{{Luppino} \& {Kaiser}(1997)}]{1997ApJ...475...20L}
{Luppino}, G.~A. \& {Kaiser}, N. 1997, \apj, 475, 20

\bibitem[{{Mahdavi} {et~al.}(2008){Mahdavi}, {Hoekstra}, {Babul}, \&
  {Henry}}]{2008MNRAS.384.1567M}
{Mahdavi}, A., {Hoekstra}, H., {Babul}, A., \& {Henry}, J.~P. 2008, \mnras,
  384, 1567

\bibitem[{{Mantz} {et~al.}(2009{\natexlab{a}}){Mantz}, {Allen}, \&
  {Rapetti}}]{2009arXiv0911.1788M}
{Mantz}, A., {Allen}, S.~W., \& {Rapetti}, D. 2009{\natexlab{a}}, ArXiv
  e-prints 0911.1788

\bibitem[{{Mantz} {et~al.}(2009{\natexlab{b}}){Mantz}, {Allen}, {Rapetti}, \&
  {Ebeling}}]{2009arXiv0909.3098M}
{Mantz}, A., {Allen}, S.~W., {Rapetti}, D., \& {Ebeling}, H.
  2009{\natexlab{b}}, ArXiv e-prints 0909.3098

\bibitem[{{Markevitch} {et~al.}(2004){Markevitch}, {Gonzalez}, {Clowe},
  {Vikhlinin}, {Forman}, {Jones}, {Murray}, \& {Tucker}}]{2004ApJ...606..819M}
{Markevitch}, M., {Gonzalez}, A.~H., {Clowe}, D., {et~al.} 2004, \apj, 606, 819

\bibitem[{{Maturi} {et~al.}(2005){Maturi}, {Meneghetti}, {Bartelmann}, {Dolag},
  \& {Moscardini}}]{MA05.2}
{Maturi}, M., {Meneghetti}, M., {Bartelmann}, M., {Dolag}, K., \& {Moscardini},
  L. 2005, \aap, 442, 851

\bibitem[{{Medezinski} {et~al.}(2007){Medezinski}, {Broadhurst}, {Umetsu},
  {Coe}, {Ben{\'{\i}}tez}, {Ford}, {Rephaeli}, {Arimoto}, \&
  {Kong}}]{2007ApJ...663..717M}
{Medezinski}, E., {Broadhurst}, T., {Umetsu}, K., {et~al.} 2007, \apj, 663, 717

\bibitem[{{Meneghetti} {et~al.}(2007){Meneghetti}, {Argazzi}, {Pace},
  {Moscardini}, {Dolag}, {Bartelmann}, {Li}, \& {Oguri}}]{2007A&A...461...25M}
{Meneghetti}, M., {Argazzi}, R., {Pace}, F., {et~al.} 2007, \aap, 461, 25

\bibitem[{Meneghetti {et~al.}(2003{\natexlab{a}})Meneghetti, Bartelmann, \&
  Moscardini}]{ME03.2}
Meneghetti, M., Bartelmann, M., \& Moscardini, L. 2003{\natexlab{a}}, MNRAS,
  346, 67

\bibitem[{Meneghetti {et~al.}(2003{\natexlab{b}})Meneghetti, Bartelmann, \&
  Moscardini}]{ME03.1}
Meneghetti, M., Bartelmann, M., \& Moscardini, L. 2003{\natexlab{b}}, MNRAS,
  340, 105

\bibitem[{{Meneghetti} {et~al.}(2008){Meneghetti}, {Melchior}, {Grazian}, {De
  Lucia}, {Dolag}, {Bartelmann}, {Heymans}, {Moscardini}, \&
  {Radovich}}]{2008A&A...482..403M}
{Meneghetti}, M., {Melchior}, P., {Grazian}, A., {et~al.} 2008, \aap, 482, 403

\bibitem[{{Merten} {et~al.}(2009){Merten}, {Cacciato}, {Meneghetti}, {Mignone},
  \& {Bartelmann}}]{2009A&A...500..681M}
{Merten}, J., {Cacciato}, M., {Meneghetti}, M., {Mignone}, C., \& {Bartelmann},
  M. 2009, \aap, 500, 681

\bibitem[{{Mewe} {et~al.}(1985){Mewe}, {Gronenschild}, \& {van den
  Oord}}]{1985A&AS...62..197M}
{Mewe}, R., {Gronenschild}, E.~H.~B.~M., \& {van den Oord}, G.~H.~J. 1985,
  \aaps, 62, 197

\bibitem[{{Nagai} {et~al.}(2007){Nagai}, {Vikhlinin}, \&
  {Kravtsov}}]{2007ApJ...655...98N}
{Nagai}, D., {Vikhlinin}, A., \& {Kravtsov}, A.~V. 2007, \apj, 655, 98

\bibitem[{Navarro {et~al.}(1997)Navarro, Frenk, \& White}]{NA97.1}
Navarro, J., Frenk, C., \& White, S. 1997, ApJ, 490, 493

\bibitem[{{Okabe} {et~al.}(2009){Okabe}, {Takada}, {Umetsu}, {Futamase}, \&
  {Smith}}]{2009arXiv0903.1103O}
{Okabe}, N., {Takada}, M., {Umetsu}, K., {Futamase}, T., \& {Smith}, G.~P.
  2009, ArXiv e-prints 0903.1103

\bibitem[{{Ota} {et~al.}(2004){Ota}, {Pointecouteau}, {Hattori}, \&
  {Mitsuda}}]{OT04.1}
{Ota}, N., {Pointecouteau}, E., {Hattori}, M., \& {Mitsuda}, K. 2004, \apj,
  601, 120

\bibitem[{{Percival} {et~al.}(2007){Percival}, {Cole}, {Eisenstein}, {Nichol},
  {Peacock}, {Pope}, \& {Szalay}}]{2007MNRAS.381.1053P}
{Percival}, W.~J., {Cole}, S., {Eisenstein}, D.~J., {et~al.} 2007, \mnras, 381,
  1053

\bibitem[{Perlmutter {et~al.}(1999)Perlmutter, Aldering, Goldhaber, Knop,
  {et~al.}}]{PE99.1}
Perlmutter, S., Aldering, G., Goldhaber, G., Knop, R., {et~al.} 1999, ApJ, 517,
  565

\bibitem[{{Piffaretti} \& {Valdarnini}(2008)}]{2008A&A...491...71P}
{Piffaretti}, R. \& {Valdarnini}, R. 2008, \aap, 491, 71

\bibitem[{{Pointecouteau} {et~al.}(2005){Pointecouteau}, {Arnaud}, \&
  {Pratt}}]{2005A&A...435....1P}
{Pointecouteau}, E., {Arnaud}, M., \& {Pratt}, G.~W. 2005, \aap, 435, 1

\bibitem[{Press \& Schechter(1974)}]{PR74.1}
Press, W. \& Schechter, P. 1974, ApJ, 187, 425

\bibitem[{{Puchwein} {et~al.}(2005){Puchwein}, {Bartelmann}, {Dolag}, \&
  {Meneghetti}}]{PU05.1}
{Puchwein}, E., {Bartelmann}, M., {Dolag}, K., \& {Meneghetti}, M. 2005, \aap,
  442, 405

\bibitem[{{Rasia} {et~al.}(2006){Rasia}, {Ettori}, {Moscardini}, {Mazzotta},
  {Borgani}, {Dolag}, {Tormen}, {Cheng}, \& {Diaferio}}]{2006MNRAS.369.2013R}
{Rasia}, E., {Ettori}, S., {Moscardini}, L., {et~al.} 2006, \mnras, 369, 2013

\bibitem[{{Rasia} {et~al.}(2008){Rasia}, {Mazzotta}, {Bourdin}, {Borgani},
  {Tornatore}, {Ettori}, {Dolag}, \& {Moscardini}}]{2008ApJ...674..728R}
{Rasia}, E., {Mazzotta}, P., {Bourdin}, H., {et~al.} 2008, \apj, 674, 728

\bibitem[{Rasia {et~al.}(2004)Rasia, Tormen, \& Moscardini}]{RA04.1}
Rasia, E., Tormen, G., \& Moscardini, L. 2004, MNRAS, 351, 237

\bibitem[{Reblinsky \& Bartelmann(1999)}]{RE99.1}
Reblinsky, K. \& Bartelmann, M. 1999, A\&A, 345, 1

\bibitem[{{Refregier}(2003)}]{RE03.1}
{Refregier}, A. 2003, \mnras, 338, 35

\bibitem[{{Riemer-S{\o}rensen} {et~al.}(2009){Riemer-S{\o}rensen}, {Paraficz},
  {Ferreira}, {Pedersen}, {Limousin}, \& {Dahle}}]{2009ApJ...693.1570R}
{Riemer-S{\o}rensen}, S., {Paraficz}, D., {Ferreira}, D.~D.~M., {et~al.} 2009,
  \apj, 693, 1570

\bibitem[{Riess {et~al.}(2004)Riess, Strolger, Tonry, Casertano,
  {et~al.}}]{RI04.1}
Riess, A., Strolger, L.-G., Tonry, J., Casertano, S., {et~al.} 2004, ApJ, 607,
  665

\bibitem[{Riess {et~al.}(1998)Riess, Filippenko, Challis, {et~al.}}]{RI98.1}
Riess, A.~G., Filippenko, A.~V., Challis, P., {et~al.} 1998, AJ, 116, 1009

\bibitem[{{Salpeter}(1955)}]{1955ApJ...121..161S}
{Salpeter}, E.~E. 1955, \apj, 121, 161

\bibitem[{{Sand} {et~al.}(2008){Sand}, {Treu}, {Ellis}, {Smith}, \&
  {Kneib}}]{2008ApJ...674..711S}
{Sand}, D.~J., {Treu}, T., {Ellis}, R.~S., {Smith}, G.~P., \& {Kneib}, J.-P.
  2008, \apj, 674, 711

\bibitem[{{Sarazin}(1988)}]{1988xrec.book.....S}
{Sarazin}, C.~L. 1988, {X-ray emission from clusters of galaxies}, ed. C.~L.
  Sarazin

\bibitem[{{Saro} {et~al.}(2006){Saro}, {Borgani}, {Tornatore}, {Dolag},
  {Murante}, {Biviano}, {Calura}, \& {Charlot}}]{2006MNRAS.373..397S}
{Saro}, A., {Borgani}, S., {Tornatore}, L., {et~al.} 2006, \mnras, 373, 397

\bibitem[{{Schmidt} \& {Allen}(2007)}]{2007MNRAS.379..209S}
{Schmidt}, R.~W. \& {Allen}, S.~W. 2007, \mnras, 379, 209

\bibitem[{Sheth \& Tormen(2002)}]{SH02.1}
Sheth, R. \& Tormen, G. 2002, MNRAS, 329, 61

\bibitem[{{Smith} {et~al.}(2005){Smith}, {Kneib}, {Smail}, {Mazzotta},
  {Ebeling}, \& {Czoske}}]{2005MNRAS.359..417S}
{Smith}, G.~P., {Kneib}, J.-P., {Smail}, I., {et~al.} 2005, \mnras, 359, 417

\bibitem[{{Springel}(2005)}]{SP05.1}
{Springel}, V. 2005, \mnras, 364, 1105

\bibitem[{{Springel} {et~al.}(2001){Springel}, {White}, {Tormen}, \&
  {Kauffmann}}]{2001MNRAS.328..726S}
{Springel}, V., {White}, S.~D.~M., {Tormen}, G., \& {Kauffmann}, G. 2001,
  \mnras, 328, 726

\bibitem[{Tormen {et~al.}(1997)Tormen, Bouchet, \& White}]{TO97.2}
Tormen, G., Bouchet, F., \& White, S. 1997, MNRAS, 286, 865

\bibitem[{{Tornatore} {et~al.}(2007){Tornatore}, {Borgani}, {Dolag}, \&
  {Matteucci}}]{2007MNRAS.382.1050T}
{Tornatore}, L., {Borgani}, S., {Dolag}, K., \& {Matteucci}, F. 2007, \mnras,
  382, 1050

\bibitem[{{Tornatore} {et~al.}(2004){Tornatore}, {Borgani}, {Matteucci},
  {Recchi}, \& {Tozzi}}]{2004MNRAS.349L..19T}
{Tornatore}, L., {Borgani}, S., {Matteucci}, F., {Recchi}, S., \& {Tozzi}, P.
  2004, \mnras, 349, L19

\bibitem[{{Vikhlinin} {et~al.}(2009){Vikhlinin}, {Kravtsov}, {Burenin},
  {Ebeling}, {Forman}, {Hornstrup}, {Jones}, {Murray}, {Nagai}, {Quintana}, \&
  {Voevodkin}}]{2009ApJ...692.1060V}
{Vikhlinin}, A., {Kravtsov}, A.~V., {Burenin}, R.~A., {et~al.} 2009, \apj, 692,
  1060

\bibitem[{{Vikhlinin} {et~al.}(2005){Vikhlinin}, {Markevitch}, {Murray},
  {Jones}, {Forman}, \& {Van Speybroeck}}]{2005ApJ...628..655V}
{Vikhlinin}, A., {Markevitch}, M., {Murray}, S.~S., {et~al.} 2005, \apj, 628,
  655

\bibitem[{{Warren} {et~al.}(2006){Warren}, {Abazajian}, {Holz}, \&
  {Teodoro}}]{2006ApJ...646..881W}
{Warren}, M.~S., {Abazajian}, K., {Holz}, D.~E., \& {Teodoro}, L. 2006, \apj,
  646, 881

\bibitem[{{White} \& {Vale}(2004)}]{2004APh....22...19W}
{White}, M. \& {Vale}, C. 2004, Astroparticle Physics, 22, 19

\bibitem[{Wu \& Mao(1996)}]{WU96.1}
Wu, X.-P. \& Mao, S. 1996, ApJ, 463, 404

\bibitem[{Yoshida {et~al.}(2001)Yoshida, Sheth, \& Diaferio}]{YO01.1}
Yoshida, N., Sheth, R., \& Diaferio, A. 2001, MNRAS, 328, 669

\bibitem[{{Yoshida} {et~al.}(2000){Yoshida}, {Springel}, {White}, \&
  {Tormen}}]{2000ApJ...544L..87Y}
{Yoshida}, N., {Springel}, V., {White}, S.~D.~M., \& {Tormen}, G. 2000, \apjl,
  544, L87

\bibitem[{{Zhang} {et~al.}(2008){Zhang}, {Finoguenov}, {B{\"o}hringer},
  {Kneib}, {Smith}, {Kneissl}, {Okabe}, \& {Dahle}}]{2008A&A...482..451Z}
{Zhang}, Y.-Y., {Finoguenov}, A., {B{\"o}hringer}, H., {et~al.} 2008, \aap,
  482, 451

\bibitem[{{Zhang} {et~al.}(2009){Zhang}, {Okabe}, {Finoguenov}, {Smith},
  {Piffaretti}, {Valdarnini}, {Babul}, {Evrard}, {Mazzotta}, {Sanderson}, \&
  {Marrone}}]{zhang09}
{Zhang}, Y.-Y., {Okabe}, N., {Finoguenov}, A., {et~al.} 2009, in prep

\end{thebibliography}
\bibliographystyle{aa}

\appendix
\section{Lensing definitions}
\label{sect:lensing}
In this Sect., we summarize some lensing definitions which will be useful in the rest of the paper. 

We start with an isolated
lens whose surface-mass density is $\Sigma(\vec\theta)$ at the angular
position $\vec\theta$ on the sky. Its lensing potential is
\begin{equation}
  \psi(\vec\theta)=
    \frac{4G}{c^2}\frac{D_{\rm l}D_{\rm s}}{D_{\rm ls}}
    \int\d^2\theta'\Sigma(\vec\theta')\,
    \ln\left|\vec\theta-\vec\theta'\right| \;,
\end{equation}
where $D_{\rm l,s,ls}$ are the usual angular-diameter distances
between the observer and the lens, the observer and the source, and
the lens and the source, respectively. The reduced deflection angle
experienced by a light ray crossing the lens plane at $\vec\theta$ is
the gradient of the potential,
\begin{equation}
  \vec\alpha(\vec\theta)=\vec\nabla\psi(\vec\theta) \;.
\label{eq:alpha}
\end{equation}
The image positions $\vec\theta$ for a source located at $\vec\beta$
are given by the lens equation
\begin{equation}
  \vec\beta=\vec\theta-\vec\alpha(\vec\theta) \;.
  \label{eq:leq}
\end{equation}

For sources much smaller than the typical scales on which the lens
properties vary, the lens mapping can be linearised. The deformation
of images with respect to the source is then given by the Jacobian
matrix
\begin{equation}
  \mathcal{A}\equiv
  \frac{\partial\vec\beta}{\partial\vec\theta}=
  \left(\delta_{ij}-
  \frac{\partial^2\psi(\vec\theta)}{\partial\theta_i\partial\theta_j}
  \right)=\left(
  \begin{array}{cc}
    1-\kappa-\gamma_1 & -\gamma_2 \\
    -\gamma_2 & 1-\kappa+\gamma_1 \\
  \end{array} \right) \;.
\label{eq:jacobian}
\end{equation} 
Here, $\kappa$ is the convergence
\begin{equation}
  \kappa(\vec\theta)=\frac{\Sigma(\vec\theta)}{\Sigma_{\rm cr}}=
  \frac{1}{2}\left(\psi_{11}+\psi_{22}\right) \;,
\label{eq:k}
\end{equation}
i.e.~the surface-mass density scaled by its critical value
\begin{equation}
  \Sigma_{\rm cr}=
  \frac{c^2}{4\pi G}\frac{D_{\rm ls}}{D_{\rm l}D_{\rm s}} \;.
\label{eq:sigmacr}
\end{equation}
The distortion is described by the two components of the shear, 
\begin{equation}
  \gamma_1=\frac{1}{2}\left(\psi_{11}-\psi_{22}\right)\;,\quad
  \gamma_2=\psi_{12} \;,
  \label{eq:g}
\end{equation}
which are combined into the complex shear $\gamma\equiv\gamma_1+{\rm
i}\gamma_2$. We further use the common abbreviation
\begin{equation}
  \frac{\partial^2 \psi(\vec{\theta})}{\partial \theta_i
    \partial \theta_j} \equiv \psi_{ij} \ .
\end{equation}

The inverse of the Jacobian determinant defines the magnification factor 
\begin{equation}
\mu(\vec\theta)=\frac{1}{\det \mathcal{A}(\vec\theta)}
\end{equation}
The loci on the image plane where $\det \mathcal{A}=0$, i.e. where the magnification diverges, are the tangential and the radial {\em critical lines}. These form where the tangential and the radial eigenvalues of the Jacobian determinant nullifies:
\begin{eqnarray}
\lambda_{\rm t} & = & 1-\kappa-\gamma = 0 \\
\lambda_{\rm r} & = & 1-\kappa+\gamma = 0 \;. 
\end{eqnarray}
The inverses of the eigenvalues of the Jacobian matrix define the tangential and the radial magnifications: images forming close to the tangential or to the radial critical lines are highly elongated tangentially or perpendicularly to the critical lines, respectively.

The critical lines are mapped onto the {\em caustics} on the source plane, via the lens equation. The caustics encompass regions of the source plane which are characterized by different multiplicities of the images. 

Outside critical curves, image ellipticities are determined by the
complex reduced shear
\begin{equation}
  g(\vec\theta)\equiv
  \frac{\gamma(\vec\theta)}{1-\kappa(\vec\theta)} \;.
\label{eq:rg}
\end{equation}
In the weak-lensing limit, $\kappa\ll1$, and the reduced shear
approximates the shear, $g\approx\gamma$, to first order.

Source and image shapes are quantified by the complex ellipticity
\begin{equation}
  \epsilon=\frac{a-b}{a+b}{\rm e}^{2{\rm i}\vartheta}\;,
\end{equation}
where $a$ and $b$ are the semi-major and semi-minor axes of an ellipse
fitting the object's surface-brightness distribution. The position
angle of the ellipse's major axis is $\vartheta$. The expectation
value of the intrinsic source ellipticity $\epsilon_{\rm s}$ is
assumed to vanish.

A sufficiently small source with ellipticity $\epsilon_{\rm s}$ is
imaged to have an ellipticity
\begin{equation}
  \epsilon=\frac{\epsilon_{\rm s}+g}{1+g^*\epsilon_\mathrm{s}}\;,
\label{eq:ell}
\end{equation}
where the asterisk denotes complex conjugation. This equation
illustrates that the lensing distortion is determined by the reduced
shear, which is the only lensing quantity directly accessible through
measurements of galaxy ellipticities alone.

{ 
\section{Individual mass profiles}
\label{sect:indmasses}
In this Sect. we show the 3D-mass profiles (normalized to the true mass profiles) of each individual cluster projection in our sample. This allows the reader to compare the mass estimates at any radius different from $R_{2500}$, $R_{500}$, and $R_{200}$, which are already discussed in the paper. The profiles are displayed in Fig.~\ref{fig:indmasses} for all the lensing and the X-ray methods. Additionally, we also show the mass profiles derived from the true gas and temperature profiles under the assumption of hydrostatic equilibrium. These last curves are quite noisy, especially in the case of $g72$, because we have not applied any smoothing to the data. In the case of $g51$ the extended cooling region in the cluster center causes the hydrostatic mass $M_{\rm HEQ}$ to be larger than the true mass. The X-ray derived profiles reproduce well the trends seen in the $M_{\rm HEQ}$ profiles although some differences are present. Note that, while we mask many cold blobs in the simulated observations, we are not removing these structures from the input clusters when calculating $M_{\rm HEQ}$. Thus, some differences between the profiles are expected.

\begin{figure*}[t!]
  \includegraphics[width=0.33\hsize]{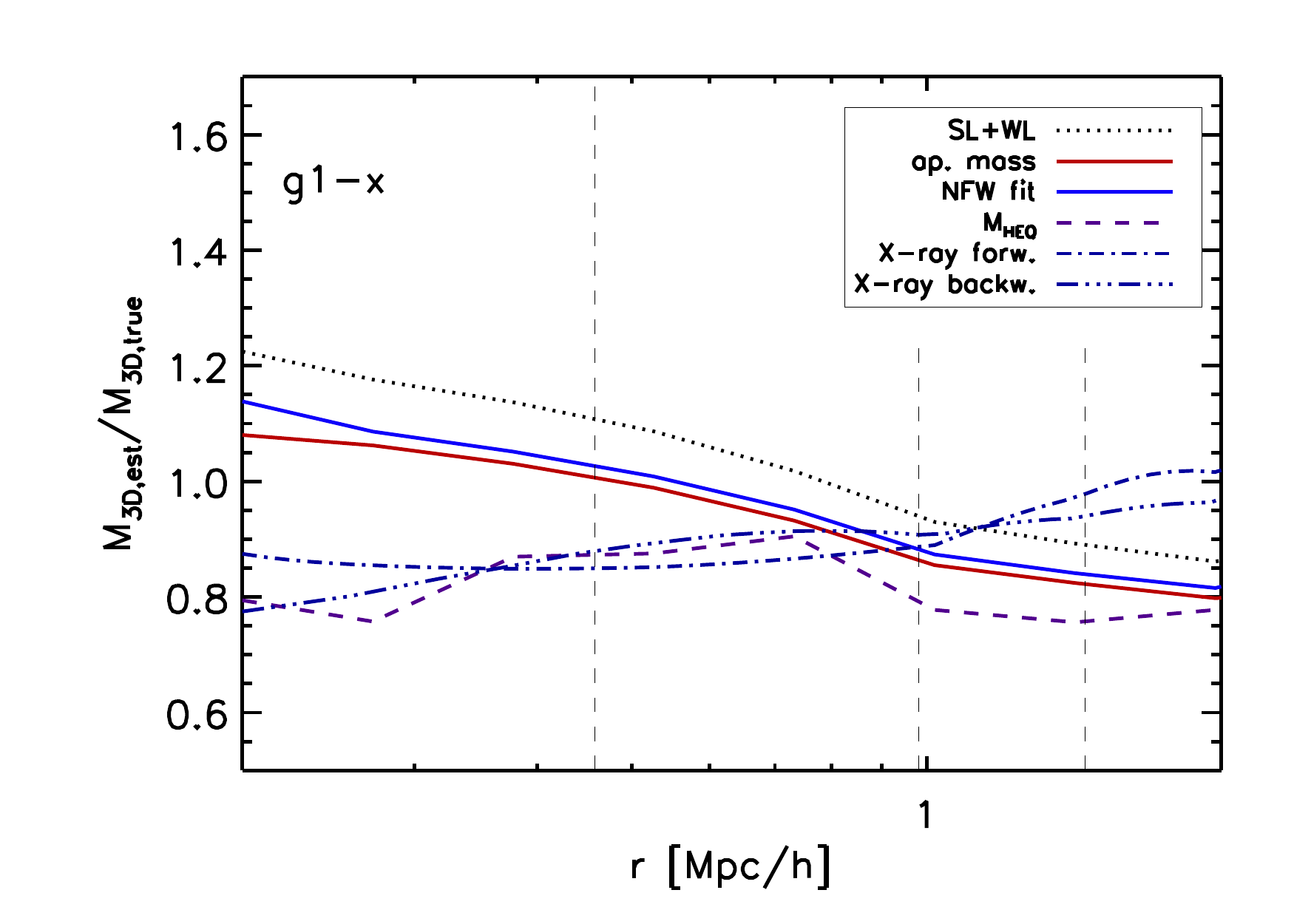}
  \includegraphics[width=0.33\hsize]{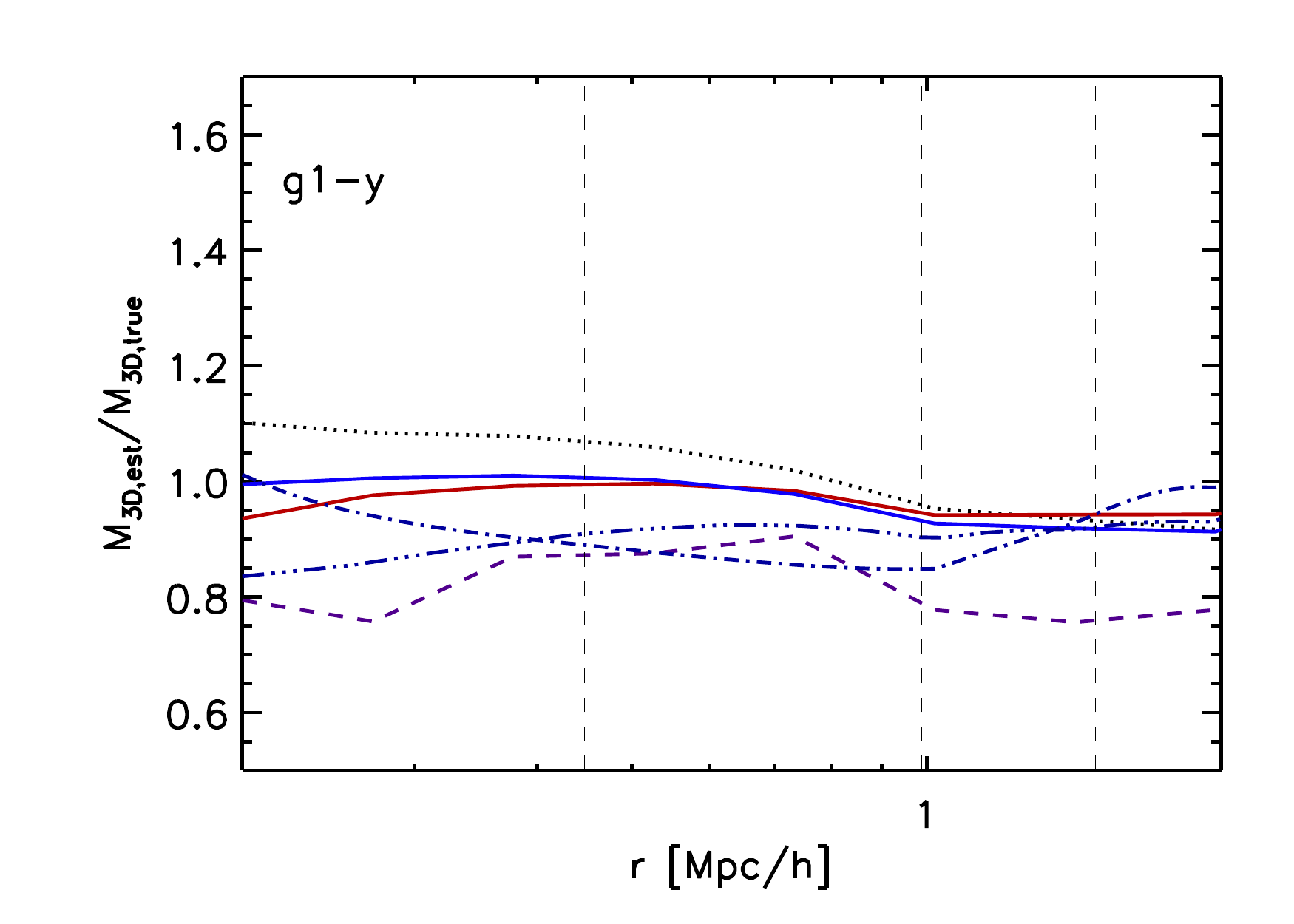}
  \includegraphics[width=0.33\hsize]{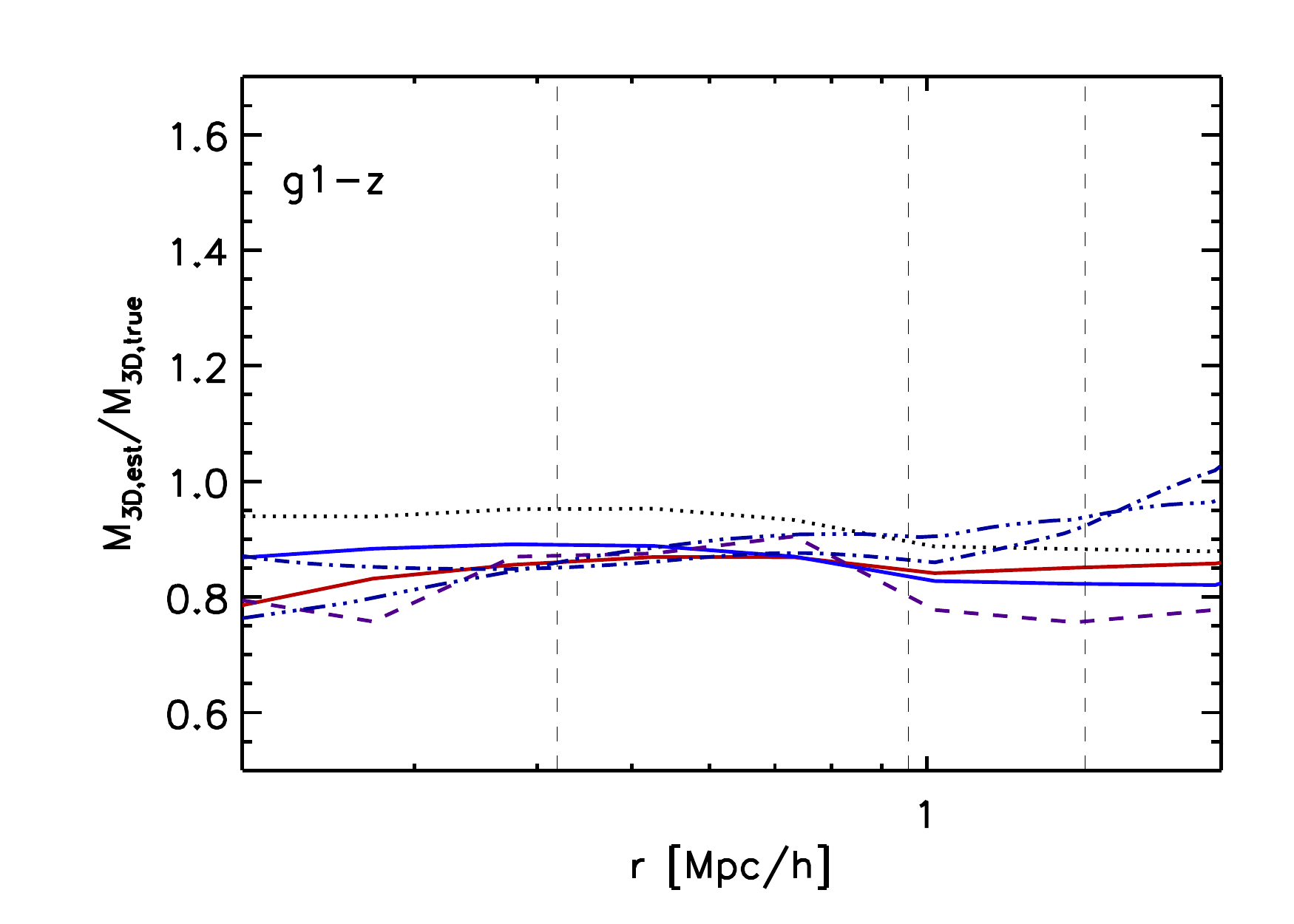} 
  \includegraphics[width=0.33\hsize]{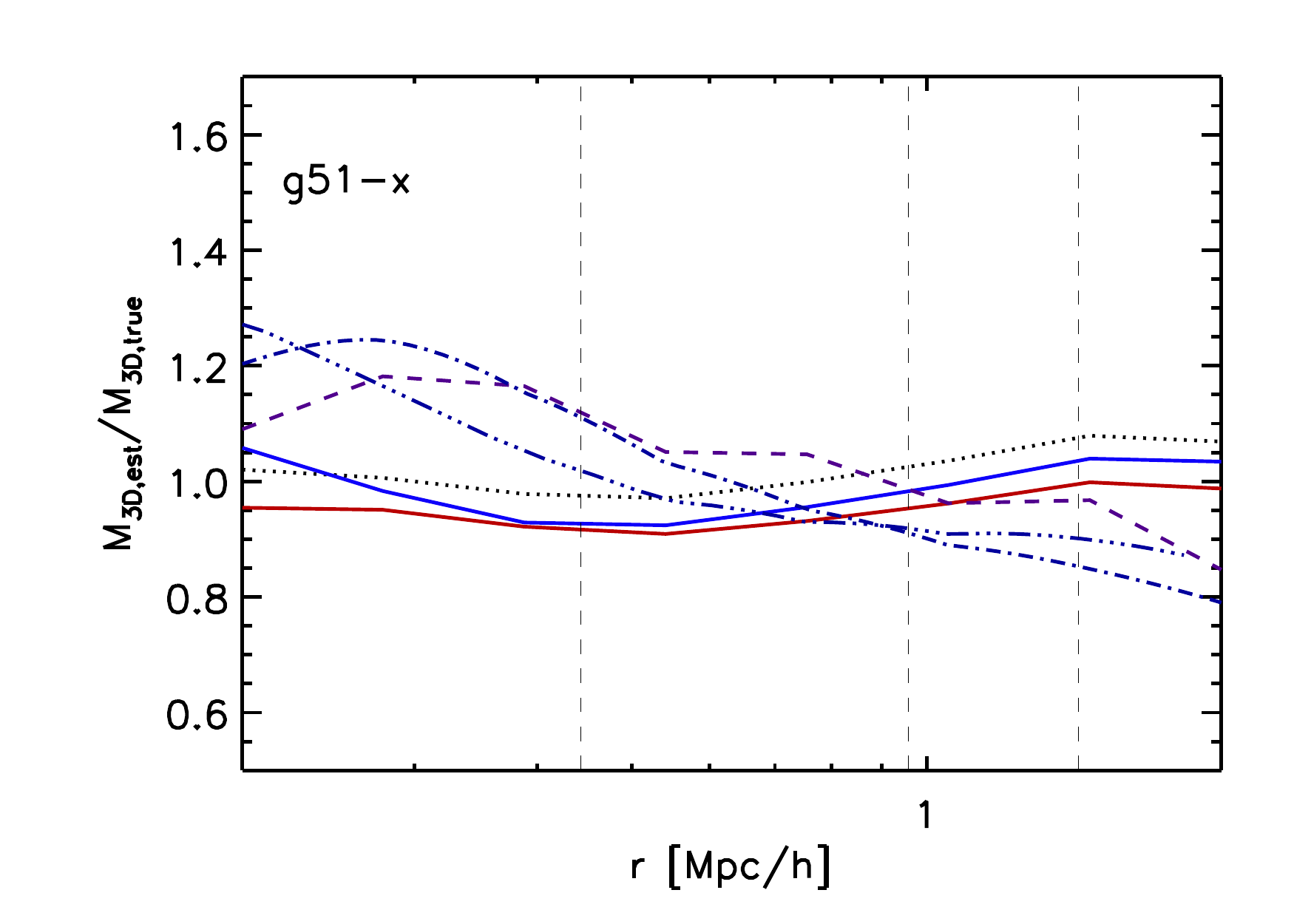}
  \includegraphics[width=0.33\hsize]{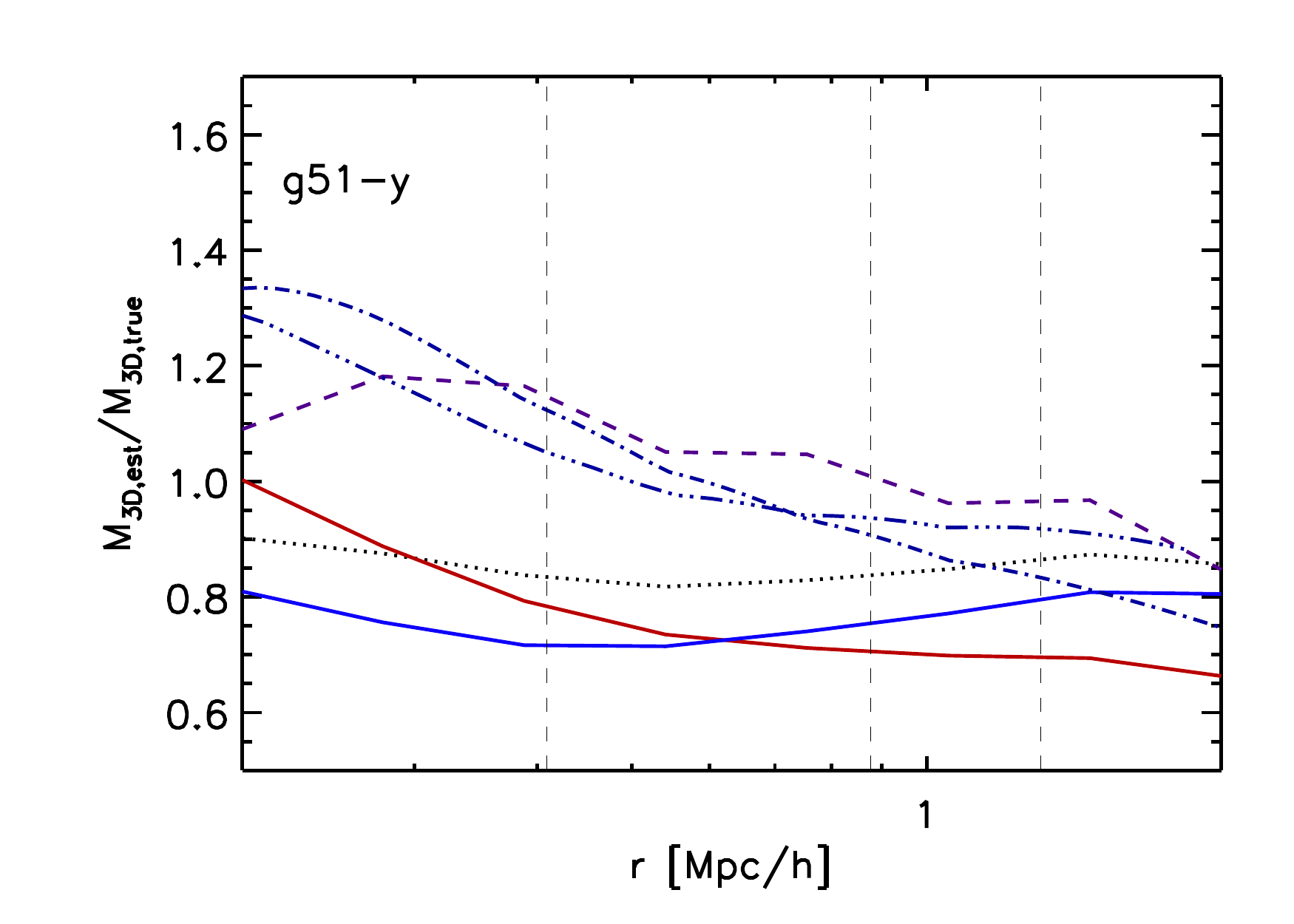}
  \includegraphics[width=0.33\hsize]{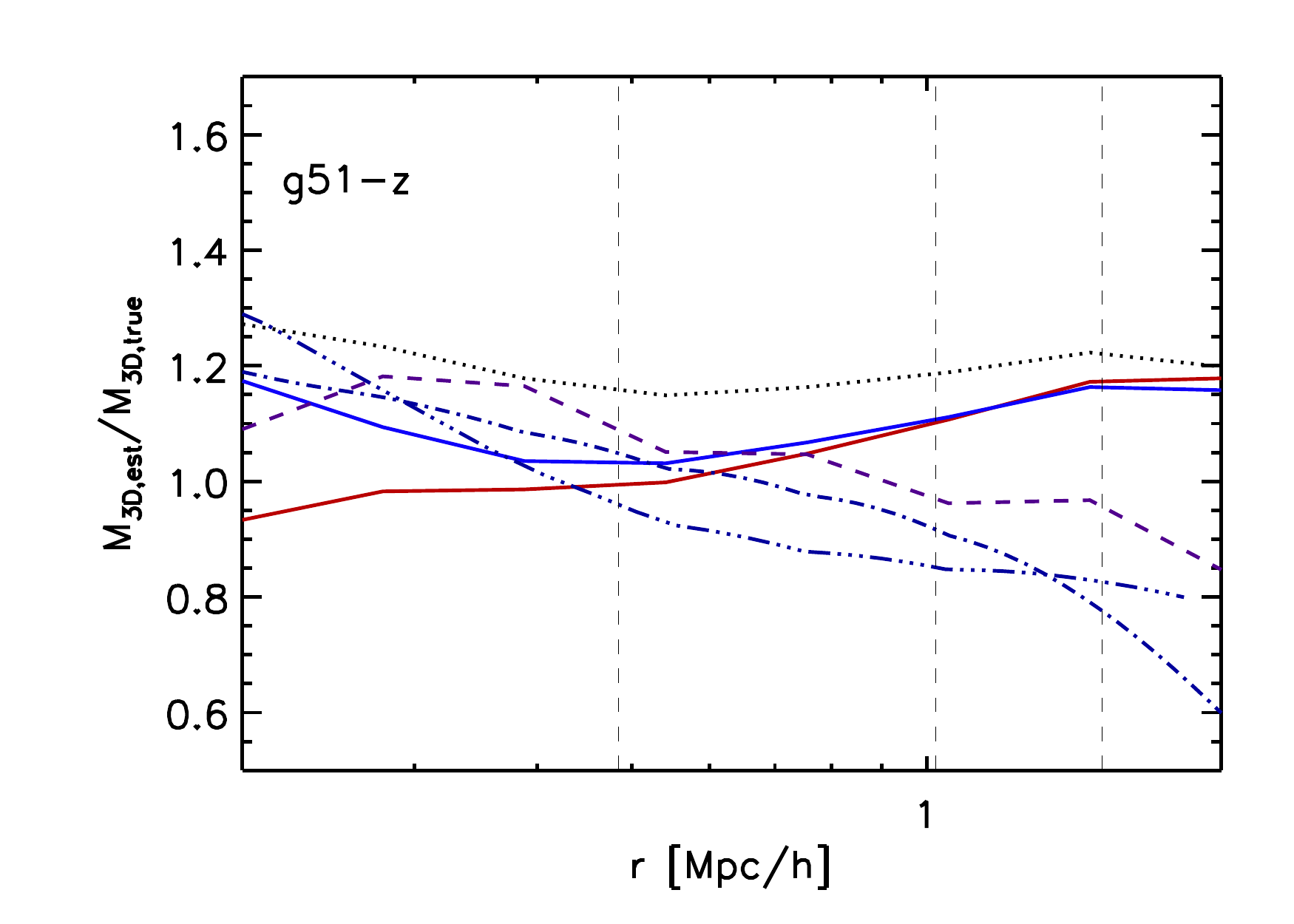}	
  \includegraphics[width=0.33\hsize]{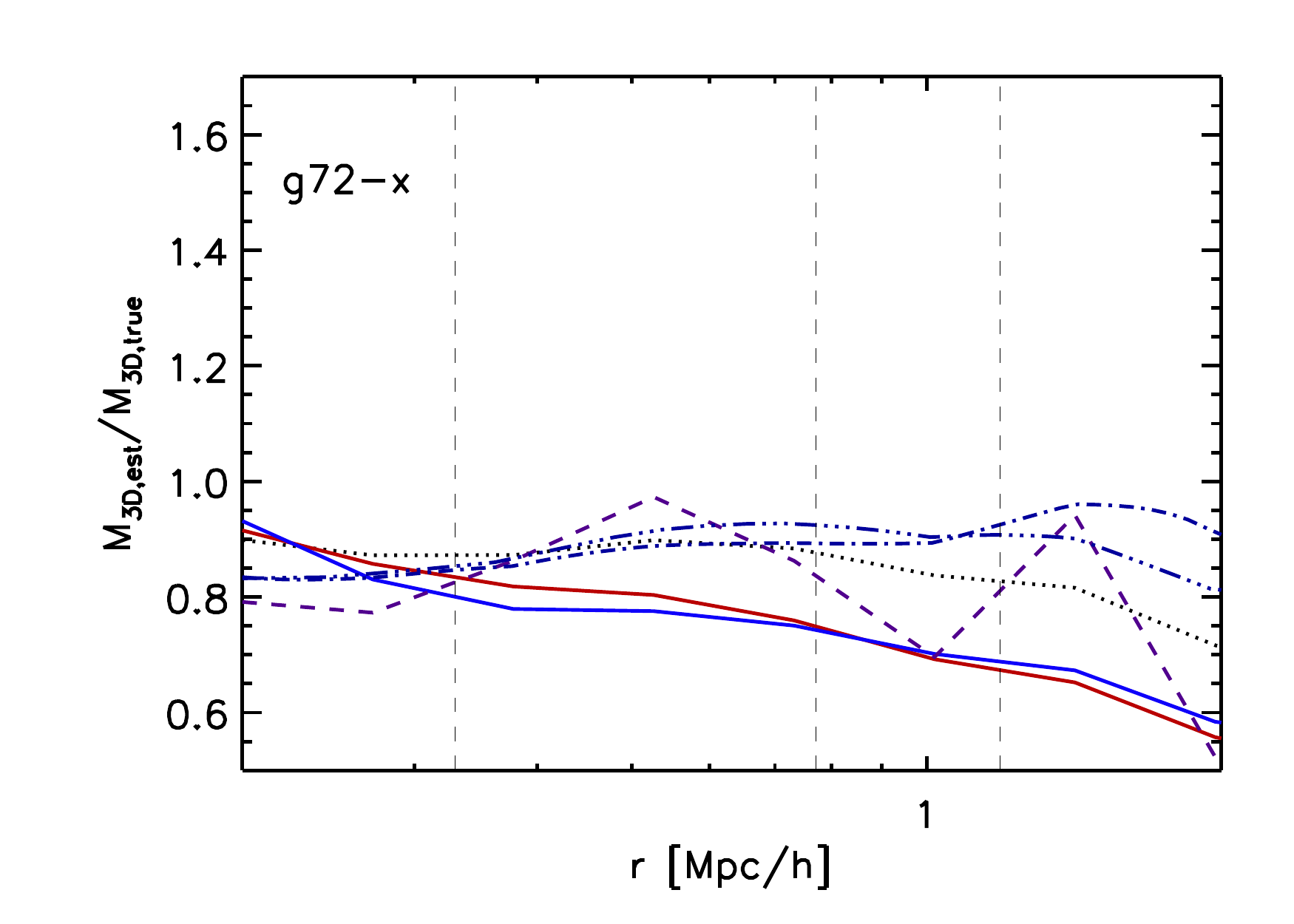}
  \includegraphics[width=0.33\hsize]{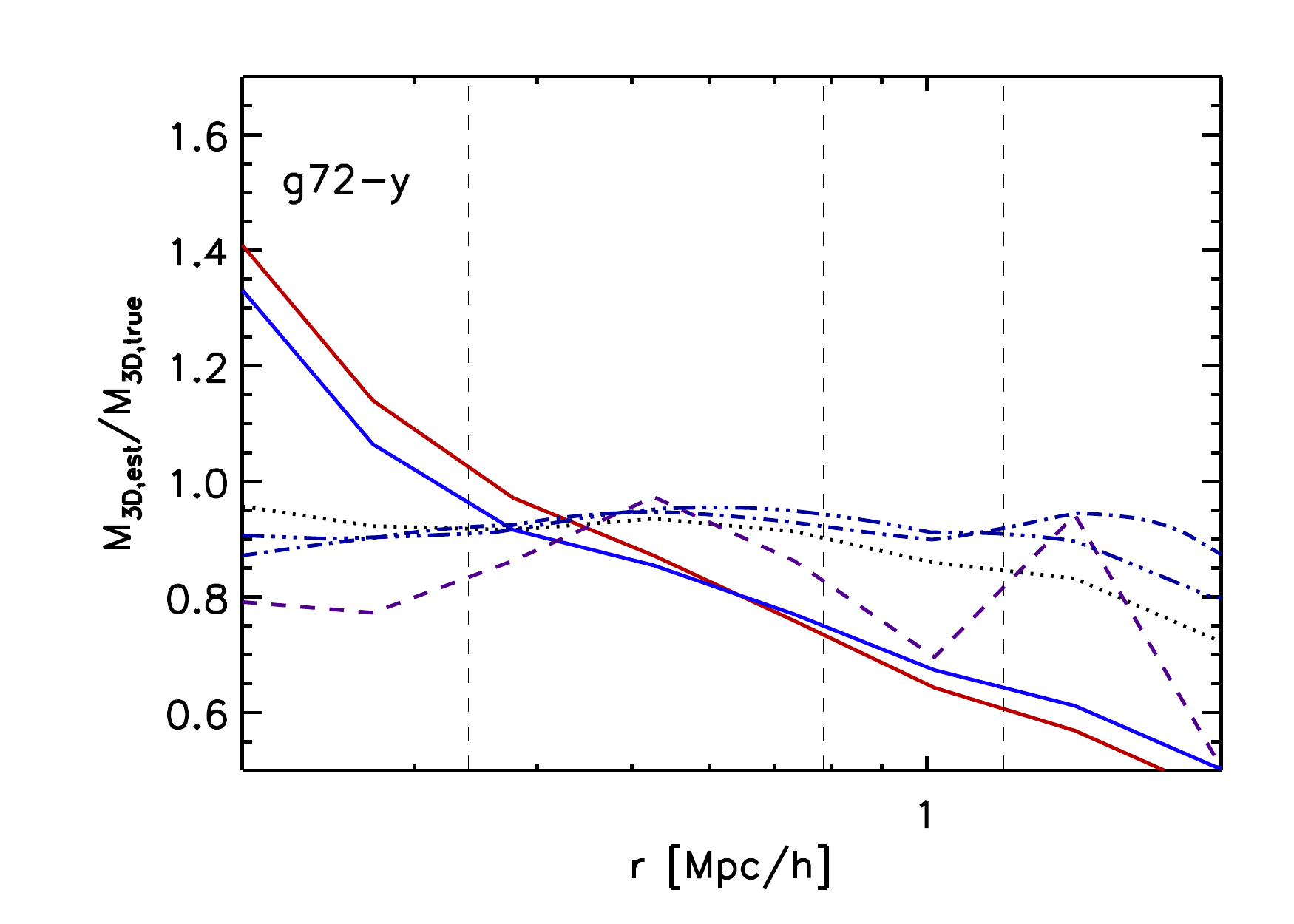}
  \includegraphics[width=0.33\hsize]{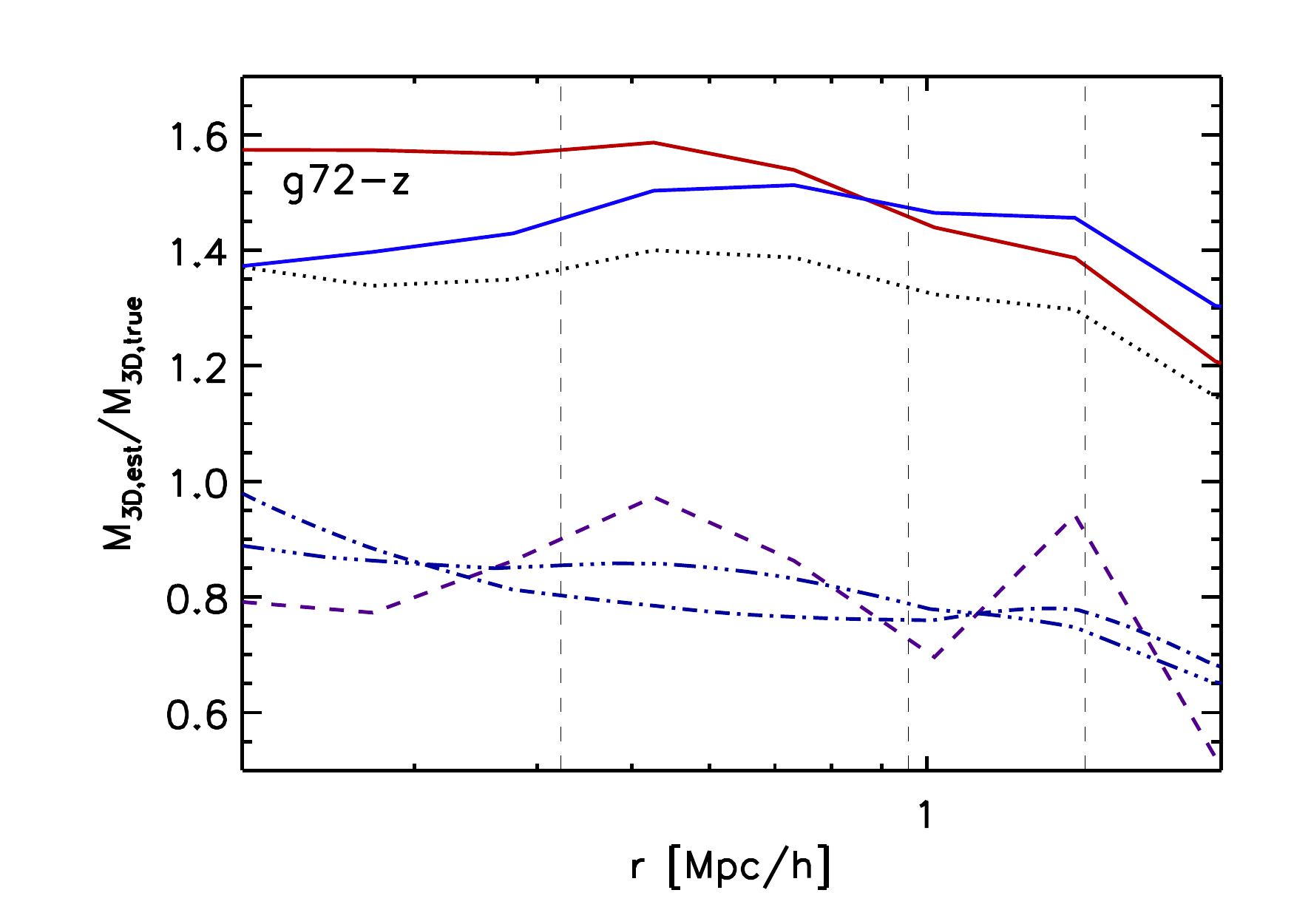}  
  \caption{Ratios between the mass profiles recovered from the lensing and the X-ray analyses and the true 3D mass profiles of each cluster. The meaning of each line is explained in the legend. The vertical dashed lines in each panel mark the positions of $R_{2500}$, $R_{500}$, and $R_{200}$, as derived from the lensing analysis (see text for more details).}
\label{fig:indmasses}
\end{figure*} 
}

\end{document}